\documentclass[traditabstract]{aa} 
\usepackage{graphicx}
\usepackage{txfonts}
\usepackage{natbib}
\newcommand{\reduceme}{\mbox{R\raisebox{-0.35ex}{E}D%
\hspace{-0.05em}\raisebox{0.85ex}{uc}\hspace{-0.90em}%
\raisebox{-.35ex}{{m}}\hspace{0.05em}E}}

\begin{document}
   \title{Formation and evolution of dwarf early-type galaxies in the Virgo cluster}

   \subtitle{I. Internal kinematics}

   \author{E. Toloba \inst{1} \and
    A. Boselli \inst{2} \and
  A. J. Cenarro \inst{3} \and
   R. F. Peletier \inst{4} \and
   J. Gorgas \inst{1} \and
   A. Gil de Paz \inst{1} \and
  J. C. Mu\~{n}oz-Mateos \inst{1}\inst{5}}
   \institute{Departamento de Astrof\'{i}sica y CC. de la Atm\'{o}sfera, Universidad Complutense de Madrid,
              28040, Madrid, Spain\\
              \email{etj@astrax.fis.ucm.es} 
              \email{jgorgas@fis.ucm.es}
              \email{gildepaz@gmail.com}
              \email{jcmunoz@astrax.fis.ucm.es} \and
          Laboratoire d'Astrophysique de Marseille, UMR 6110 CNRS, 38 rue F. Joliot-Curie, F-13388 Marseille, France\\
               \email{alessandro.boselli@oamp.fr} \and
Centro de Estudios de F\'isica del Cosmos de Arag\'on, E-44001, Teruel, Spain \\
\email{cenarro@cefca.es} \and
Kapteyn Astronomical Institute, Rijksuniversiteit Groningen, Postbus 800, 9700 AV Groningen, the Netherlands\\
\email{peletier@astro.rug.nl} \and
National Radio Astronomy Observatory, 520 Edgemont Road, Charlottesville, VA 22903-2475\\
             }

   \date{Received ; accepted }

 
\abstract{We present new medium resolution kinematic data for a
sample of 21 dwarf early-type galaxies (dEs) mainly in the Virgo cluster, obtained with the WHT and INT telescopes at the Roque de los Muchachos
Observatory (La Palma, Spain). These data are used to study the origin of the
dwarf elliptical galaxy population inhabiting clusters. We confirm that
dEs are not dark matter dominated galaxies, at least up to the half-light radius.
We also find that the
observed galaxies in the outer parts of the cluster are mostly rotationally
supported systems with disky morphological shapes. Rotationally supported dEs have rotation curves similar to
those of star forming galaxies of similar luminosity and follow the Tully-Fisher relation. This is expected if dE
galaxies are the descendant of low luminosity star forming systems which
recently entered the cluster environment and lost their gas due to a ram
pressure stripping event, quenching their star formation activity and
transforming into quiescent systems, but conserving their angular
momentum.}


            
\keywords{Galaxies: clusters: individual: Virgo
                Galaxies: dwarf
                Galaxies: elliptical and lenticular, cD
                Galaxies: kinematics and dynamics
                Galaxies: evolution
               Galaxies: dark matter}

   \maketitle
%

\section{Introduction}

The processes involved in galaxy formation and evolution through cosmic time are still poorly understood. It is indeed still unclear how matter assembled to form the present day galaxy population, whether it followed a passive evolution after the collapse of the primordial density fluctuations (secular evolution),  through a subsequent merging of growing structures (hierarchical formation) or a combination of the two. A way of quantifying the relative role of these different mechanisms
is to study dwarf galaxies, the most numerous objects in the universe
\citep{FB94}. Their importance resides in the fact that these low-luminosity systems are expected to be the building blocks of massive galaxies in lambda cold dark matter ($\Lambda$CDM) hierarchical merging scenarios
\citep[e.g.,][]{WR78,WF91}.

Among dwarf galaxies,  quiescent dwarfs (which we here define to be all quiescent galaxies with $M_B$ $>$ $-$18, including both dwarf ellipticals and
spheroidals, hereafter indicated as dEs) are of particular interest since they are the most
numerous population in clusters  \citep{FB94}. These objects were originally
thought to be the low luminosity extension of giant ellipticals (Es).
Since the 1990s it is known that 
dEs are composed of several families of objects (e.g. compact and low surface brightness dwarfs) (\citet{Bender92}; \citet{Kormendy09}. 
Later on, it was shown that dEs were no longer small Es with simple, old and
metal-poor stellar populations, but much more complex objects exhibiting a
wide range of stellar contents. For example, in the Virgo cluster, they
have stellar populations ranging from very young (around 1 Gyr old) luminosity-weighted ages to as
old as the oldest Es galaxies (14 Gyr) \citep{Mich08}. Their proximity allowed detailed studies of their structural properties which 
indicated
that, behind their elliptical appearance, dEs show a great variety of
underlying structures, like discs, spiral arms, irregular features, etc, making them a very heterogeneous class of galaxies
\citep{Lisk06b,Lisk06a,Lisk07}.

These evidences indicate a complex formation process shaping the evolution of dEs in clusters. Two main different processes have been proposed in the literature: the first  mechanism is based on the idea that dEs are formed through 
internal processes, like supernova feedback, where the
interstellar medium (ISM) of the progenitor star forming galaxy is swept away by the kinetic pressure generated
by supernovae \citep{YosAri87}, although it seems highly unlikely in
dark-matter dominated systems \citep{ST01}; the second mechanism rests upon
external processes induced by the interaction with the hostile environment in which dEs
reside \citep{Sand85,Blant05}. In a dense environment several mechanisms are affecting
galaxies. This might happen through interactions with the intergalactic medium (IGM), as
ram-pressure stripping \citep{Boselli08a,Boselli08b}, galaxy-galaxy interactions
\citep[e.g.,][]{ByrdValt90} and galaxy
harassment \citep[e.g.,][]{Moore98,Mast05}. It has been shown that all these interactions are able to reproduce
some of the observational properties of local dEs in clusters, like their structural parameters \citep{Lisk06a,Lisk07} or their stellar populations \citep{Geha02,Geha03,VZ04,Mich08,Paudel2010}, but none of them have been tested versus the kinematic properties. In this regard, whereas in the harassment scenario the system is rapidly heated, leading to an  increase of the velocity dispersion of the galaxy and a decrease of its rotation \citep{Mast05}, in a ram-pressure stripping event the angular momentum is conserved \citep{Boselli08a,Boselli08b}.

With the aim of using kinematic data to constrain dwarf galaxy evolution, we recently started an ambitious observational program at the Observatory El Roque de los Muchachos (La Palma, Spain)  to gather medium resolution spectroscopic data of dEs in the Virgo cluster. In this paper we present a detailed analysis of the internal kinematics focusing our attention into the most rapidly rotating systems. In \citet{etj09b} we have studied the kinematics as a function of local environment inside the Virgo cluster.  Several typical scaling relations of pressure supported systems, such as the Fundamental Plane as well as their stellar population properties will be analysed in a forthcoming communication.

This paper is structured as follows: in Sections 2, 3 and 4 we describe the sample selection, the observations and the
data reduction process. In Section 5 we report the kinematic measurements
paying special attention to the method employed and the stellar templates used. We
also describe the procedure followed to obtain the radial kinematic profiles
(Section 5.1), the central velocity dispersion and the
maximum rotational speed of the selected galaxies (Section 5.2), making comparisons with previous works (Section 5.3). Combined with photometric data (Section 6), the present kinematic observations are used to study the properties of rotationally supported systems in the framework of various models of galaxy evolution (Sections 7, 8 and 9).


\section{The sample}

The sample analysed in this work is composed of galaxies  with $M_r> -16$ classified as dE or dS0  in the Virgo Cluster Catalog (VCC) by \citet{Bing85}. All galaxies have been selected to have SDSS imaging and to be within the GALEX MIS fields \citep{Boselli05}, thus to have a measured UV magnitude or an upper limit. To these we added a few field quiescent dwarfs (originally used as fillers in our observing runs) useful for comparison in a statistical study. Out of the 43 Virgo galaxies satisfying these requirements in the VCC, 18 have been observed for this work. To make the observations accessible to 2.5-4.2m telescopes, we chose those objects with the highest surface brightness.

The field sample consists of early-type dwarfs in low density regions with magnitudes between $-$18.5 $<M_r<$ $-$14.5 and distances similar to Virgo (375 km$~$s$^{-1}$$<v<$ 1875 km$~$s$^{-1}$, 5-25 Mpc). Quiescent objects have been selected assuming the colour criterion FUV-NUV $>$ 0.9 or $u-g>$ 1.2 when UV detections were not available. We observed only 3 out of 10 field dEs candidates.
To these 18 Virgo and 3 field dEs we added M32, selected to test the setup of the instruments. Thanks to the large amount of available data, M32 is also an ideal target for comparison with other data available in the literature. 

Although we can consider it representative of the bright end of the Virgo Cluster dE population, the observed sample is not complete in any sense.

\section{Observations}

The observing time that we obtained for this work was part of the International Time Program (ITP 2005-2007) at El Roque de los Muchachos Observatory. Here
we focus on the medium resolution ($R\simeq3800$), long-slit spectroscopy carried out during three observing runs. In runs 1 and 3 (December 2005, February 2007) we used the ISIS double-arm spectrograph at the 4.2m WHT, and in run 2 (January 2007) we used the IDS spectrograph at the INT (2.5m telescope).

The advantage of ISIS over IDS is that it allows us to use a dichroic ({\it 5300 dichroic} in our case)  to split the light into two beams to observe simultaneously two wavelength ranges, one in the blue optical part of the spectra and another in the red. This technique allowed us to cover, in 3 settings in the first run, the full wavelength range from 3500 \AA~ to 8950 \AA, using a mirror to cover 5000-5600  \AA, the only range that we could not cover with this dichroic. In the third run we used 2 settings to cover the same wavelength range except for the dichroic gap. 

The wavelength range covered by the IDS was smaller (4600-5960 \AA), since detector and grating are the same as on the blue arm of ISIS, the data obtained had similar resolution. The spectral resolution ($R\simeq3800$) is high enough to obtain reliable kinematics for dwarf galaxies.

All the details of the configurations used in each run are specified in Table ~\ref{t1}.

\begin{table*}
\begin{center}
\caption{Observational configurations\label{t1}}
\begin{tabular}{|c|c|c|c|c|c|}
\hline \hline
   & \multicolumn{2}{c|}{Run 1} & Run 2 & \multicolumn{2}{c|}{Run 3} \\
\hline 
Date       & \multicolumn{2}{c|}{Dec.24-27 2005} & Jan.21-23 2007 & \multicolumn{2}{c|}{Feb.10-12 2007} \\
Telescope  & \multicolumn{2}{c|}{WHT 4.2m}       & INT 2.5m  & \multicolumn{2}{c|}{WHT 4.2}      \\
Spectrograph & \multicolumn{2}{c|}{ISIS}      & IDS       & \multicolumn{2}{c|}{ISIS}      \\
Detector        & EEV12(blue)  & Marconi(red)& EEV10  & EEV12(blue) & RedPlus(red) \\
Grating              & R1200B(blue) & R600R (red) & R1200B & R1200B(blue)& R600R(red) \\
Wavelength range 1 (\AA) & 3500-4300     & 5500-6700 & 3700-4790 & 3500-4300 & 5500-6700  \\
Wavelength range 2 (\AA) & 4100-4900     & 7750-8950 & 4600-5690 & 4100-4900 & 7750-8950  \\
Wavelength range 3 (\AA) & 4800-5600     & ---             & ---             & ---             & ---               \\
Dispersion (\AA/pixel)     &   0.44              &  0.87           &    0.48          &         0.44    &   0.97          \\
Spectral Resolution (FWHM, \AA) & 1.56 & 3.22  & 1.80 & 1.56 & 3.23  \\
Instrumental Resolution (km s$^{-1}$) & 40 & 58  & 46 & 40 & 58  \\
Spatial scale ($"/$pix)&0.40 &  0.44 &  0.40 & 0.40 & 0.44 \\
Slit width ($"$)& 1.95  & 1.95  & 1.94  &  1.91 & 1.91 \\
\hline
\end{tabular}
\end{center}
\end{table*}

In Table \ref{t2} we list the observed sample. Column 5 presents the morphological type classification according to \citet{Lisk06b} and \citet{Lisk06a}: dE(di) indicates dwarf ellipticals with a certain, probable or possible underlying disk (i.e. showing spiral arms, edge-on disks and$/$or a  bar) or other structures (such as irregular central features (VCC21)); dE(bc) refers to galaxies with a blue center; dE to galaxies with no evident underlying structure. Four out of our 21 dwarf galaxies were not in the Lisker et al. sample (NGC3073, PGC1007217, PGC1154903 and VCC1947), therefore we classified them as described in Section 6 attending only to their boxyness/diskyness. Column 6 gives the Virgo substructure to which the galaxy belongs, taken from GOLDMine Database \citep{GOLDMine} and defined as in \citet{Gavazzi99}. Columns 7 and 8 refer to the observational campaign (see Table \ref{t1}) and the exposure time for each setting, which allowed us to get a typical signal-to-noise ratio\footnote{ The S$/$N per $\AA^{-1}$ is obtained dividing the S$/$N per pixel by the square root of the spatial scale along the slit. These measurement is therefore independent of the instrument used.} for the central spectra of $\sim$60  $\AA^{-1}$, enough to obtain reliable central kinematics. The galaxies were observed along their major axis. Their position angles (PA) from the HyperLEDA Database \citep{Paturel03} are given in column 9. Column 10 gives the Galactic colour excess from \citet{Schlegel98}.

A total of 37 B to M stars in common with the MILES library \citep{SB06lib} and the CaT library \citep{cen01} were observed to flux-calibrate our data and to use them as templates for velocity dispersion measurements.

\begin{table*}
\begin{center}
\caption{The observed galaxies. \label{t2}}
\begin{tabular}{|c|c|c|c|c|c|c|c|c|c|c|c|c|}
\hline \hline
Galaxy     & Other name  &  RA(J2000)  & Dec.(J2000)  &  Type  & Env. & Run &$\rm{t_{exp}}$ & PA & E(B-V)\\
               &                      &  (h:m:s)      & ($^{\circ}$:':'') &            &        &         &  (sec)    & ($^{\circ}$) & (mag) \\
\hline
M 32                & NGC 221   & 00:42:41.84&+40:51:57.4& cE2       & M31 group    & 1 &1350& 170 & 0.062\\
NGC 3073        &UGC 05374&10:00:52.10&+55:37:08.0& dE(di)     & Field              & 1 &3200& 120 & 0.010 \\
PGC 1007217   &2MASX J02413514-0810243&02:41:35.8&-08:10:24.8& dE(di) &Field &1 &3600& 126 &0.024\\
PGC 1154903   &2MASX J02420036+0000531&02:42:00.3&+00:00:52.3& dE &Field&1 &3600& 126 &0.031\\
VCC 21            &IC 3025     &12:10:23.14&+10:11:18.9 &  dE(di,bc)& Virgo N Cloud & 2 &3600&  99 & 0.021 \\
VCC 308          &IC 3131     &12:18:50.77&+07:51:41.3 &  dE(di,bc)& Virgo B Cluster& 3 &2400& 109 & 0.021\\
VCC 397          &CGCG 042-031&12:20:12.25&+06:37:23.6&dE(di) & Virgo B Cluster& 3 &3600& 133 & 0.020\\
VCC 523          &NGC 4306 &12:22:04.13&+12:47:15.1 &  dE(di)     & Virgo A Cluster& 1 &3400& 144 & 0.044\\
VCC 856          &IC 3328     &12:25:57.93&+10:03:13.8&   dE(di)     & Virgo B Cluster& 2 &2740&  72  & 0.024\\
VCC 917          &IC 3344     &12:26:32.40&+13:34:43.8&   dE          & Virgo A Cluster& 3 &3600&  57  & 0.032\\
VCC 990          &IC 3369     &12:27:16.91&+16:01:28.4&   dE(di)     & Virgo A Cluster& 2 &3000& 135 & 0.028\\
VCC 1087        &IC 3381     &12:28:17.88&+11:47:23.7&   dE          & Virgo A Cluster& 3 &3600&  106 & 0.026\\
VCC 1122        &IC 3393     &12:28:41.74&+12:54:57.3&   dE          & Virgo A Cluster& 3 &3600&  132 & 0.021\\
VCC 1183        &IC 3413     &12:29:22.49&+11:26:01.8&   dE(di)     & Virgo A Cluster& 2 &3600&  144 & 0.031\\
VCC 1261        &NGC 4482 &12:30:10.35&+10:46:46.3&   dE          & Virgo A Cluster& 2 &6930&  133 & 0.028\\
VCC 1431        &IC 3470     &12:32:23.39&+11:15:47.4&   dE          & Virgo A Cluster& 2 &3000&  135 & 0.054\\
VCC 1549        &IC 3510     &12:34:14.85&+11:04:18.1&   dE          & Virgo A Cluster& 2 &3300&    13 & 0.030\\
VCC 1695        &IC 3586     &12:36:54.79&+12:31:12.3&   dE(di)     & Virgo A Cluster& 3 &3600&    39 & 0.045\\
VCC 1861        &IC 3652     &12:40:58.60&+11:11:04.1&   dE          & Virgo E Cloud   & 3 &3600&  109 & 0.030\\
VCC 1910        &IC 809       &12:42:08.68&+11:45:15.9&   dE(di)     & Virgo E Cloud   & 2 &3800&  135 & 0.030\\
VCC 1912        &IC 810       &12:42:09.12&+12:35:48.8&   dE(bc)    & Virgo E Cloud   & 2 &3600&  166 & 0.032\\
VCC 1947        &CGCG 043-003&12:42:56.36&+03:40:35.6& dE(di)& Virgo S Cloud & 2 &3060& 126 & 0.027\\
\hline
\end{tabular}
\end{center}
\end{table*}

\section{Data reduction}

The data reduction was performed with \reduceme~ \citep{Car99}, a package specially designed to reduce long-slit spectroscopy with particular attention to the treatment of errors. This package is ideal for treating in parallel the data and error frames, producing an error spectrum associated with each individual data spectrum, which means that the errors are controlled at all times.

Due to the similar instrumental configurations used on all observing runs, the reduction process for both telescopes was the same.
The standard procedure for long-slit spectroscopy data reduction consists of  bias and dark current subtraction, flat-fielding (using observations of tungsten lamps and twilight sky to correct for high and low frequency variations respectively), cosmic ray cleaning, C-distortion correction, wavelength calibration, S-distortion correction, sky subtraction, atmospheric and interstellar extinction correction and flux calibration. We give below some comments on steps of particular importance:

{\it{Flat-fielding.}}
The flat-fielding correction is a delicate step at near infrared wavelengths due to the fringing effects. In the first run, the Marconi CCD suffered from significant fringing that varied with the telescope position.  Since complete removal of the fringing in run 1 was not possible, we did not use the red Marconi-CCD data to determine the galaxy kinematics. The fringing produced by RedPlus, the new CCD optimised to avoid these patterns, was much lower, with an amplitude of only $\sim$1$\%$  independent of position.

{\it{Wavelength calibration.}}
The wavelength calibration was performed using between 65-100 arc lines depending on the instrumental configuration. They were fitted with a 5$^{th}$ order polynomial that led to a typical RMS dispersion of 0.1-0.25 \AA.

{\it{S-distortion, alignment of the spectra.}}
During the spectroscopic observations, the galaxies were not perfectly aligned with the rows of the detector. This effect is crucial when measuring gradients of any type  (rotation curves, velocity dispersion profiles or line-strength indices). The correction of this effect was performed using a routine that found the position of the galaxy center as a function of wavelength,  fitted all these positions with a low order polynomial and straightened the spectra using that polynomial. This alignment was done with a technique that minimised the errors due to the discretization of the signal.  This technique consists of adopting a more realistic distribution of the light in each pixel than just assuming it to be constant. To achieve this the signal in each pixel is fitted with a second order polynomial using the available information in the adjacent pixels.

{\it{Sky subtraction.}}
The sky subtraction is critical for studies where the spectra are analysed at light levels corresponding to only a few per cent of the sky signal, as in our case. For each galaxy observation a sky image  was generated fitting the data at each wavelength with a first order polynomial in regions at both sides of the galaxy close to the ends of the slit (which has a length of 3.7 arcmin on the WHT and 3.3 arcmin on the INT). This was possible since for all targets except M32 the galaxy filled only a small region of the slit, so this synthetic sky image was free from contamination from the galaxy. For M32, we observed a separate sky frame moving the telescope from the coordinates of the galaxy to a position $\Delta\alpha=$-416$"$ (West), $\Delta\delta=$-459$"$ (South) far enough from M32 to avoid its light but with the same level of contamination from M31.

{\it{Extinction correction.}} 
Atmospheric extinction was calculated using the extinction curve for El Roque de los Muchachos Observatory (www.ing.iac.es$/$Astronomy$/$observing$/$manuals$/$ps$/$tech$\_$notes$/$ tn031.pdf). The Galactic extinction was corrected using the curve of \citet{Fitzpatrick99} and the reddening from \citet{Schlegel98} listed in Table \ref{t2}. 

{\it{Flux calibration.}}
The relative flux calibration of the spectra was performed using the observed stars in common with the MILES library \citep{SB06lib} for the optical spectra, and with the CaT library \citep{cen01} for the near infrared. For each observed star we obtained a flux calibration curve. All of them were averaged to obtain one unique flux curve for each run and instrumental configuration. The deviations of each flux calibration curve from the averaged one were introduced as uncertainties in the error spectra.  The typical deviation was of 2$\%$ reaching  $\sim$7$\%$ in the first and last $\sim$150 \AA~ of each setup spectra where the noise is the highest.

\section{Measurement of the kinematic parameters}

The stellar kinematics of galaxies (radial velocities and velocity dispersions) were calculated using the routine MOVEL included in \reduceme~ package \citep{Car99}. This routine is based on the Fourier quotient method described by \citet{Sarg77} and refined with the OPTEMA algorithm \citep{Gon93} that allows us to overcome the typical template mismatch problem. In order to do this, a number of stars of different spectral types and luminosity classes were introduced in the program to create a model galaxy. These stars were of spectral type B9, A0, A3V, G0, G2III, G5III, G8III, G9III, K0III, K0I, K2III, K3III, M0III and M2III. The model galaxy was created and processed in parallel to the galaxy spectrum.  To build the model galaxy all the template spectra were scaled, shifted and broadened according to a first guess of $\gamma$ (mean line-strength), $v$ (radial velocity) and $\sigma$ (velocity dispersion). Then the algorithm looked for the linear combination of these template stars that best matched the observed galaxy spectrum. The best linear combination of observed stars was chosen as the one that minimises the residuals between the galaxy spectrum and the broadened optimal template. This provided a first model galaxy with a first kinematic output ($\gamma$, $v$ and $\sigma$). This model galaxy was then improved using this new guess of kinematic parameters. The process was iterated until it converged.
The emission lines, found only for the field dwarf galaxies, and some large sky line residuals, only present in some cases, were masked, so that the program did not use them for the minimisation of the residuals.


To minimize template mismatch effects, it is essential to use as
templates a variety of spectral types and luminosity classes which are
representative of the stellar population of the observed galaxy; as we
will discuss in Section \ref{sigmas_comparison}, small differences in ages and
metallicities could lead to a partial fit of the strongest lines and
therefore affect the derived velocity dispersion of the galaxy.

It is also important to check whether the observed stars were filling the slit during the observation. If they were not, the instrumental profile would not affect them in the same way as in the galaxies, and as a consequence, the $\sigma$ that one would measure for the galaxy would be $\sqrt{\sigma_{inst}^2+\sigma_{gal}^2}$, where $\sigma_{gal}$ is the intrinsic velocity dispersion of the galaxy and $\sigma_{inst}$, in this case, is the quadratic instrumental difference between the galaxies and the stars. 
 
To correct for this effect the physical slit width\footnote{During the observation an indicative width of the slit is selected by the user through a web interface.} is required.  We calculated it from the spatial scale and the FWHM (in pixels) of the arc lines, which illuminate homogeneously the slit.
 To see whether the stars were filling the slit  completely we checked that the FWHM of their spatial profile was larger than the physical slit width. If this was not the case, the spatial profile of the star was broadened accordingly. Although this introduced some uncertainties, the data quality improved by making this correction. The value of $\sigma$ measured and adopted in this work is thus the intrinsic velocity dispersion of the galaxy corrected for possible instrumental effects.

\begin{figure}
\centering
\resizebox{0.5\textwidth}{!}{\includegraphics[angle=-90]{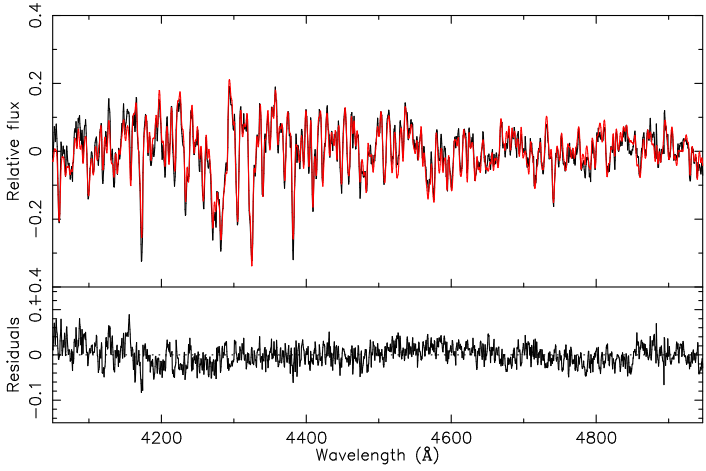}}
\caption{Example of the fit computed by MOVEL. Upper panel: in black, the central spectrum of VCC523. The spectrum has been continuum subtracted and normalised. The optimal template that fits the galaxy is shown in red, a linear combination with different weights of the stars observed with the same configuration as the galaxy. Bottom panel: residuals of the galaxy-composite template fit.}
              \label{movel}
\end{figure}

Figure \ref{movel} shows a typical fit of the observed central spectral of a galaxy and the corresponding optimal template broadened with a gaussian with the derived dynamical parameters. The errors in velocity and $\sigma$ were computed through Monte-Carlo simulations, repeating the whole process (including the derivation of the optimal template) for 100 simulated galaxy spectra created using the error spectra obtained during the reduction process. The observed and simulated spectra perfectly match.

\subsection{Kinematic parameter profiles}

To measure kinematic gradients it is important to determine the minimum S$/$N
needed to measure reliable radial velocities and velocity dispersions. To do
that we have designed and carried out a test exercise based on Monte-Carlo simulations that constraints the errors and systematic effects in the measurement of radial velocities and velocity dispersions on fake galaxy spectra with known input kinematic parameters and different ages, metallicities and S$/$N ratios. This can be outlined in the following steps: i) from the simple stellar population
models from PEGASE.HR (FWHM$\sim$ 0.55 \AA; \citet{Lb03}) we selected a subset of 9 model spectra (3 ages and 3 metallicities) representative of quiescent dwarf galaxies: ages 1 Gyr, 4 Gyr,
10 Gyr; metallicities of $Z=$ 0.0, $Z=$ $-$0.4, $Z=$ $-$0.7. The spectral resolution of these models are needed since in our bluer configuration the resolution is $\sim$ 1.6 \AA; ii) Taking into account our instrumental
resolution, each model was broadened and redshifted to match a set of input velocity dispersions, $\sigma_i$, (9 values between 20 km s$^{-1}$ and
60 km s$^{-1}$ in steps of 5 km s$^{-1}$) and radial velocities, $v_i$, (800 km s$^{-1}$ and 1500 km s$^{-1}$, typical values of
Virgo cluster members). This amounts a total of 162 model spectra. iii) For each one of the above spectra we added different levels of random noise to match S$/$N ratios of 10, 15, 20, 25, 30 and 50, hence ending up with 972 model galaxy spectra. iv) For each simulated galaxy spectrum we run exactly the same MOVEL procedure as we did for our dE galaxy sample, using the same template stars and MOVEL parameters. 100 Monte-Carlo simulations for each model galaxy were carried out to get reliable errors of the derived kinematic parameters, $\sigma_o$ and $v_o$ obtained as the mean value of the 100 Monte-Carlo simulations in each case. Since the input kinematics $\sigma_i$ and $v_i$ are set by construction, comparisons and reliability analysis are immediate to perform.

\begin{figure}
\centering
\resizebox{0.5\textwidth}{!}{\includegraphics[angle=-90]{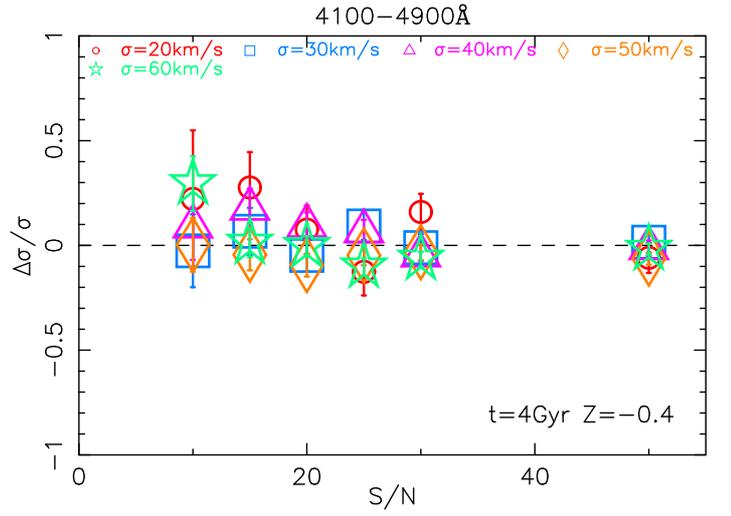}}
\caption{Simulations to study the minimum S$/$N ratio to obtain reliable measurements of the velocity dispersion. It is plotted $\Delta\sigma/\sigma$ (defined as $(\sigma_o-\sigma_i)/\sigma_i$), the relative error introduced when measuring the $\sigma$ of a galaxy as a function of S$/$N ratio. Different colours show different $\sigma$ for the simulated galaxies. Done assuming a stellar population of 4 Gyr and $Z=-$0.4.}
              \label{sig}
\end{figure}

\begin{figure}
\centering
\resizebox{0.5\textwidth}{!}{\includegraphics[angle=-90]{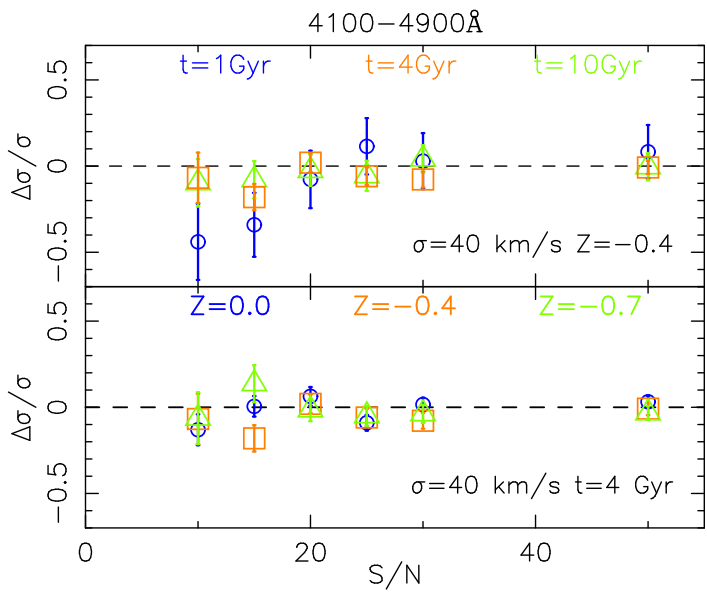}}
\caption{Simulations to study the dependence of the stellar population of the galaxy on the measurement of its velocity dispersion. We plot the relative offsets found between the measured $\sigma$ and the velocity dispersion introduced in the simulated galaxy as a function of the S$/$N ratio. In both panels a typical $\sigma$ of 40 km s$^{-1}$ has been considered. In the upper panel we fix the metallicity of the galaxy to $-$0.4 and study the influence of the age of the stellar population. In the lower panel the parameter fixed is the age to 4 Gyr, and we analyse the influence of the metallicity on $\sigma$.}
              \label{ageZ}
\end{figure}

\begin{figure}
\centering
\resizebox{0.5\textwidth}{!}{\includegraphics[angle=-90]{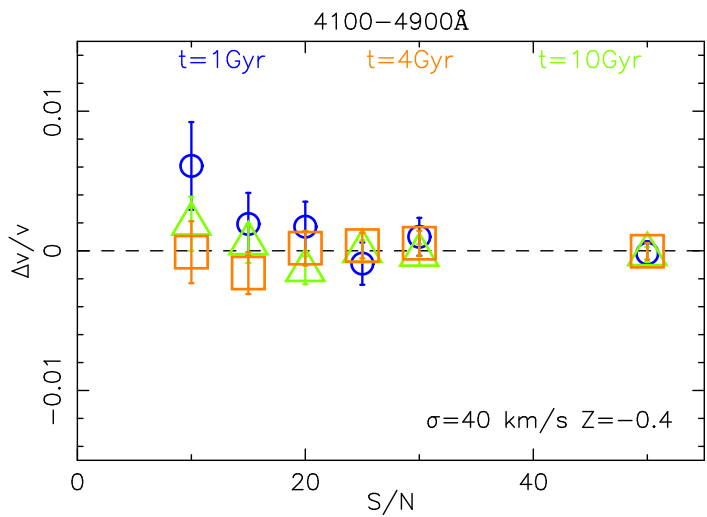}}
\caption{Simulations to study the influence of the stellar populations and the S$/$N ratio on the measurement of the radial velocities. We plot the relative uncertainty when measuring the radial velocities ($\Delta v/v$, defined as $(v_o-v_i)/v_i$) as a function of the S$/$N ratio. Only the results for the radial velocity $v=$1500 km s$^{-1}$ are shown because the offsets and errors found are independent of the radial velocity of the galaxy. The results plotted are for a typical dwarf galaxy with velocity dispersion of 40 km s$^{-1}$ and metallicity -0.4. The velocity dispersion does not have any effect on the measurement of radial velocities nor the metallicities. Note the different scale in the y-axis compared to Figures \ref{sig} and \ref{ageZ}.}
              \label{vel}
\end{figure}

The above procedure was carried out for each instrumental configuration in each of the 5 different spectral regions. The results obtained are shown
in Figures \ref{sig}, \ref{ageZ} and \ref{vel}. 

After correcting for any systematic offsets in radial velocity and velocity
dispersion due to small intrinsic differences between PEGASE.HR models and our
observed stars, we have analysed the simulations looking at the relative
differences between the measured values  and the parameters introduced in the simulated galaxies 
($\Delta v/v=(v_o-v_i)/v_i$ and
$\Delta\sigma/\sigma=\sigma_o-\sigma_i)/\sigma_i$). Figures \ref{sig}, \ref{ageZ} and \ref{vel} show these
differences as a function of S$/$N in the wavelength range 4100-4900 \AA, a range in common between the WHT and INT
observations and where lines as important as the G-band are located. The
error bars in these 3 Figures show the relative uncertainties obtained by MOVEL as
the RMS scatter resulting from the 100 Monte-Carlo simulations for each model
galaxy.

In Figure \ref{sig} we study the influence of the S$/$N ratio and the instrumental resolution on the measurement of the velocity dispersion of a galaxy.  Each point represents a galaxy of similar stellar populations (age 4 Gyr and metallicity $-$0.4) but different velocity dispersion (from 20 to 60 km s$^{-1}$). As expected, the errors increase dramatically at the lowest S$/$N ratios. For low S$/$N ratios (S$/$N=10) offsets are found even for
galaxies with high $\sigma$'s, so the velocity dispersions derived at this S$/$N
cannot be trusted.
 On the contrary, for S$/$N ratios higher than or equal to 15 we do not find statistically significant offsets, with the exception of measurements below half the instrumental resolution ($\sigma=20$ km s$^{-1}$) where special care must be taken. Only for S$/$N larger than 20 the measured $\sigma$ can be fully trusted for velocity dispersions as low as half the instrumental resolution. 

Figure \ref{ageZ} presents the influence of the stellar populations on the
measurement of the velocity dispersion of a galaxy. In the upper panel we show
the effect of the age on a dwarf galaxy of $\sigma=40$ km s$^{-1}$ and
$Z=-$0.4, while in the lower panel we show the effect of the metallicity for a
galaxy with $\sigma=$40 km s$^{-1}$ and 4 Gyr old. In this case, although
the errors barely depend on metallicity, the age-dependence is crucial, and for populations as
young as 1 Gyr the $\sigma$ measurements are underestimated for a S$/$N ratio below 15.

In Figure \ref{vel} we analyse the effect of the stellar populations on the
computation of the radial velocity. Neither a change in the velocity dispersion nor in the
metallicity have appreciable effects on this variable. Only the age of the stellar population have appreciable influence on the determination of the radial velocity whenever the age is young, around 1 Gyr. The uncertainty induced  by age variations, however, is small, $\le$1$\%$, thus rotation velocities can be accurately measured down to S$/$N$\sim$10.

\begin{figure}
\centering
\resizebox{0.5\textwidth}{!}{\includegraphics[angle=-90]{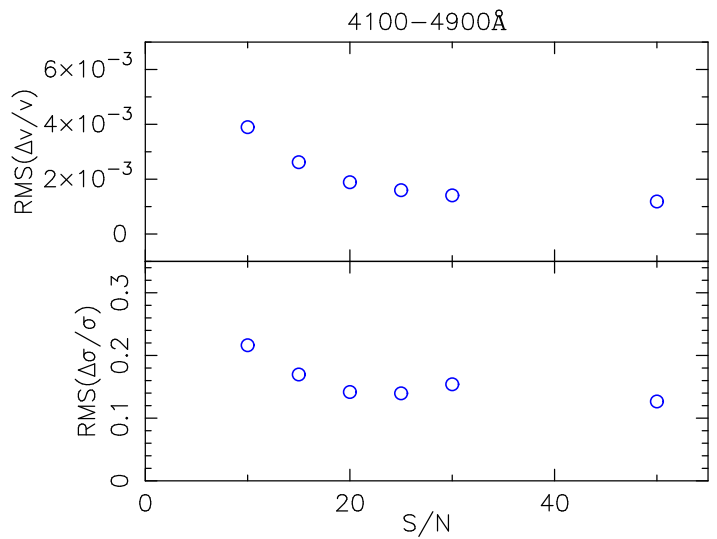}}
\caption{ Total scatter in the differences found for the 972 simulations computed as a function of the S$/$N ratio. In this Figure all models for all stellar populations and
velocity dispersion parameters have been used.}
              \label{summary}
\end{figure}

In Figure \ref{summary} we plot the expected RMS error (i.e. RMS of the average of all the simulations) as a function of S$/$N ratio in order to have a statistical estimate of the uncertainty. The average of all the simulations (corresponding to different velocity dispersions, ages and metallicities) for each S/N, takes into account that for a target galaxy we do not have a priori information about either its velocity dispersion or the parameters of the stellar population.

Figures \ref{sig}-\ref{summary} clearly show that data with S$/$N ratios below
15 might induce errors as large as 22$\%$ in the determination of $\sigma$, in particular for small velocity dispersions ($\sigma\sim$20 km s$^{-1}$), while only  0.4$\%$ for radial velocities.

All these tests have been computed for the different 
wavelength ranges covered by our survey and the results obtained are rather similar. For the 
red arm of ISIS, where the
instrumental resolution is larger, we obtain similar results as those shown in the blue arm (Figures \ref{sig}-\ref{summary}) but with slightly larger uncertainties.

These simulations  show that radial
velocities can be computed with spectra of S$/$N as low as
10 given that the uncertainty is always below 1$\%$.
However, in the study of velocity dispersions, we must discard any measurement with
S$/$N below 15 because $\sigma$ is, in this case, highly dependent on age and
not reliable for $\sigma$ as low as half the instrumental resolution.

When running the MOVEL algorithm as a function of galaxy radius, we
fitted the optimal template at every radius, rather than using the central
optimal template,  in order to improve the fit. As a result, the optimal
templates turn out to be radial dependent. The differences,
however, are not very large, because in the linear combination of templates,
G-stars always contribute with the highest weight.

Due to the different instrumental configurations used in the observation
campaigns (see Table \ref{t1}), more than one kinematic profile per galaxy was
obtained. These profiles, consistent within the errors, were averaged to produce a single, high S$/$N profile per galaxy (see Figure \ref{curves1}). 

The recessional velocity of each single galaxy, removed for the determination of the rotation curve (Figure \ref{curves1}), has been determined by averaging, with a weighted mean, the recessional velocity measured in each single position along the radius. This improved technique for measuring recessional velocities (listed on Table \ref{t4}) can be applied since the rotation curves are symmetrical.\footnote{Note that galaxies as VCC856, those with the poorest quality, have a non-zero central velocity.}

Table \ref{example} gives an example of the tables electronically available with the values of the kinematic profiles. 

\begin{table}
\begin{center}
\caption{Kinematic profiles for VCC990.\label{example}}
\begin{tabular}{|c|c|c|c|}
\hline \hline
  $R_v$ ($''$)      & $v$ (km s$^{-1}$) & $R_{\sigma}$ ($''$) & $\sigma$ (km s$^{-1}$)\\
\hline
-5.43  & 1.8 $\pm$    5.5 &  -4.98  & 36.0 $\pm$ 6.2\\
-3.56  & 16.8$\pm$   6.4 &  -2.36  & 47.2 $\pm$ 6.0\\
-2.58  &  9.8$\pm$    6.3&   -1.38  & 39.2 $\pm$ 5.4\\
-2.00  &10.5$\pm$    6.4 &  -0.80 &  35.8 $\pm$ 4.6\\
-1.60  & 1.7$\pm$     6.2&   -0.40 & 32.5 $\pm$ 3.5\\
-1.20  &-6.3$\pm$    4.4 &  0.00  & 42.2 $\pm$ 3.1\\
-0.80  &3.9 $\pm$    2.7 &  0.40  & 41.3 $\pm$ 2.8\\
-0.40  &0.1 $\pm$    2.3 &   0.80 &  39.7 $\pm$ 3.6\\
0.00    &6.8 $\pm$    2.0 &  1.38 &  37.7 $\pm$ 3.9\\
0.40    &2.2 $\pm$    2.4 &  2.36 & 36.1 $\pm$ 8.5\\
0.80    &1.0 $\pm$    3.1 &  4.97 &  37.4 $\pm$ 5.6\\
1.20    &-5.2$\pm$   5.2 &          &  \\
1.60    &-11.8$\pm$ 5.6 &          &  \\
2.00    &-25.3$\pm$ 6.4 &          &  \\
2.58    &-17.1$\pm$ 7.2 &          &   \\
3.56    &-29.6$\pm$ 5.7 &          &   \\
5.42    &-25.3$\pm$ 6.2 &          &   \\
\hline
\end{tabular}
\end{center}
NOTES: Column 1: radius for the rotation speed profile. 
            Column 2: rotation velocities. 
            Column 3: radius for the velocity dispersion profile. 
            Column 4: velocity dispersions.
            All the kinematic profiles are electronically available.
\end{table}

\begin{figure*}
\centering
\resizebox{0.4\textwidth}{!}{\includegraphics[angle=-90]{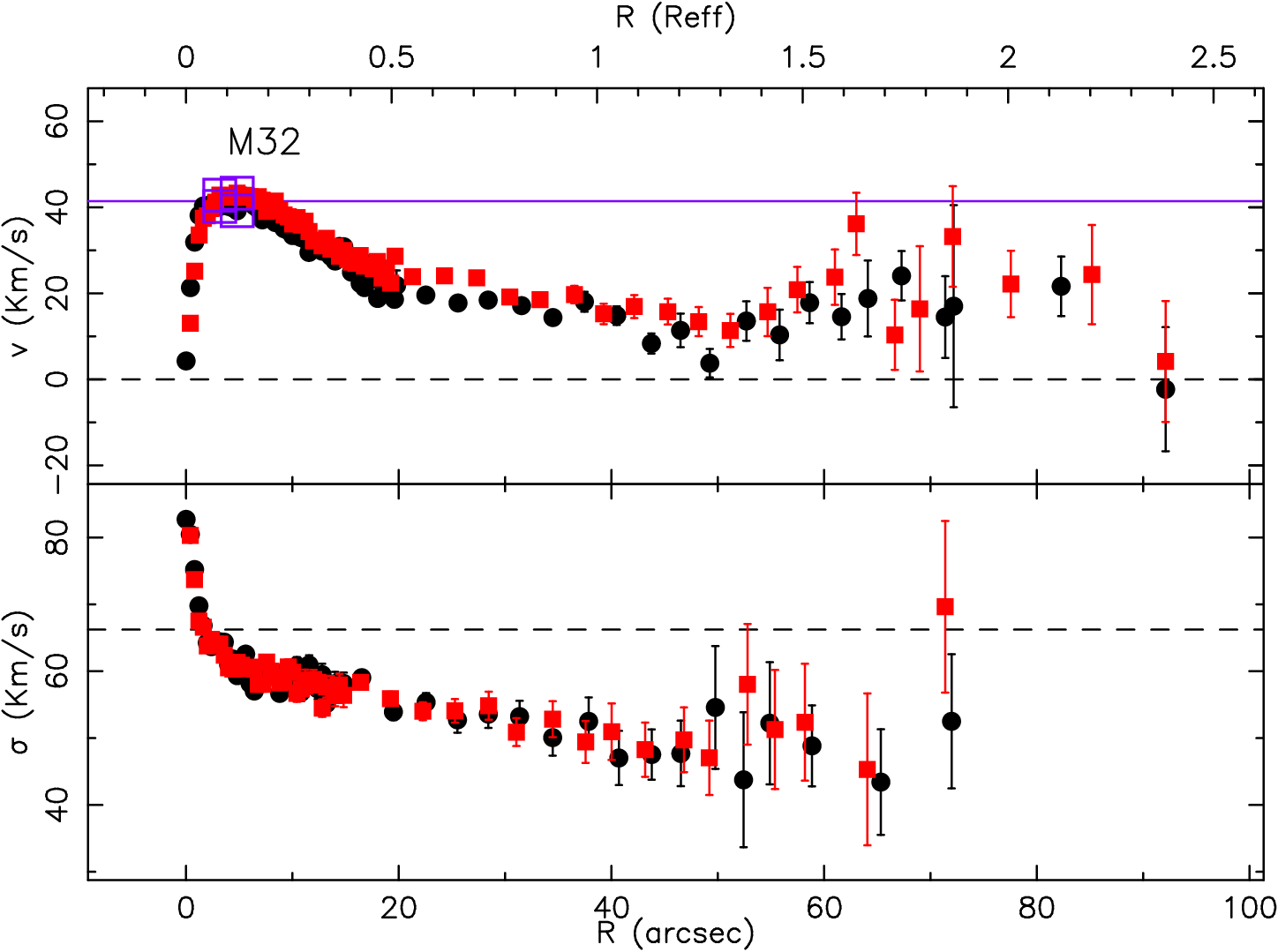}}
\resizebox{0.4\textwidth}{!}{\includegraphics[angle=-90]{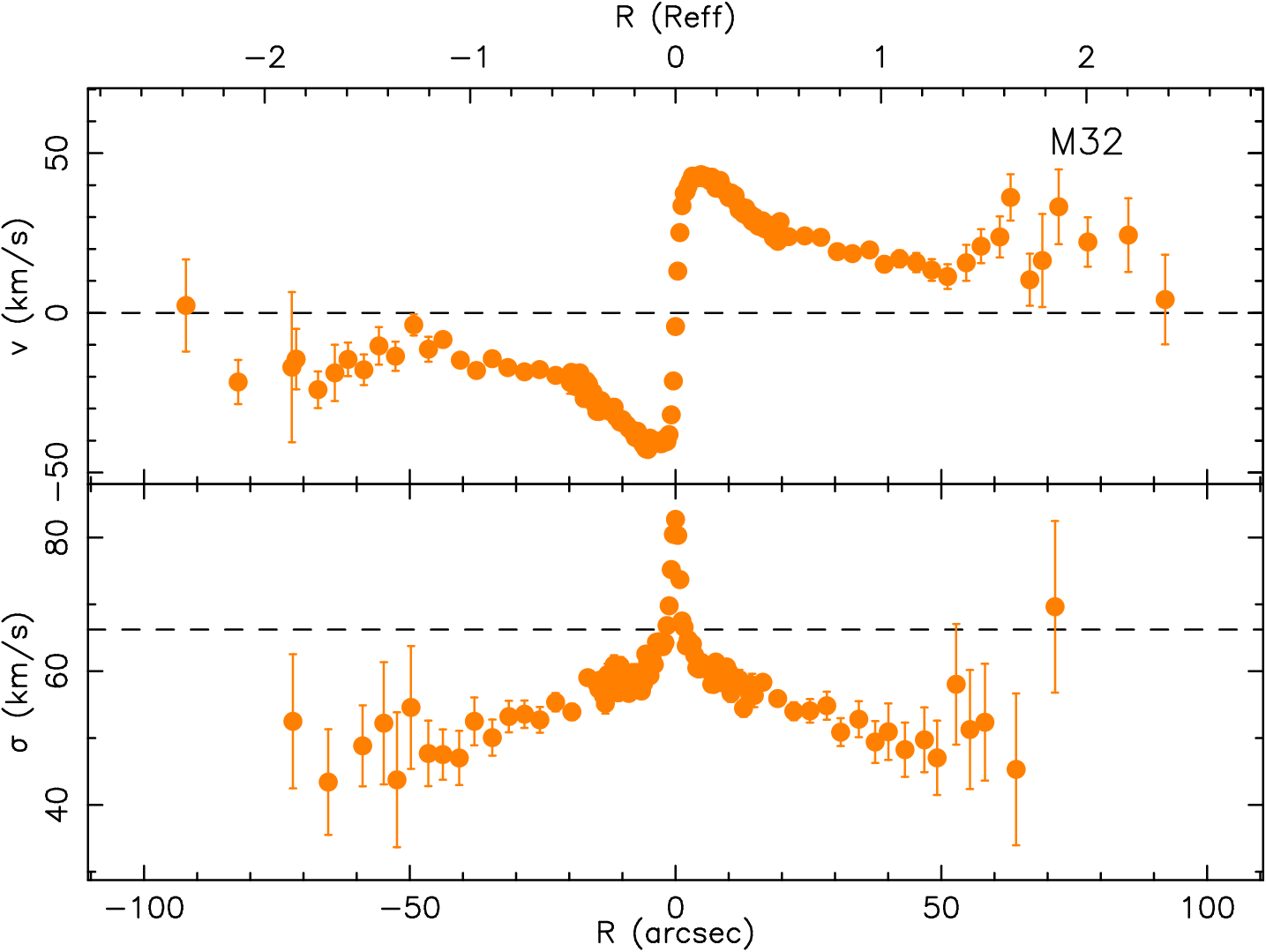}}
\resizebox{0.4\textwidth}{!}{\includegraphics[angle=-90]{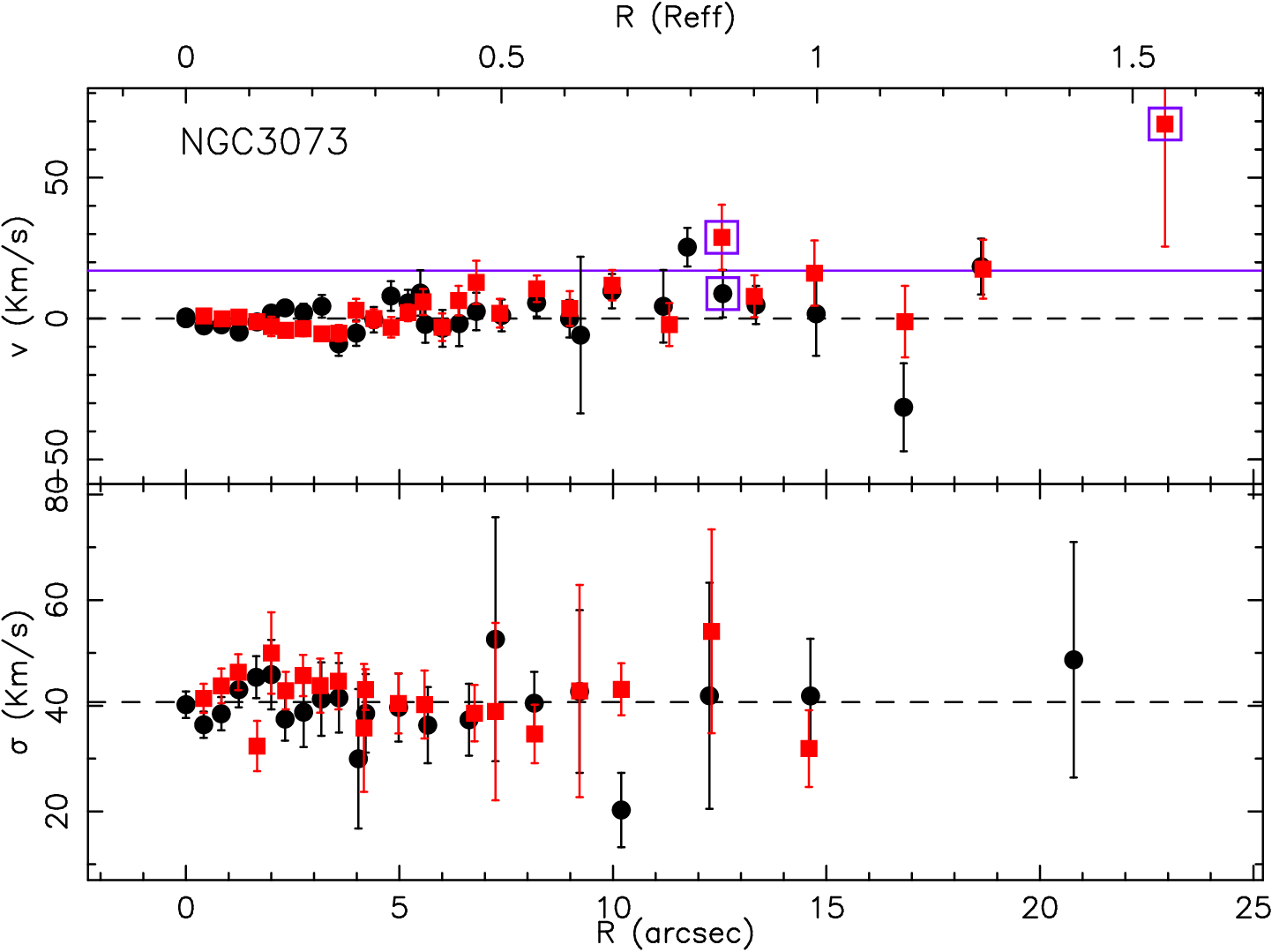}}
\resizebox{0.4\textwidth}{!}{\includegraphics[angle=-90]{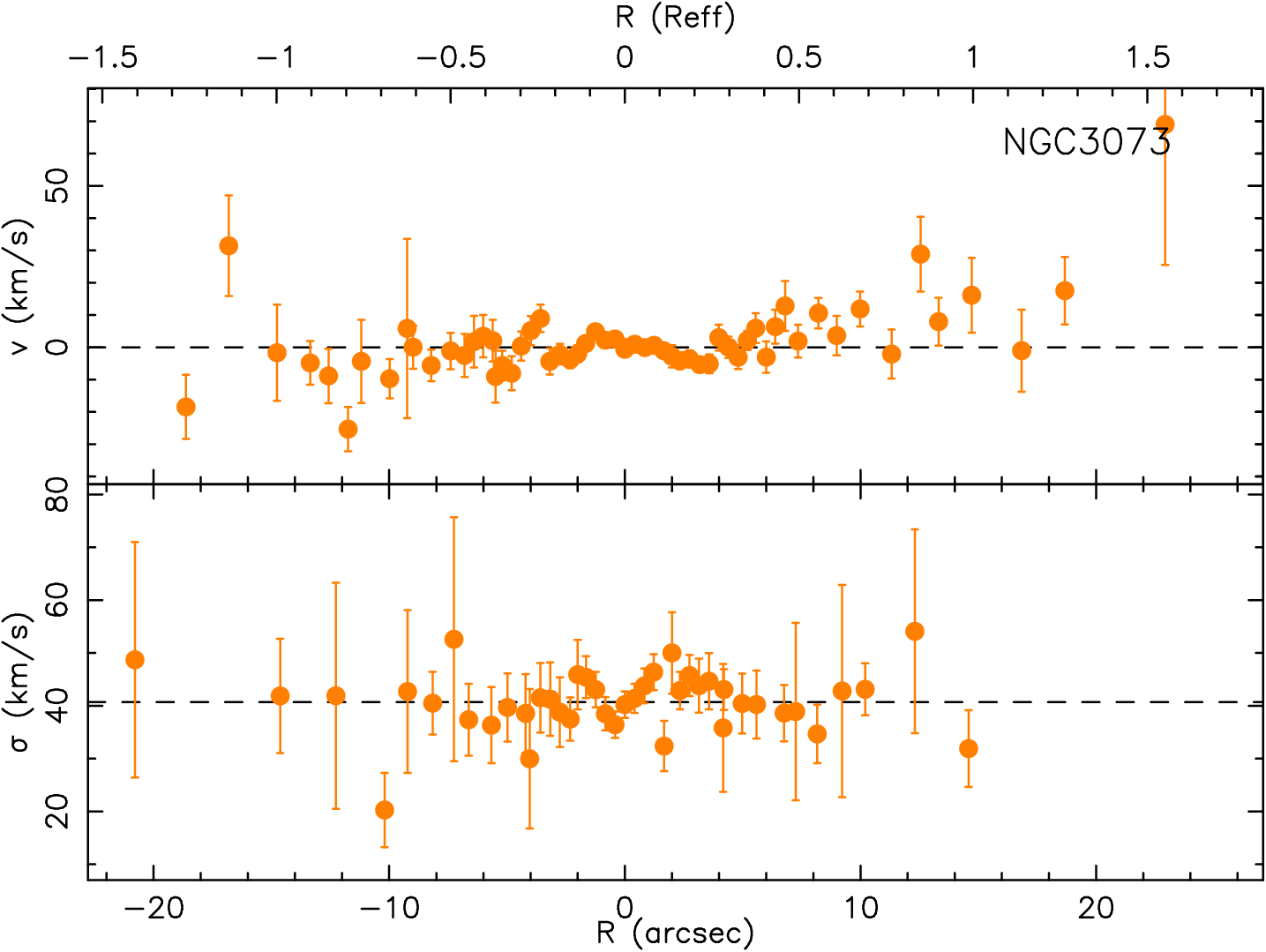}}
\resizebox{0.4\textwidth}{!}{\includegraphics[angle=-90]{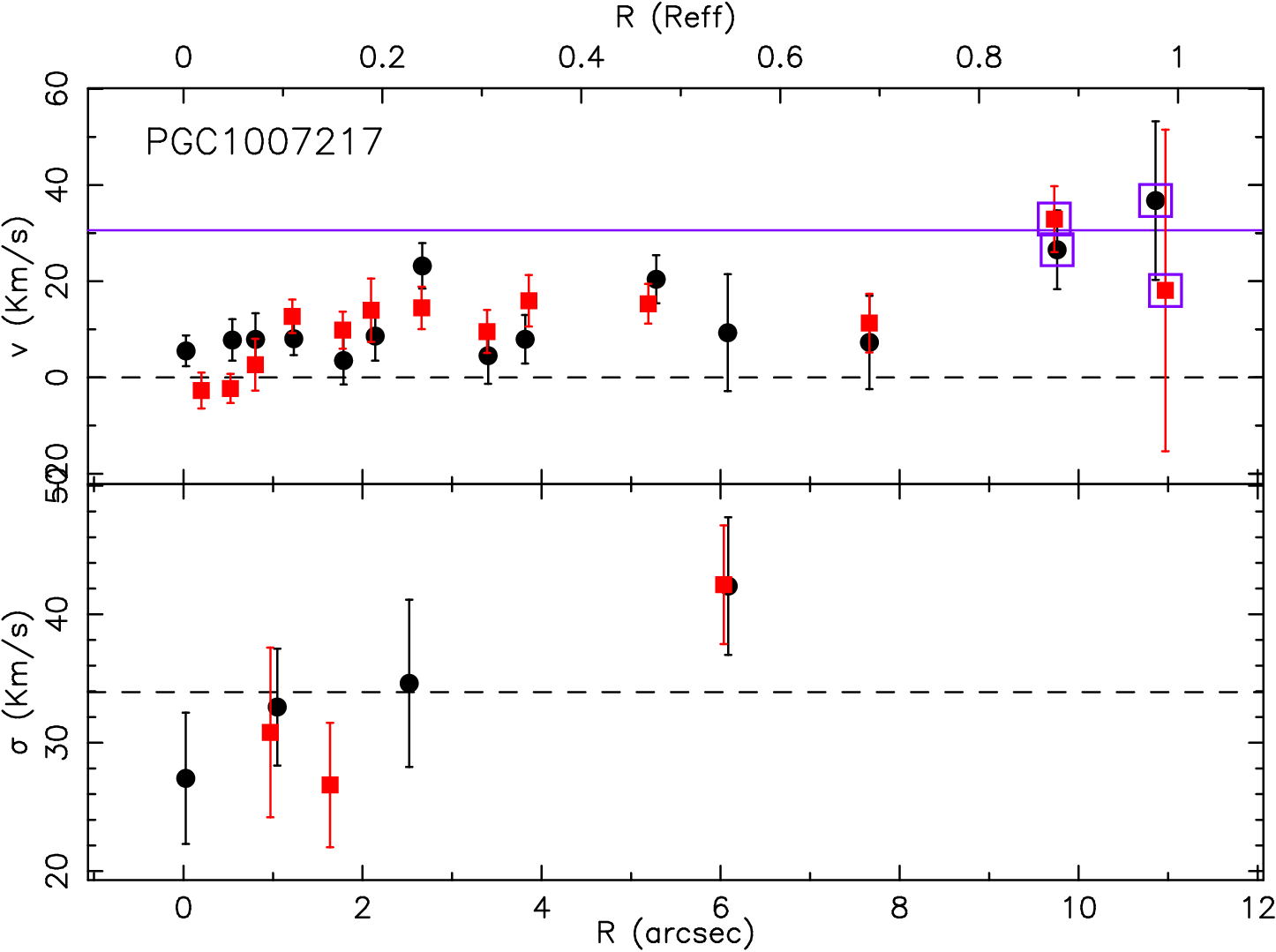}}
\resizebox{0.4\textwidth}{!}{\includegraphics[angle=-90]{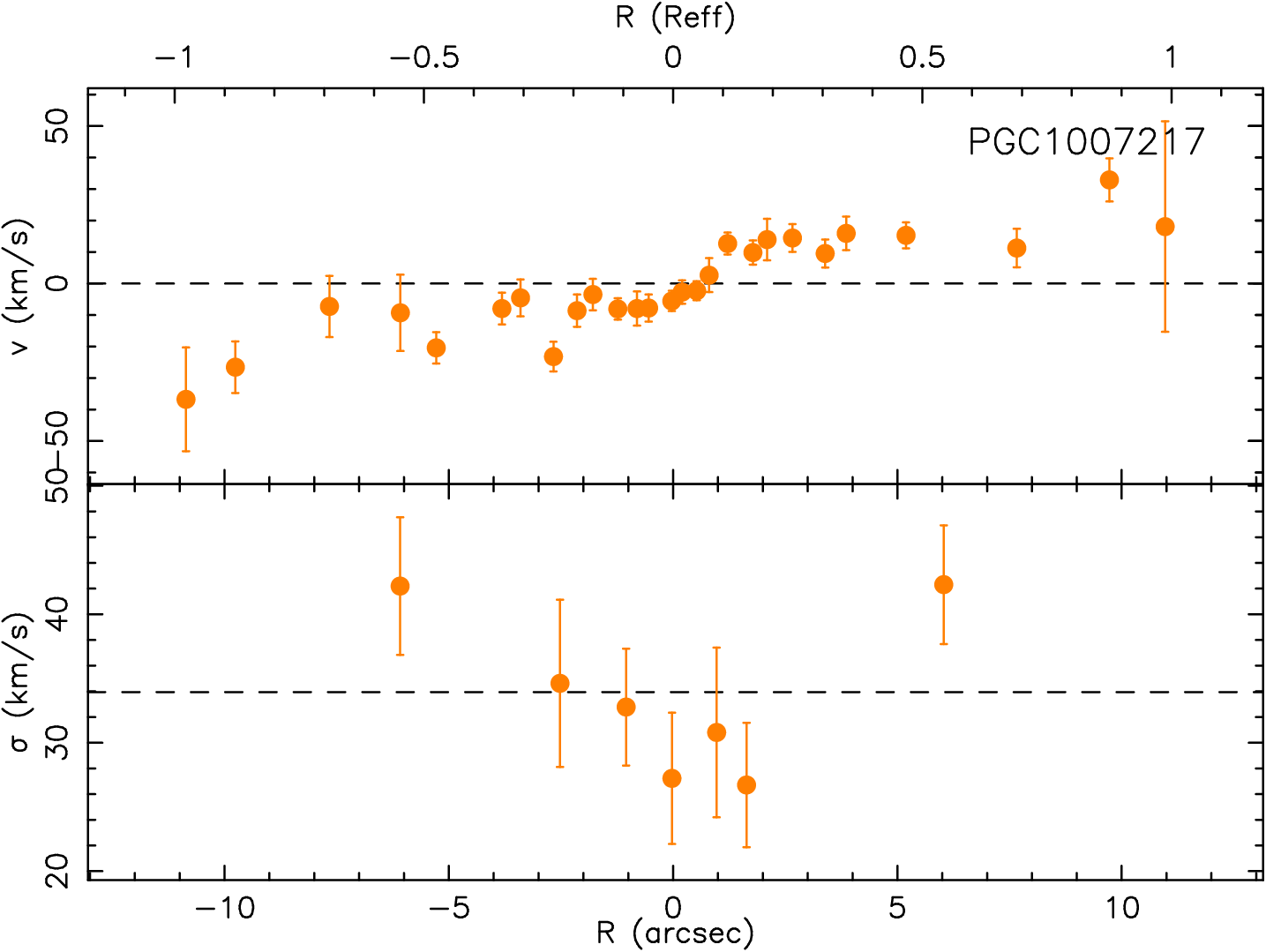}}
\resizebox{0.4\textwidth}{!}{\includegraphics[angle=-90]{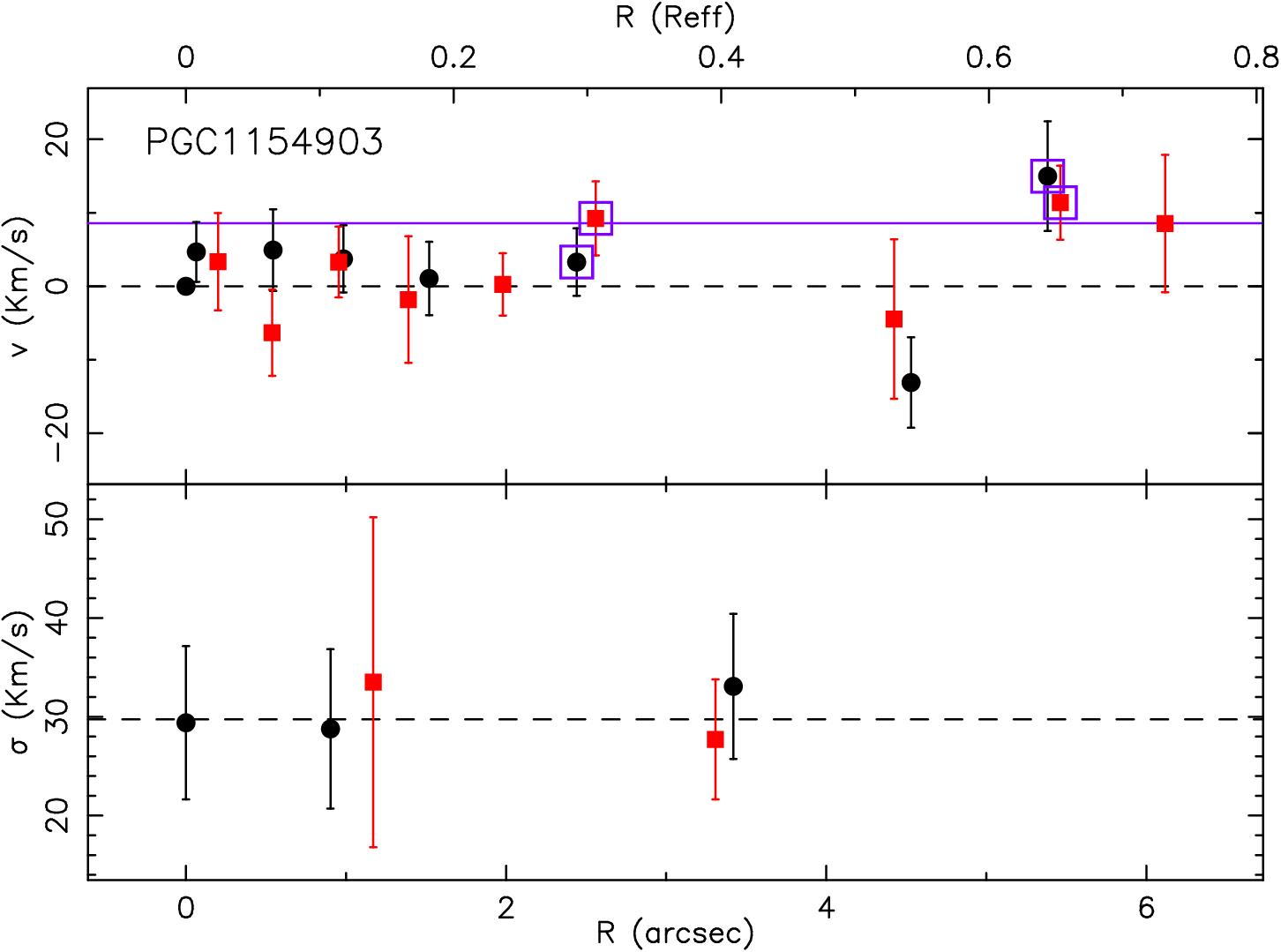}}
\resizebox{0.4\textwidth}{!}{\includegraphics[angle=-90]{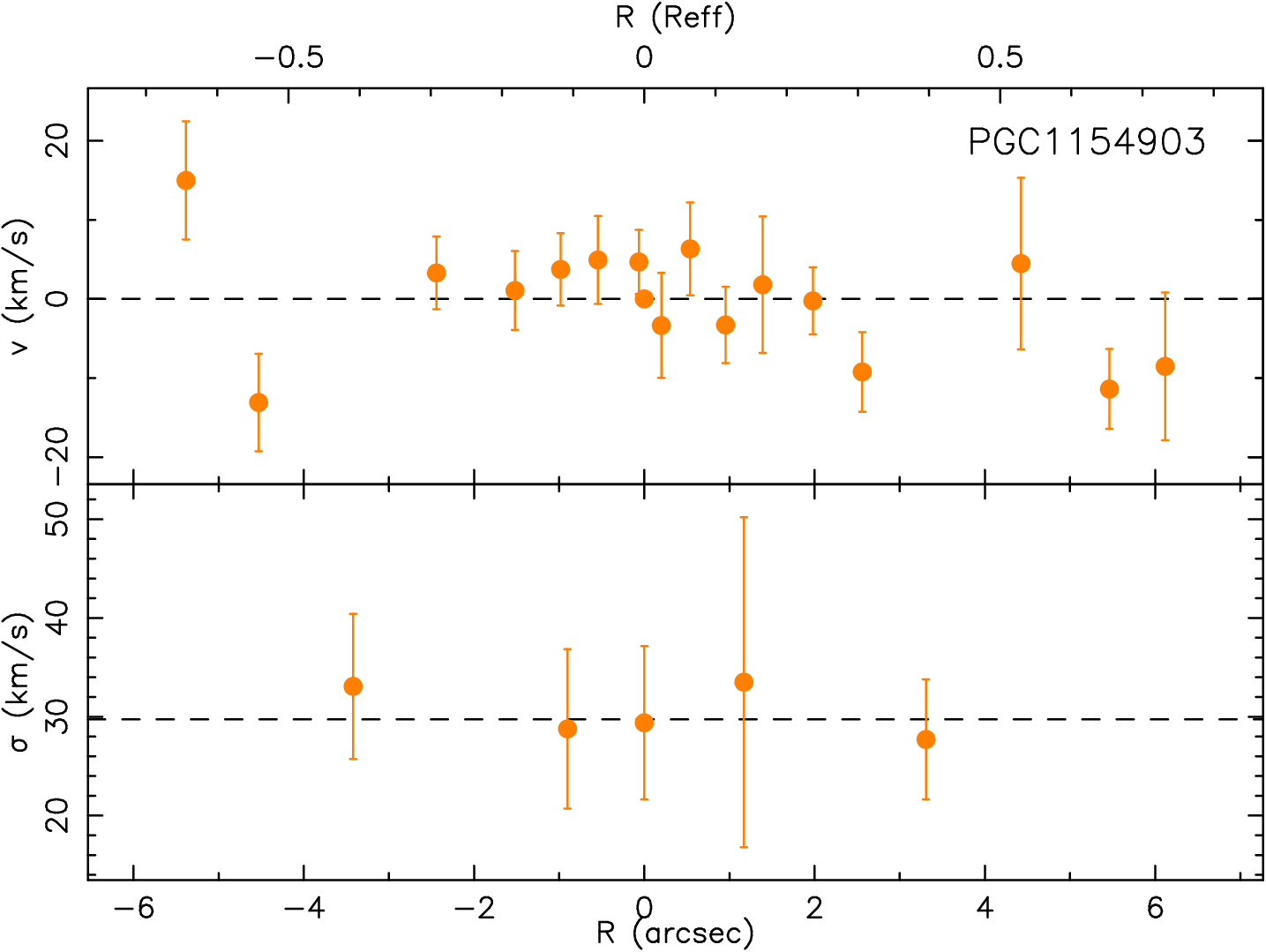}}
\caption{Kinematic profiles of the galaxy sample. Each diagram shows in the left upper panel the folded rotation curve of the galaxy and in the left bottom panel the folded velocity dispersion profile. The different sides of the galaxy are indicated with red squares and black dots. On the upper x-axis the radius is given as a  fraction of the effective radius ($R_{eff}$) of each galaxy in the I band (see Section \ref{phot}). The purple open squares show the points used to calculate the maximum rotation for each galaxy and the purple line indicates this $v_{max}$. The dashed line in the velocity dispersion profiles indicate the central $\sigma$ computed up to the $R_{eff}$ (see Table \ref{t4}). In the right panels the not-folded kinematical profiles are plotted.}
              \label{curves1}
\end{figure*}
\addtocounter{figure}{-1}
\begin{figure*}
\centering
\resizebox{0.4\textwidth}{!}{\includegraphics[angle=-90]{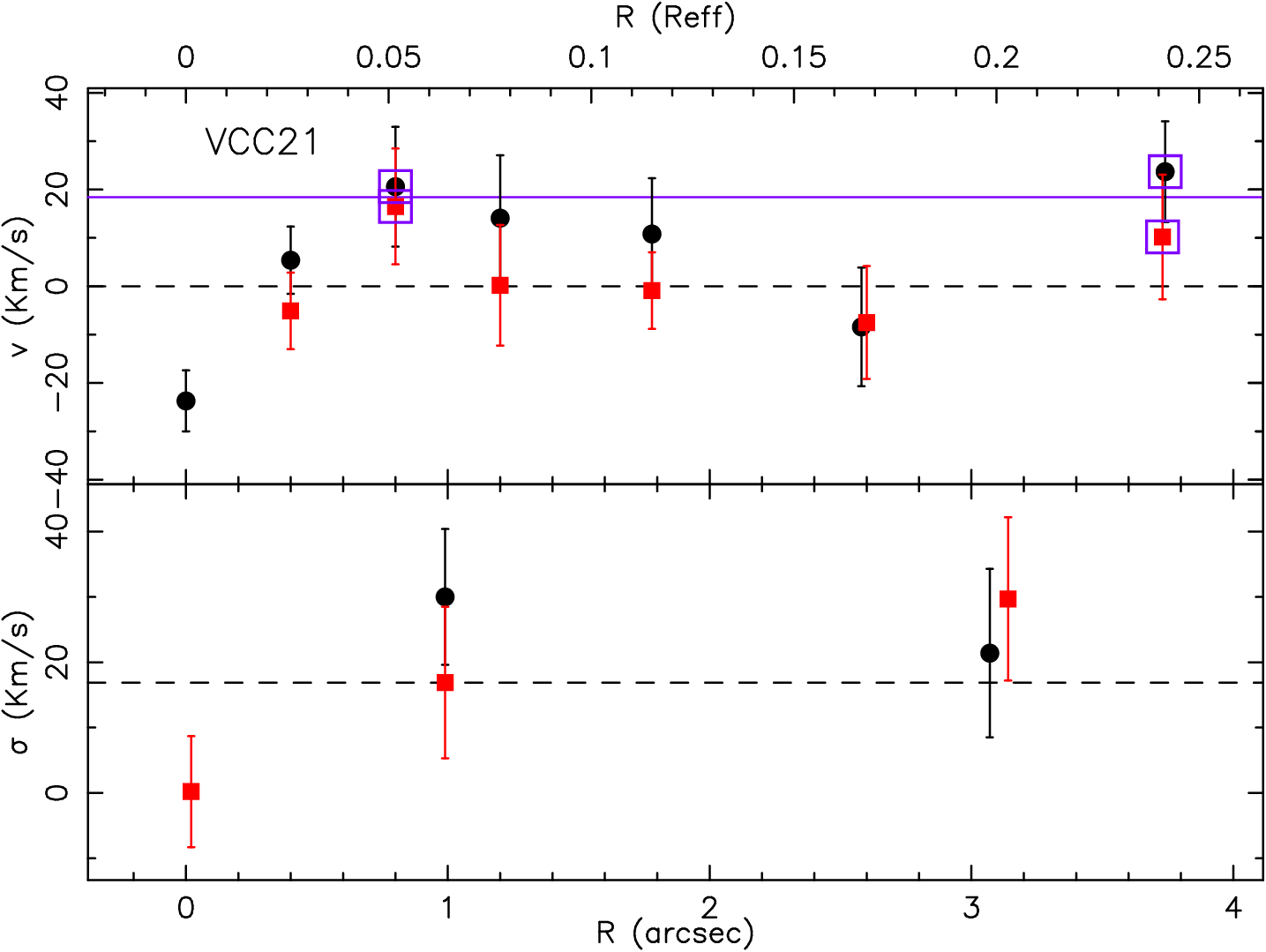}}
\resizebox{0.4\textwidth}{!}{\includegraphics[angle=-90]{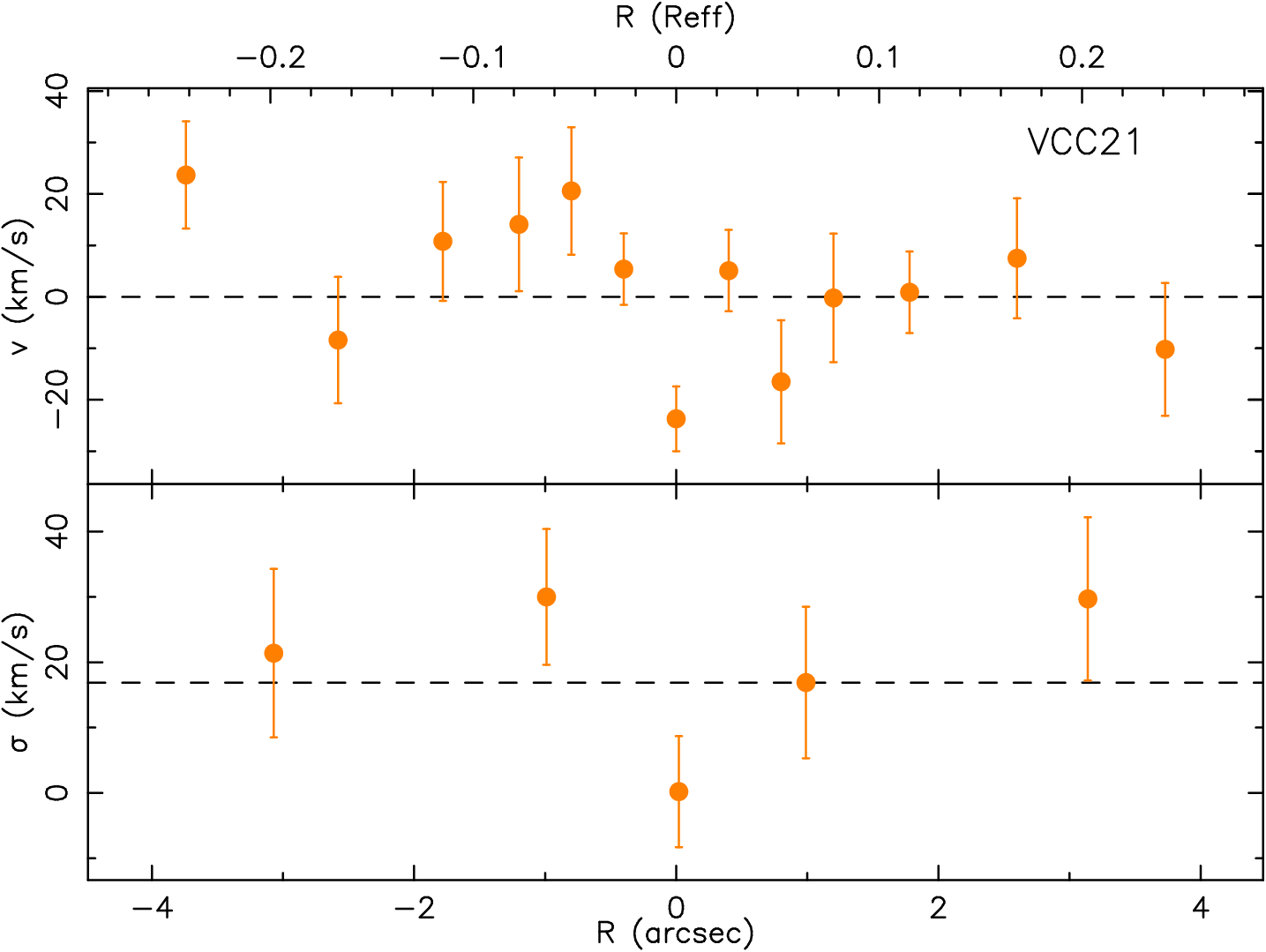}}
\resizebox{0.4\textwidth}{!}{\includegraphics[angle=-90]{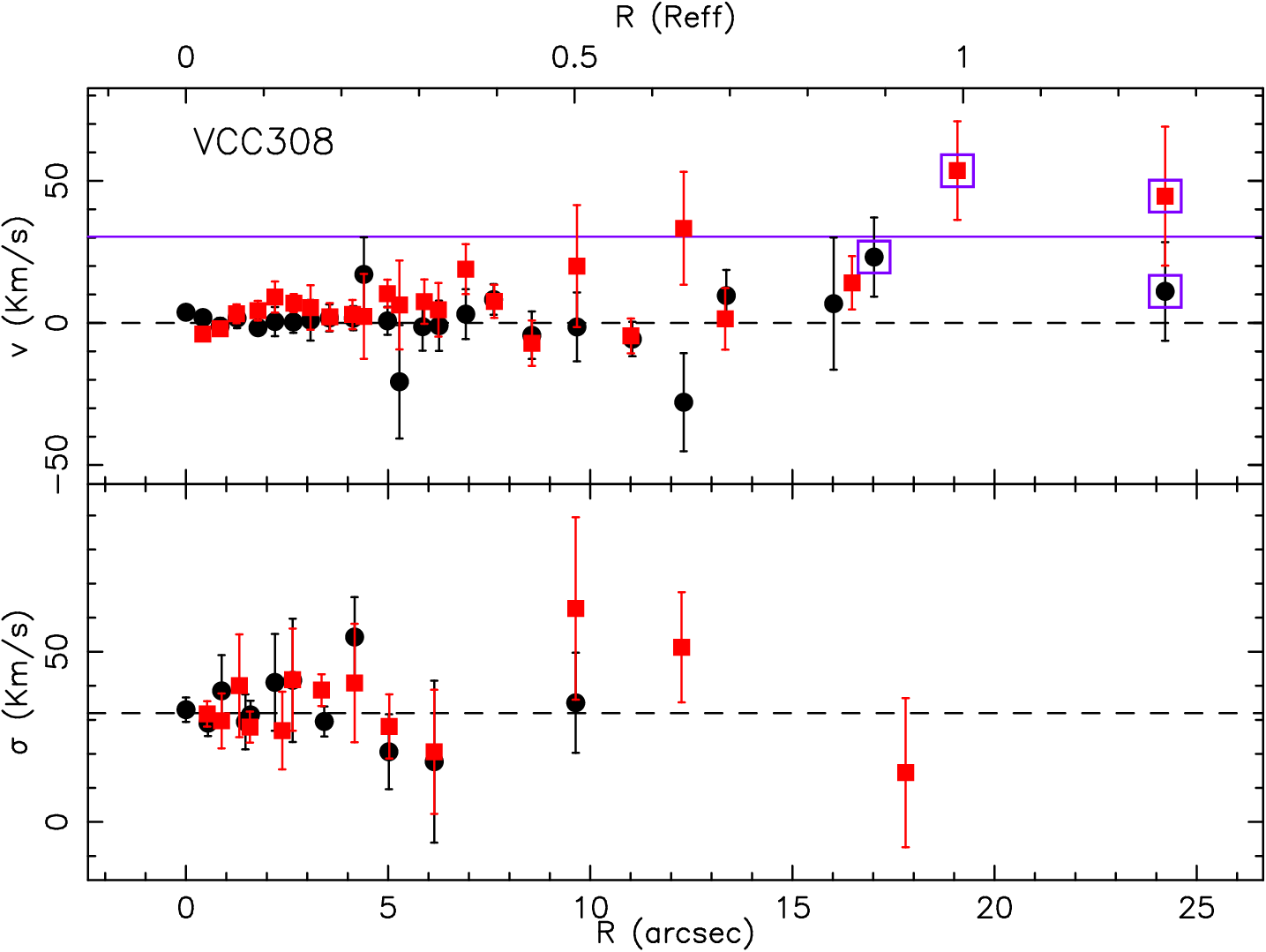}}
\resizebox{0.4\textwidth}{!}{\includegraphics[angle=-90]{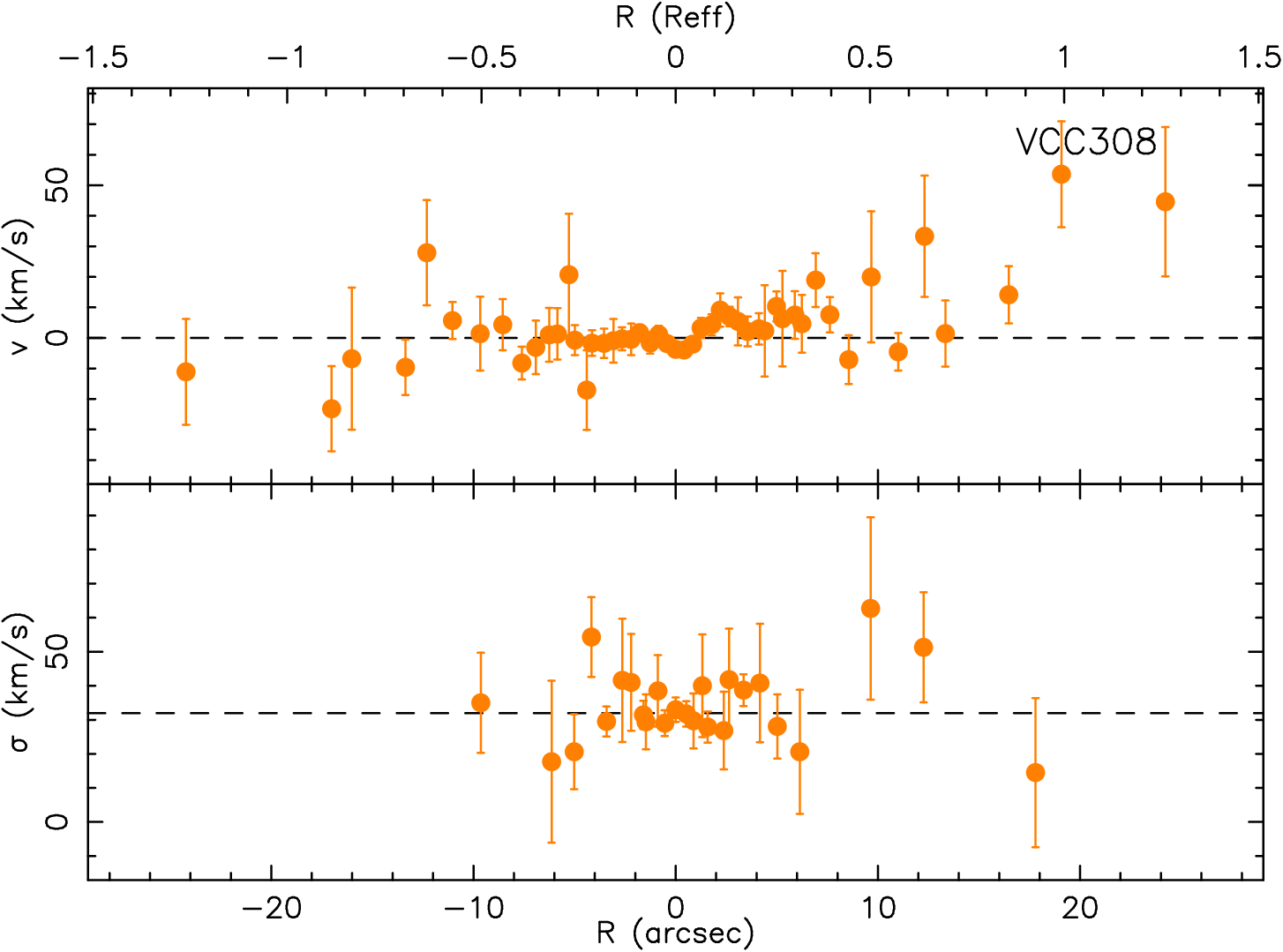}}
\resizebox{0.4\textwidth}{!}{\includegraphics[angle=-90]{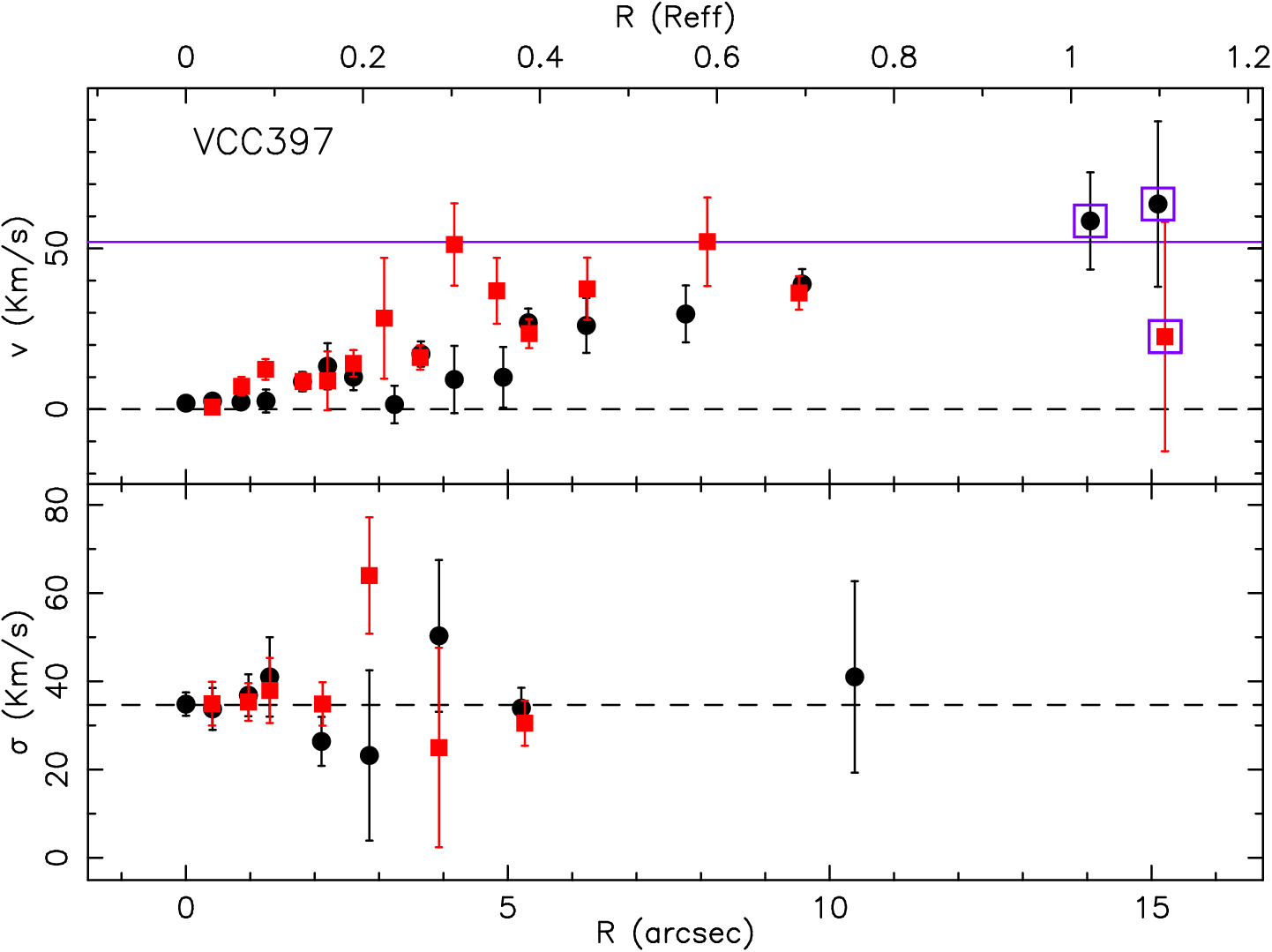}}
\resizebox{0.4\textwidth}{!}{\includegraphics[angle=-90]{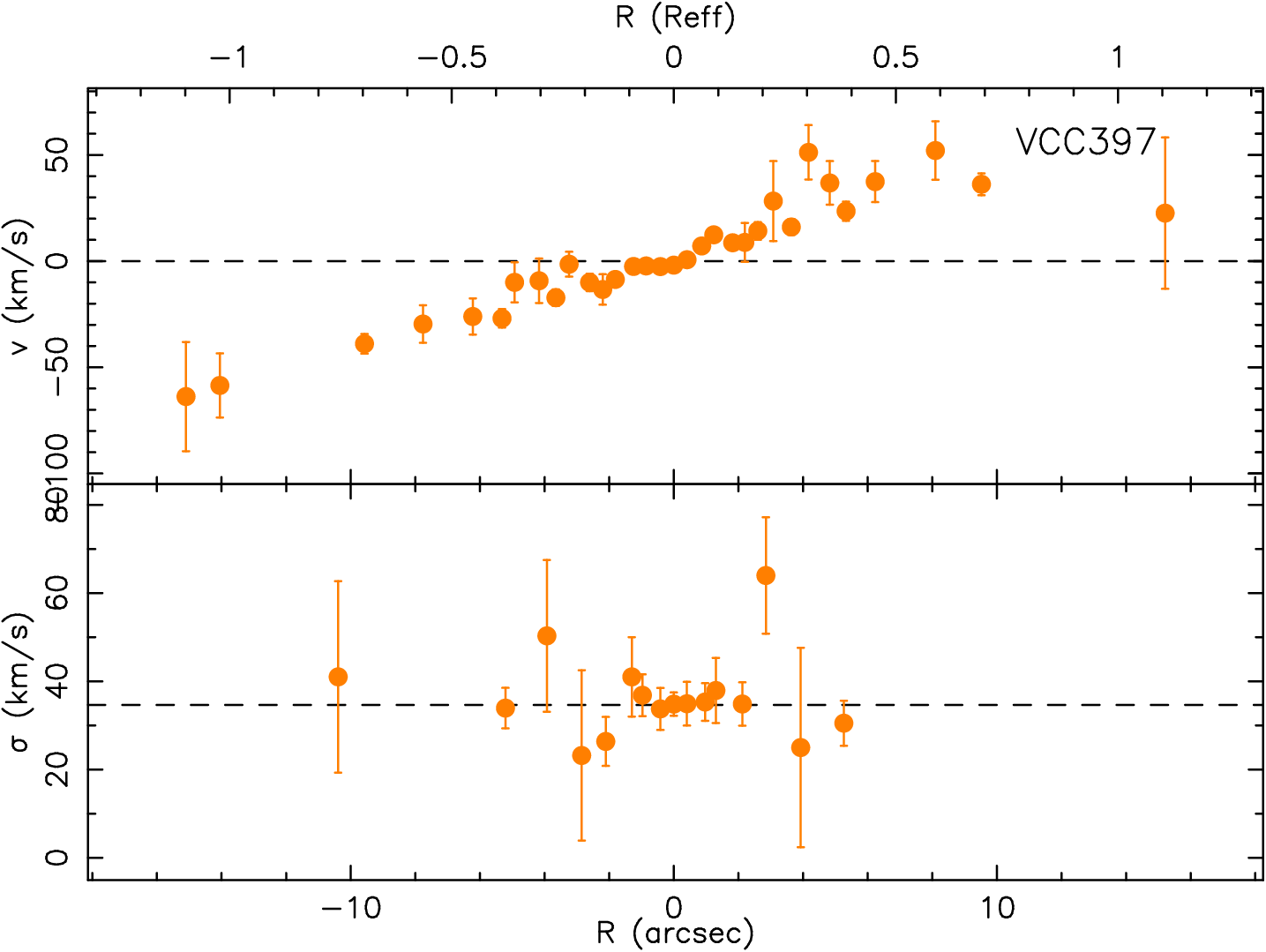}}
\resizebox{0.4\textwidth}{!}{\includegraphics[angle=-90]{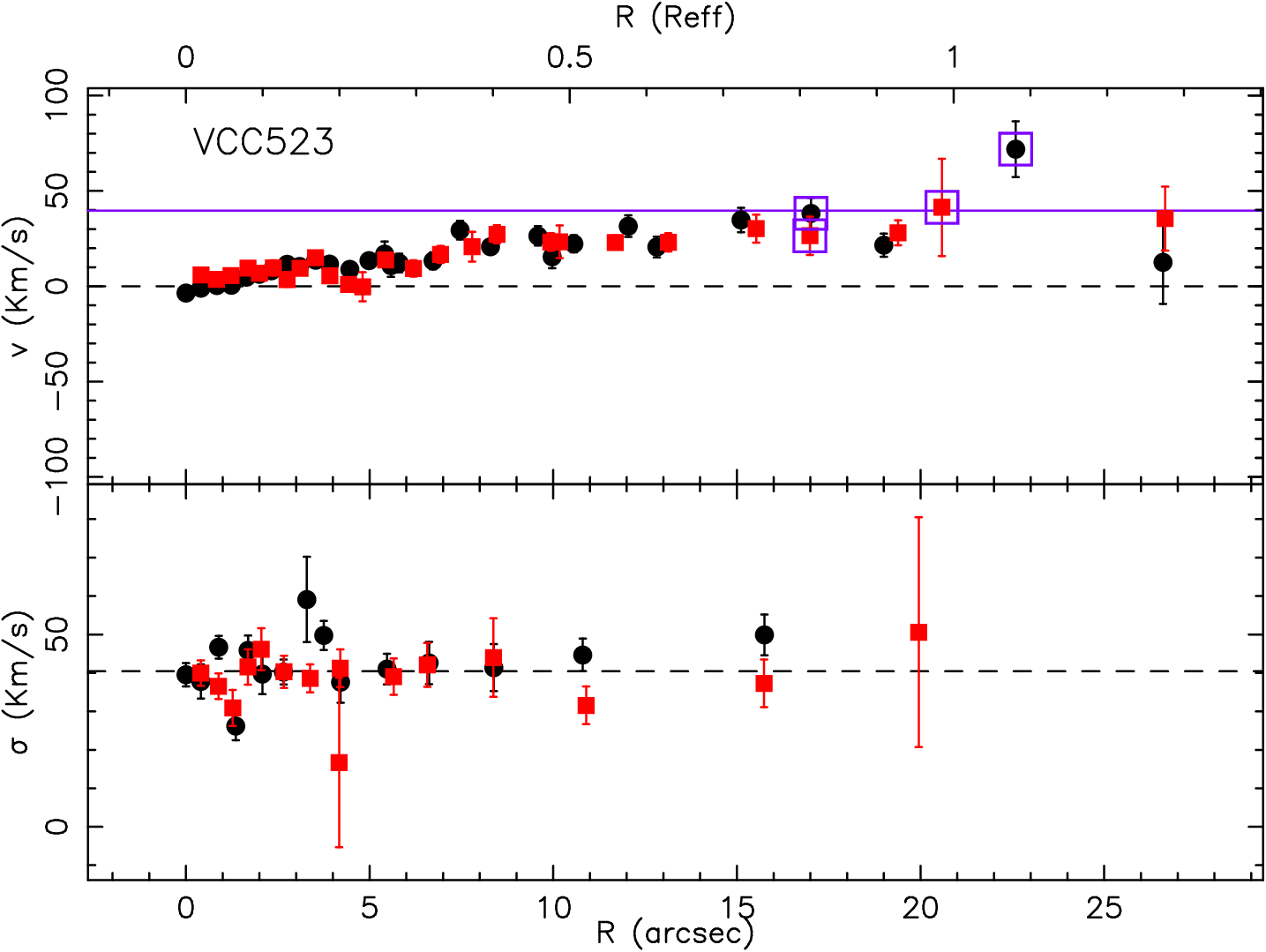}}
\resizebox{0.4\textwidth}{!}{\includegraphics[angle=-90]{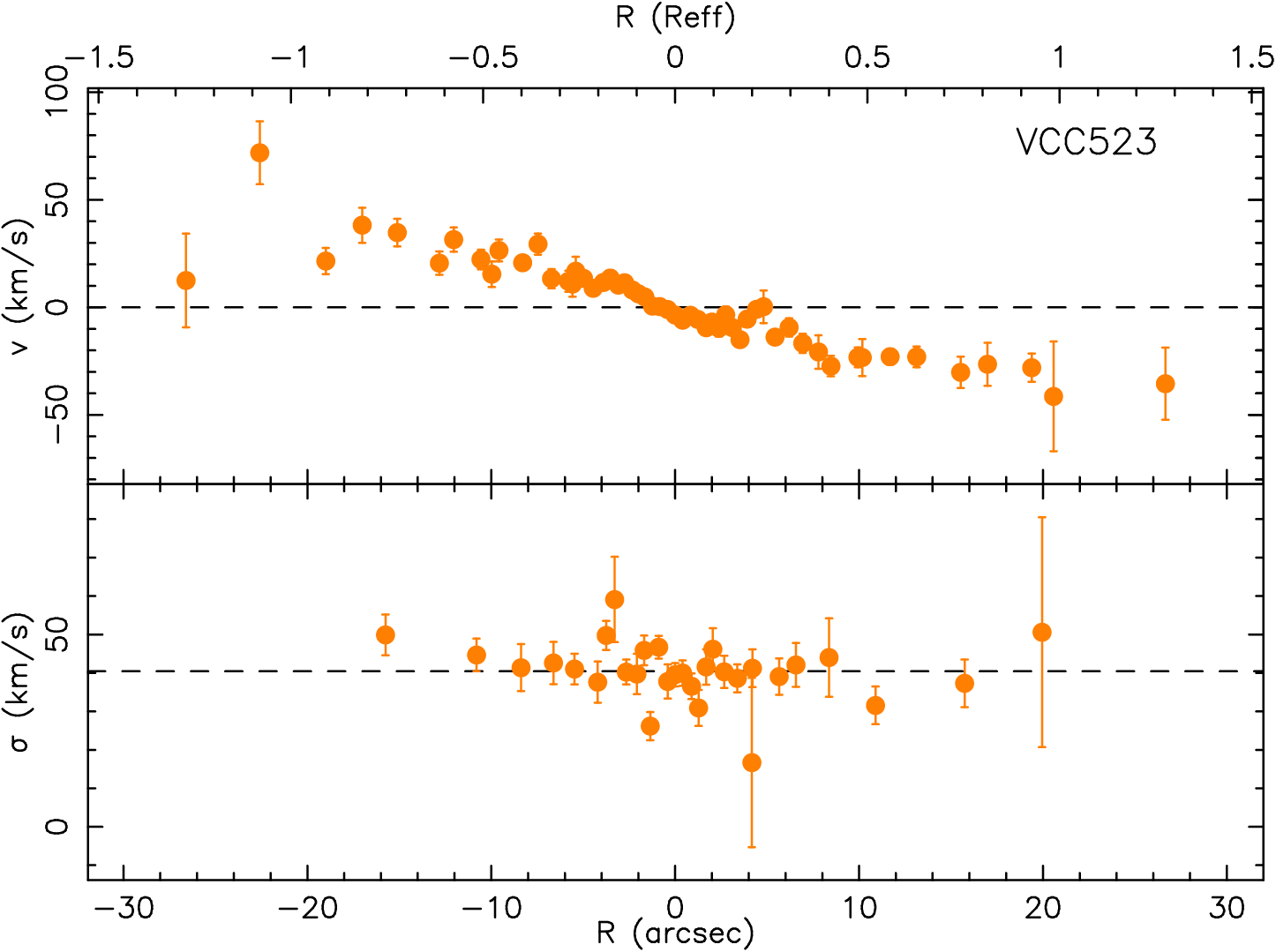}}
\caption{Continued}
\end{figure*}
\addtocounter{figure}{-1}
\begin{figure*}
\centering
\resizebox{0.4\textwidth}{!}{\includegraphics[angle=-90]{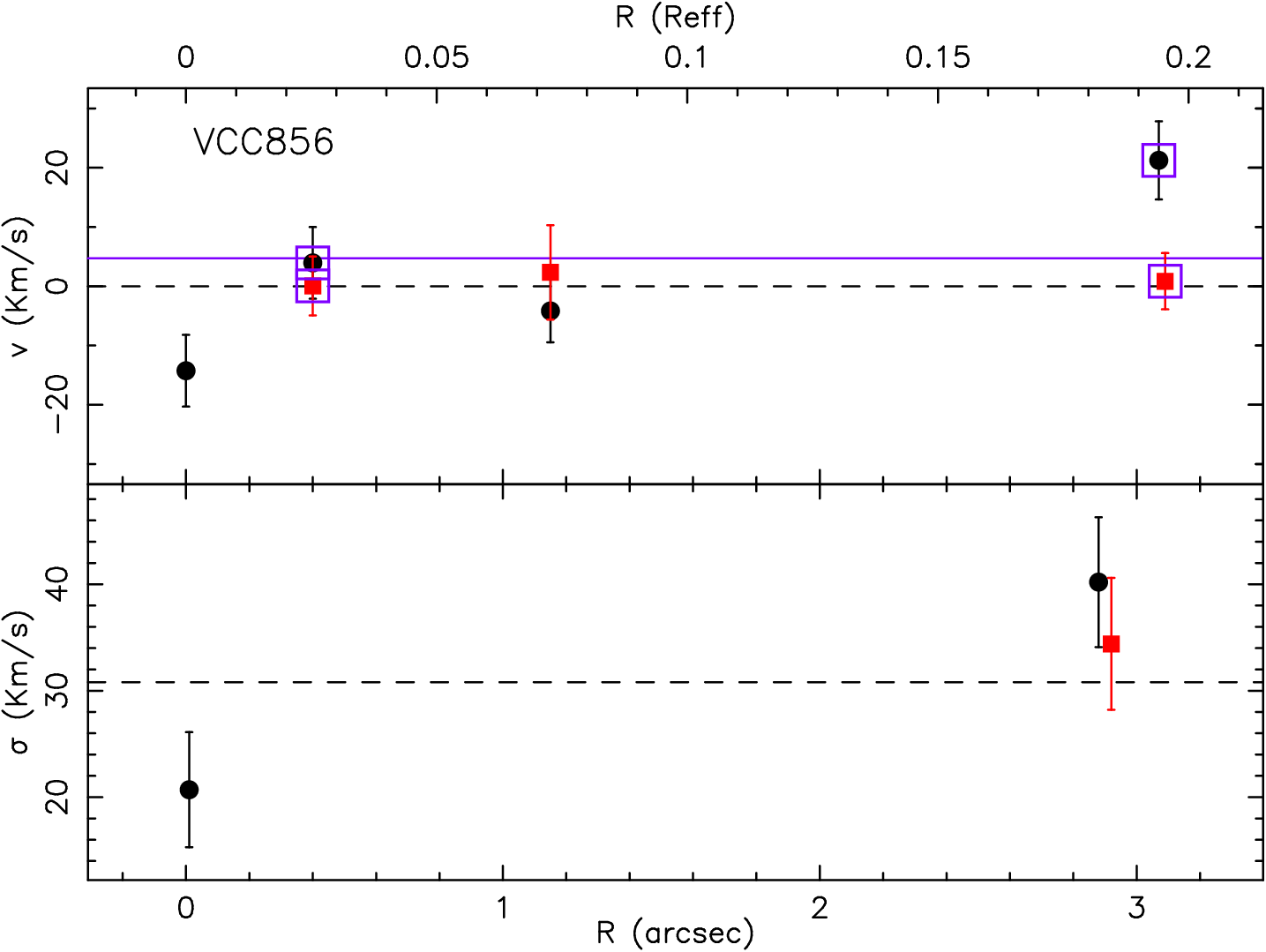}}
\resizebox{0.4\textwidth}{!}{\includegraphics[angle=-90]{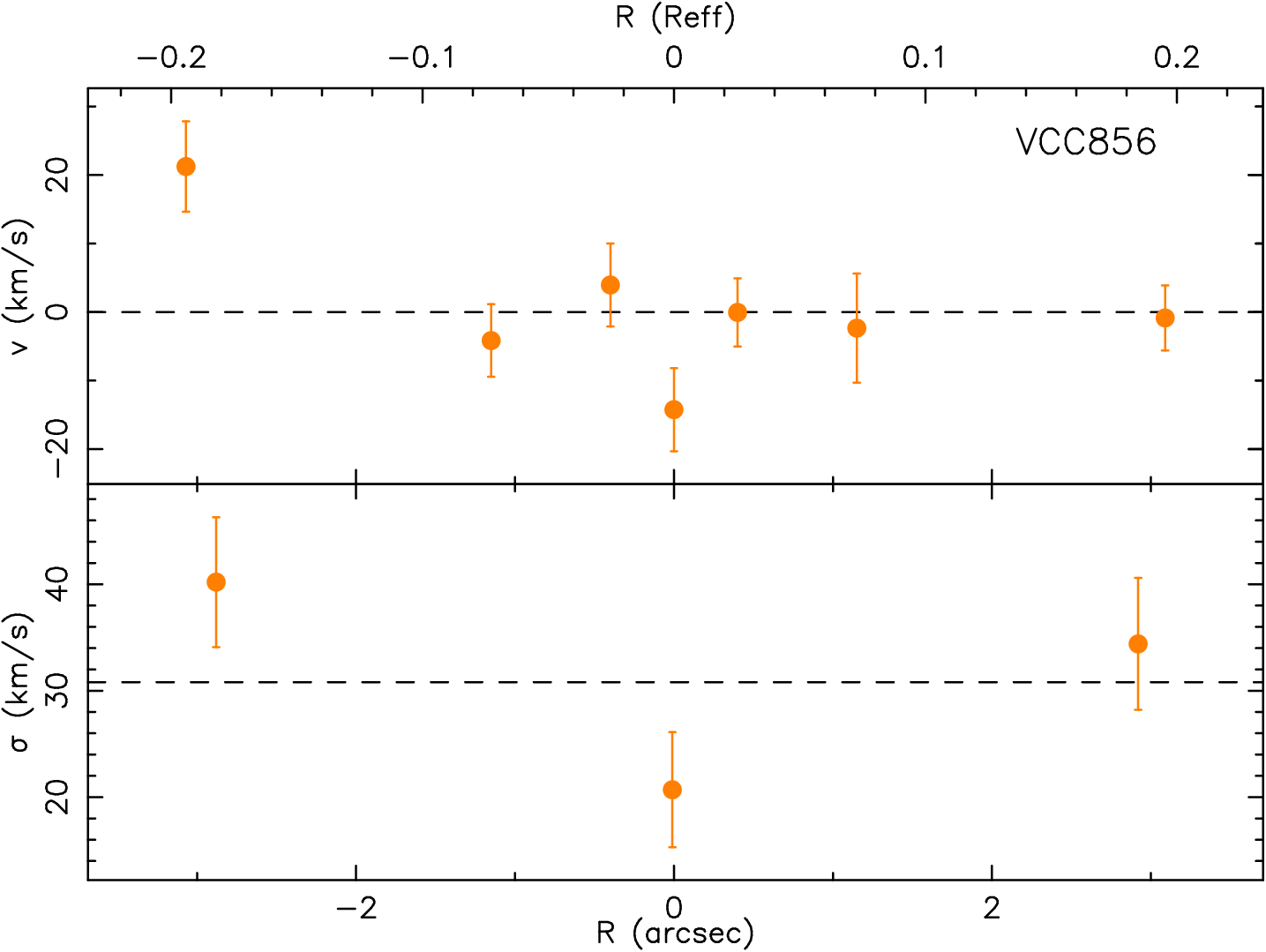}}
\resizebox{0.4\textwidth}{!}{\includegraphics[angle=-90]{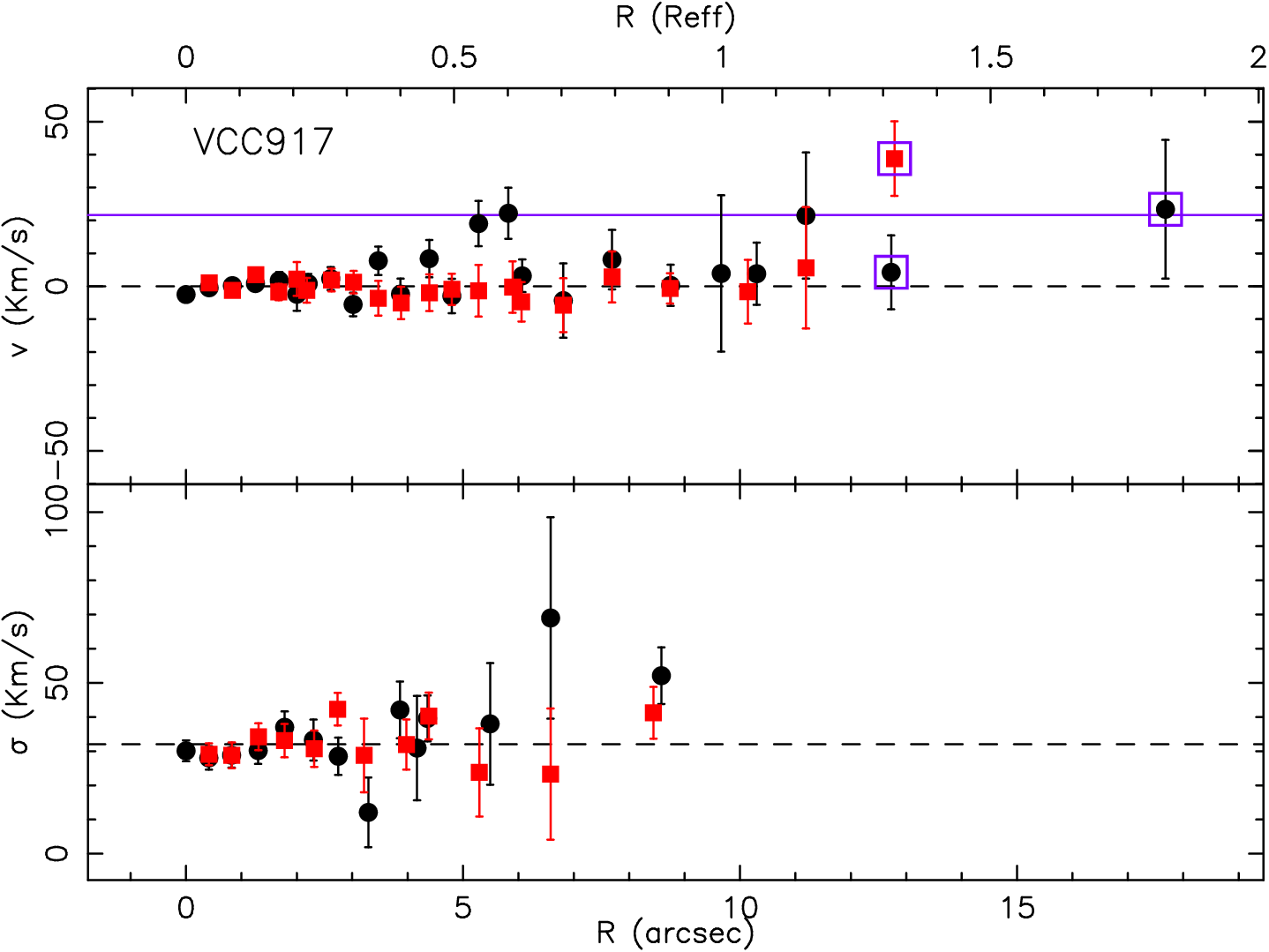}}
\resizebox{0.4\textwidth}{!}{\includegraphics[angle=-90]{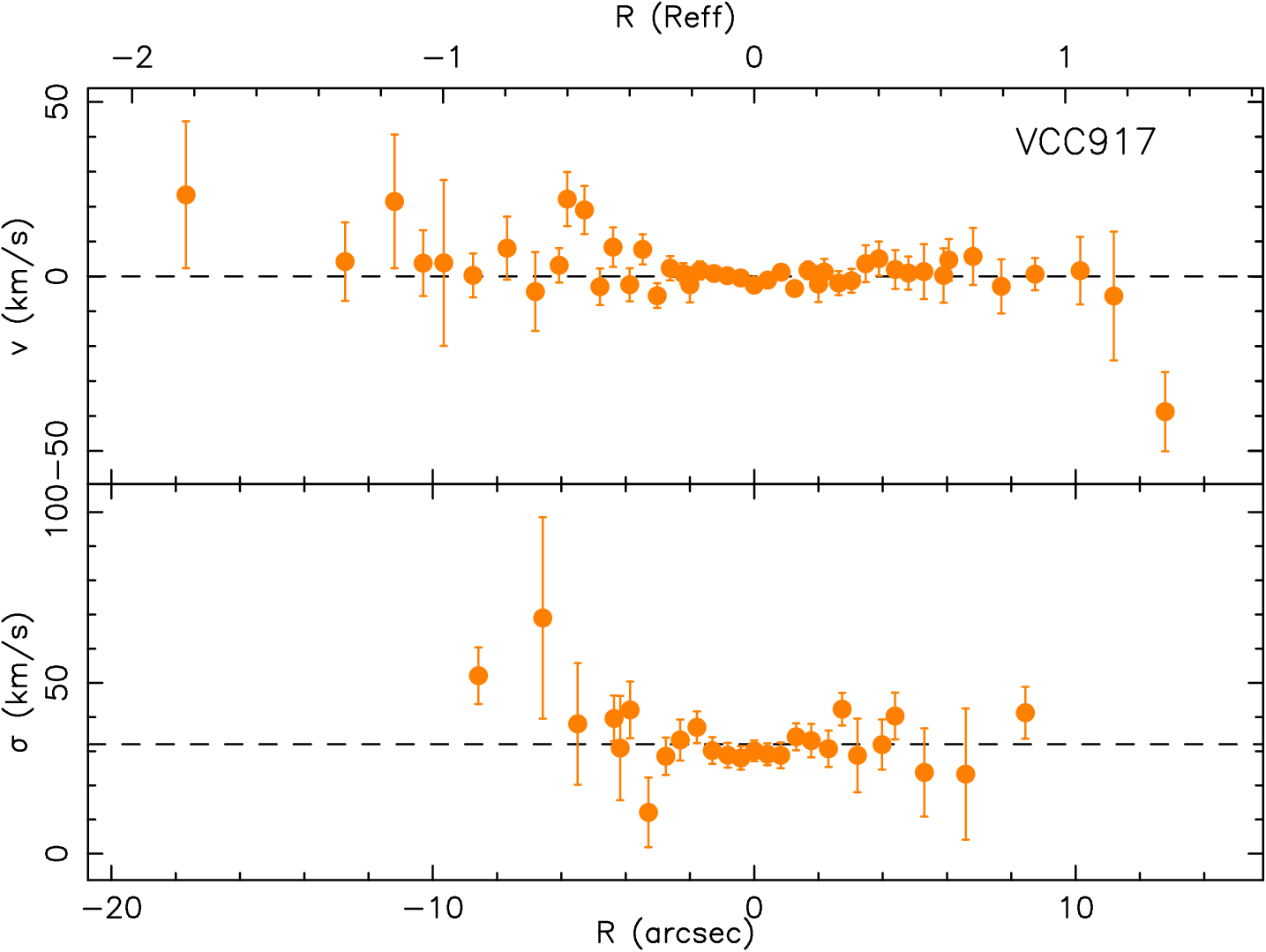}}
\resizebox{0.4\textwidth}{!}{\includegraphics[angle=-90]{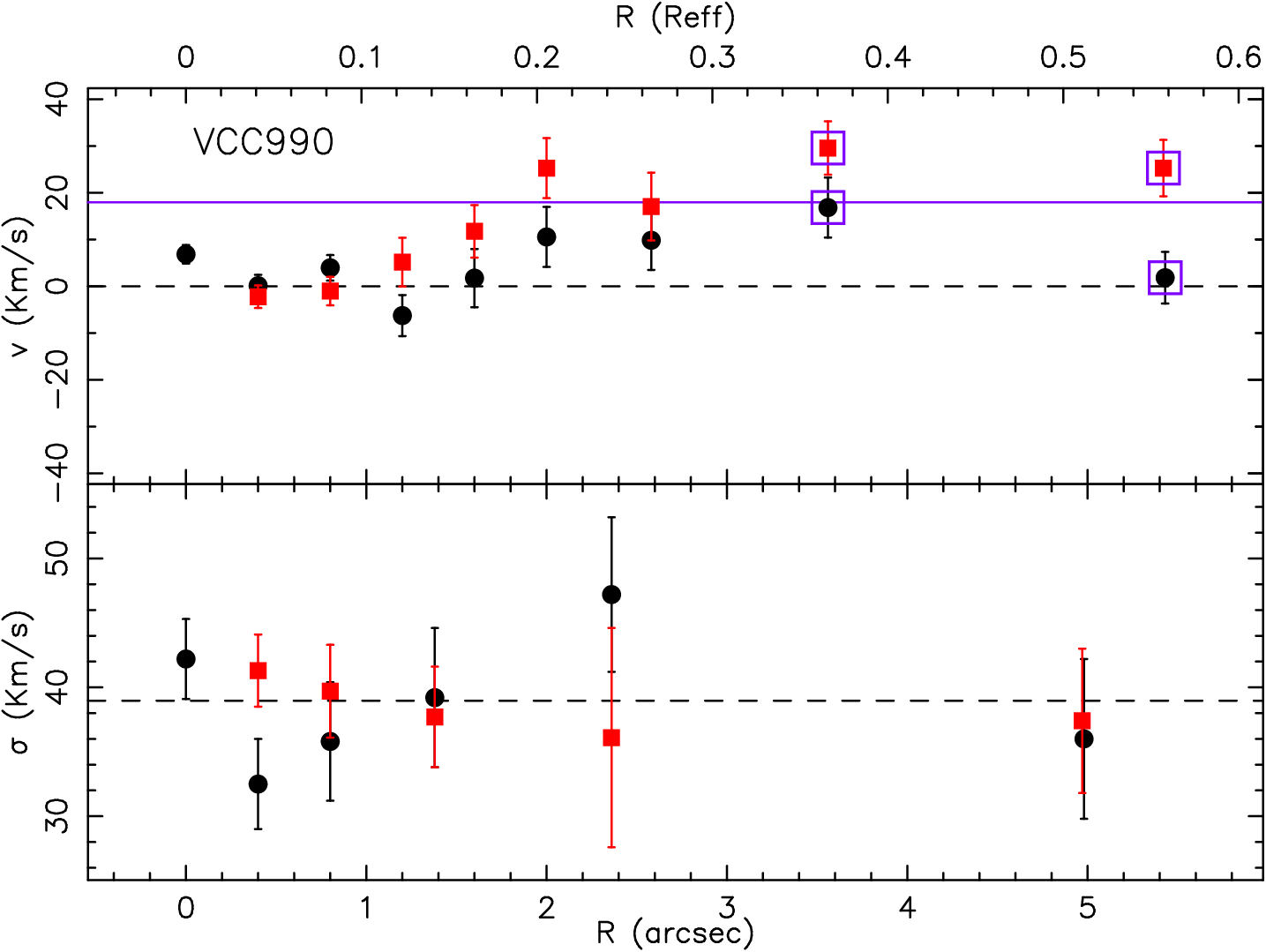}}
\resizebox{0.4\textwidth}{!}{\includegraphics[angle=-90]{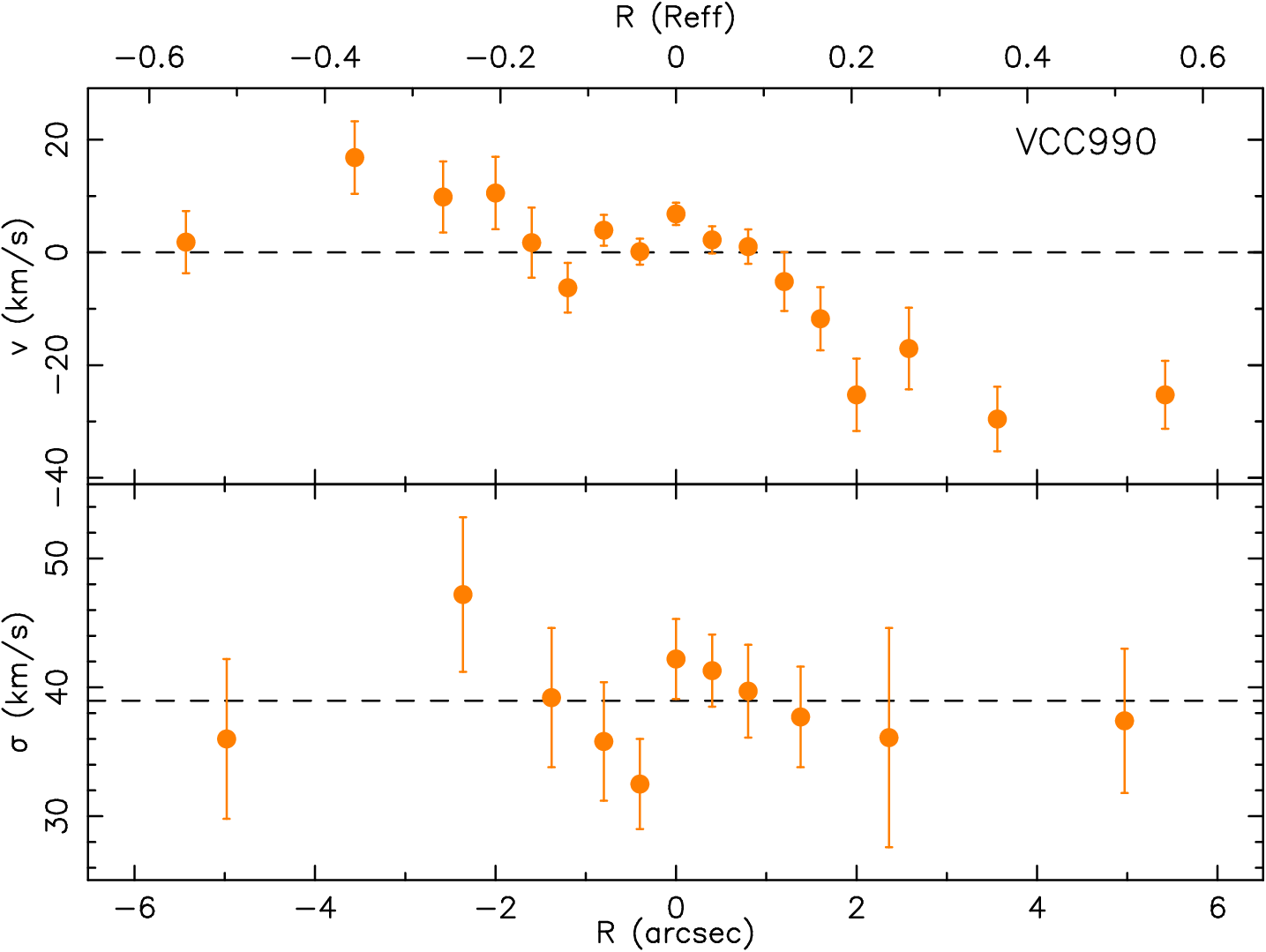}}
\resizebox{0.4\textwidth}{!}{\includegraphics[angle=-90]{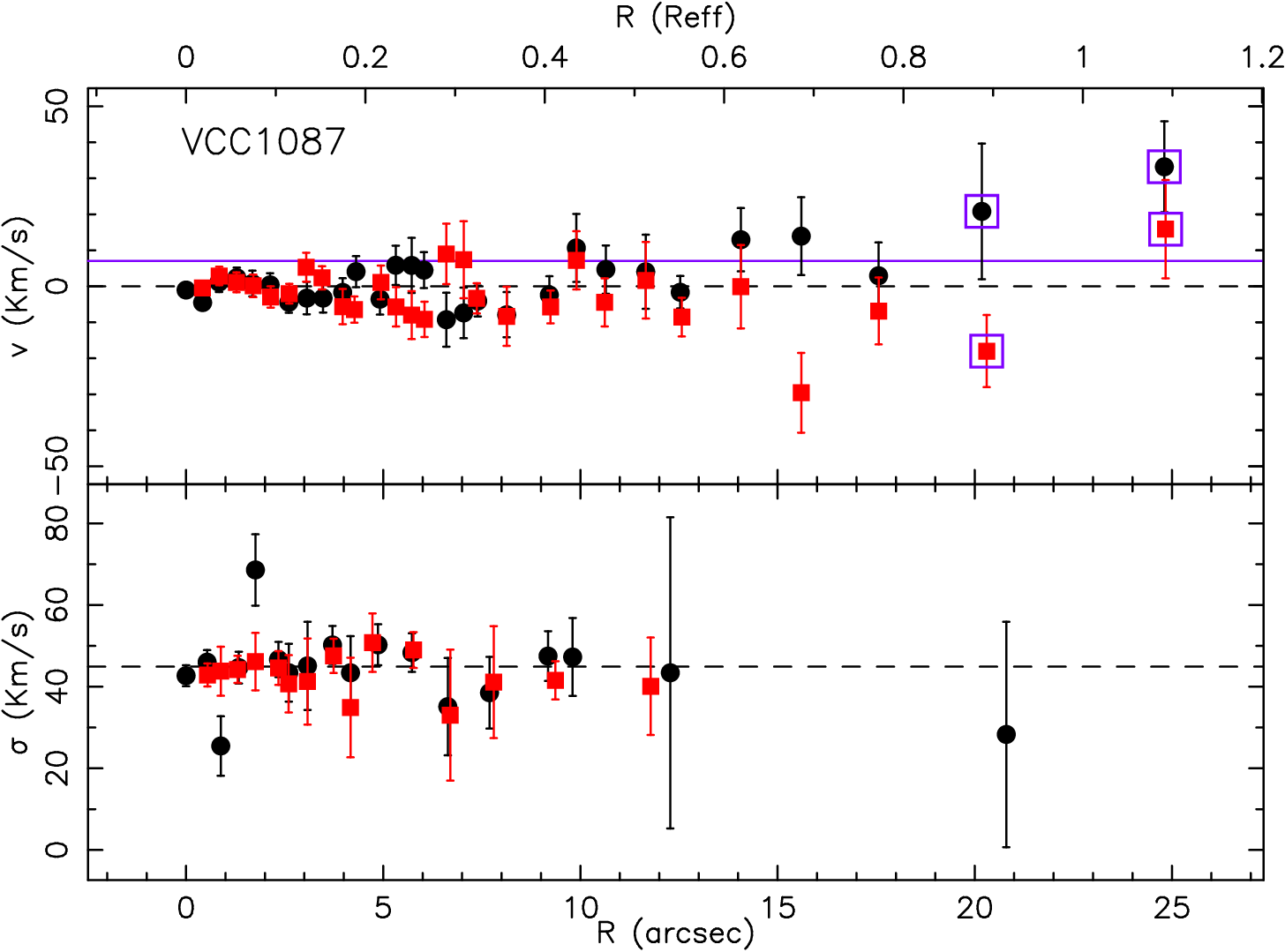}}
\resizebox{0.4\textwidth}{!}{\includegraphics[angle=-90]{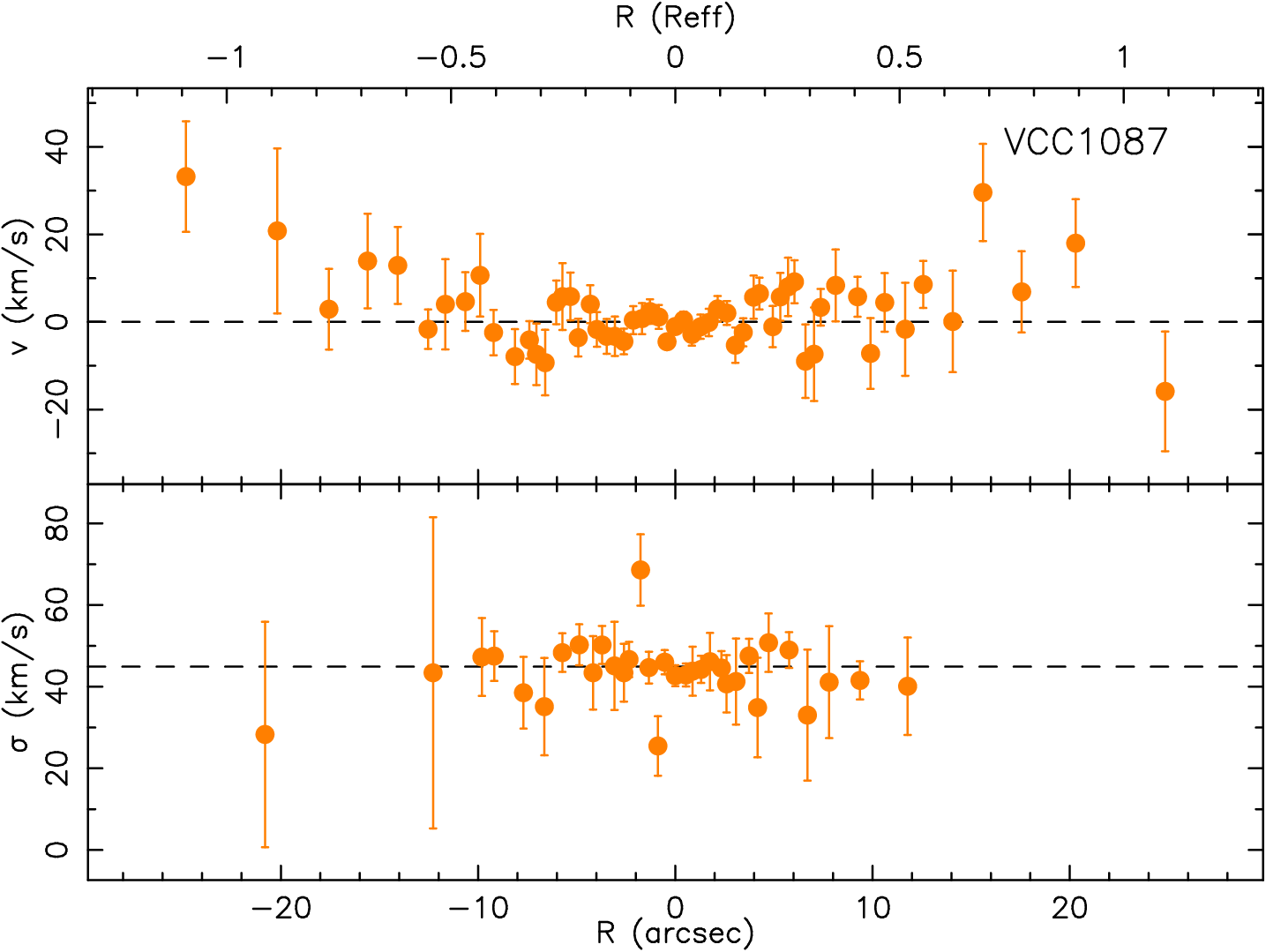}}
\caption{Continued}
\end{figure*}
\addtocounter{figure}{-1}
\begin{figure*}
\centering
\resizebox{0.4\textwidth}{!}{\includegraphics[angle=-90]{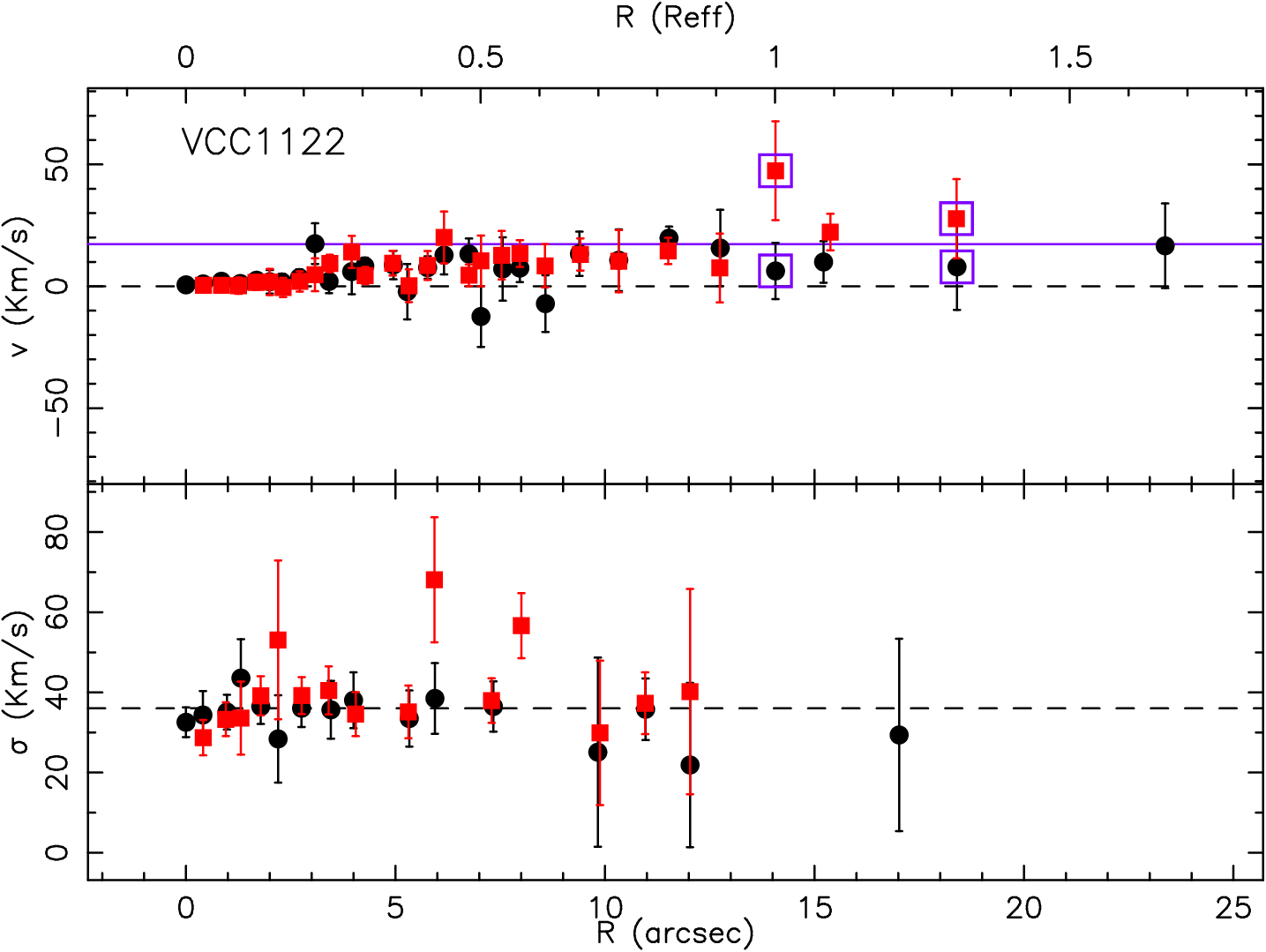}}
\resizebox{0.4\textwidth}{!}{\includegraphics[angle=-90]{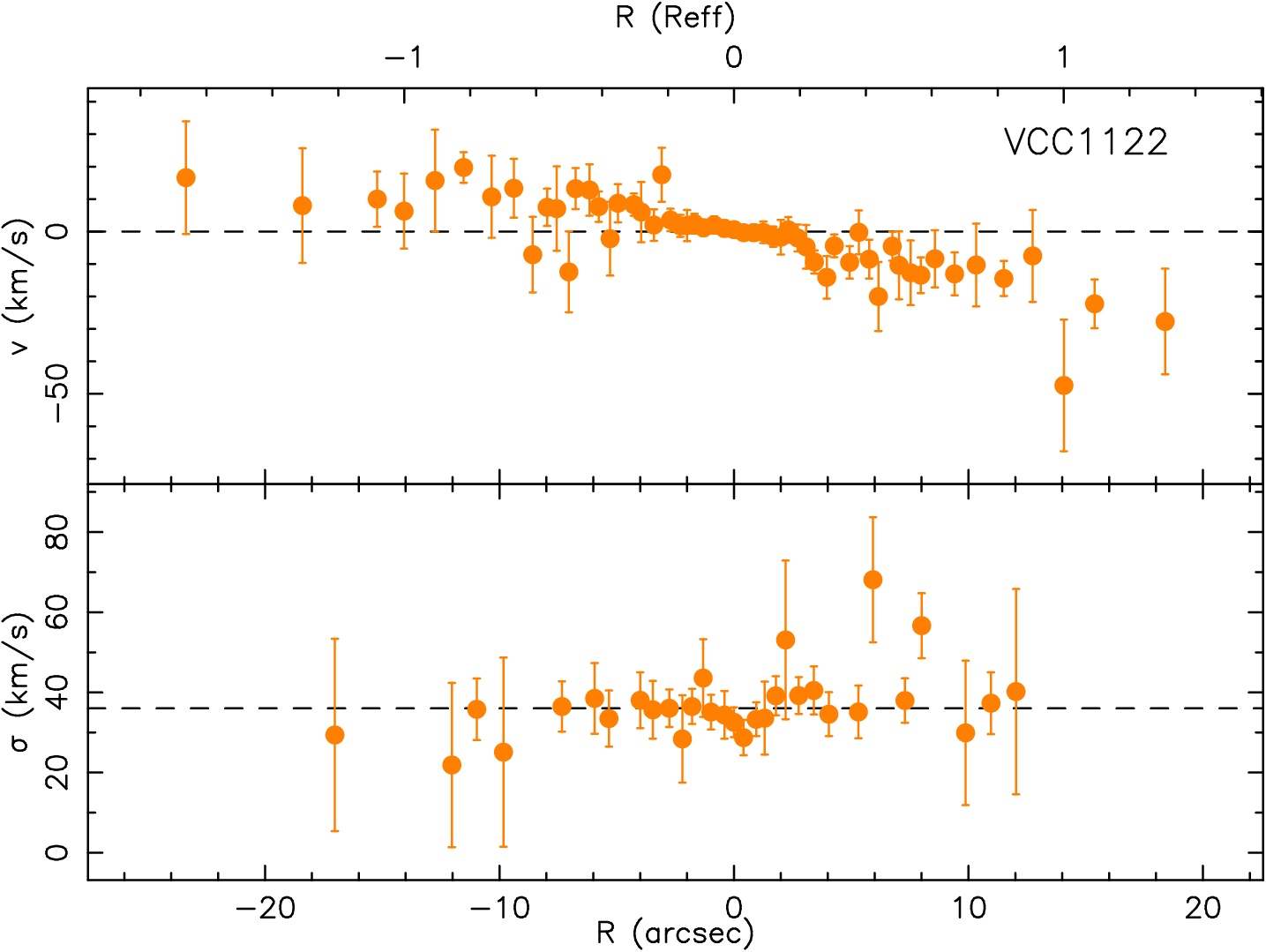}}
\resizebox{0.4\textwidth}{!}{\includegraphics[angle=-90]{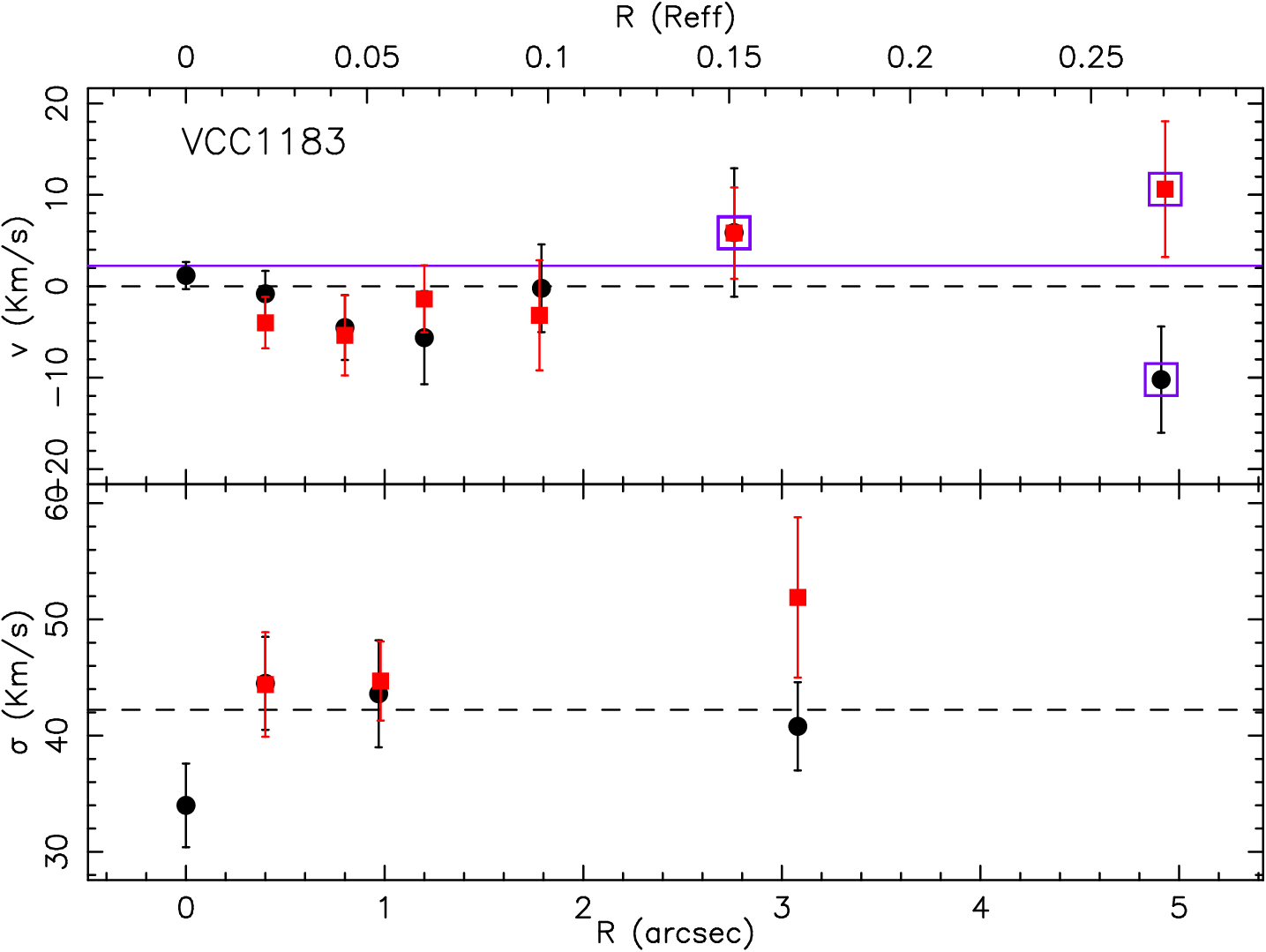}}
\resizebox{0.4\textwidth}{!}{\includegraphics[angle=-90]{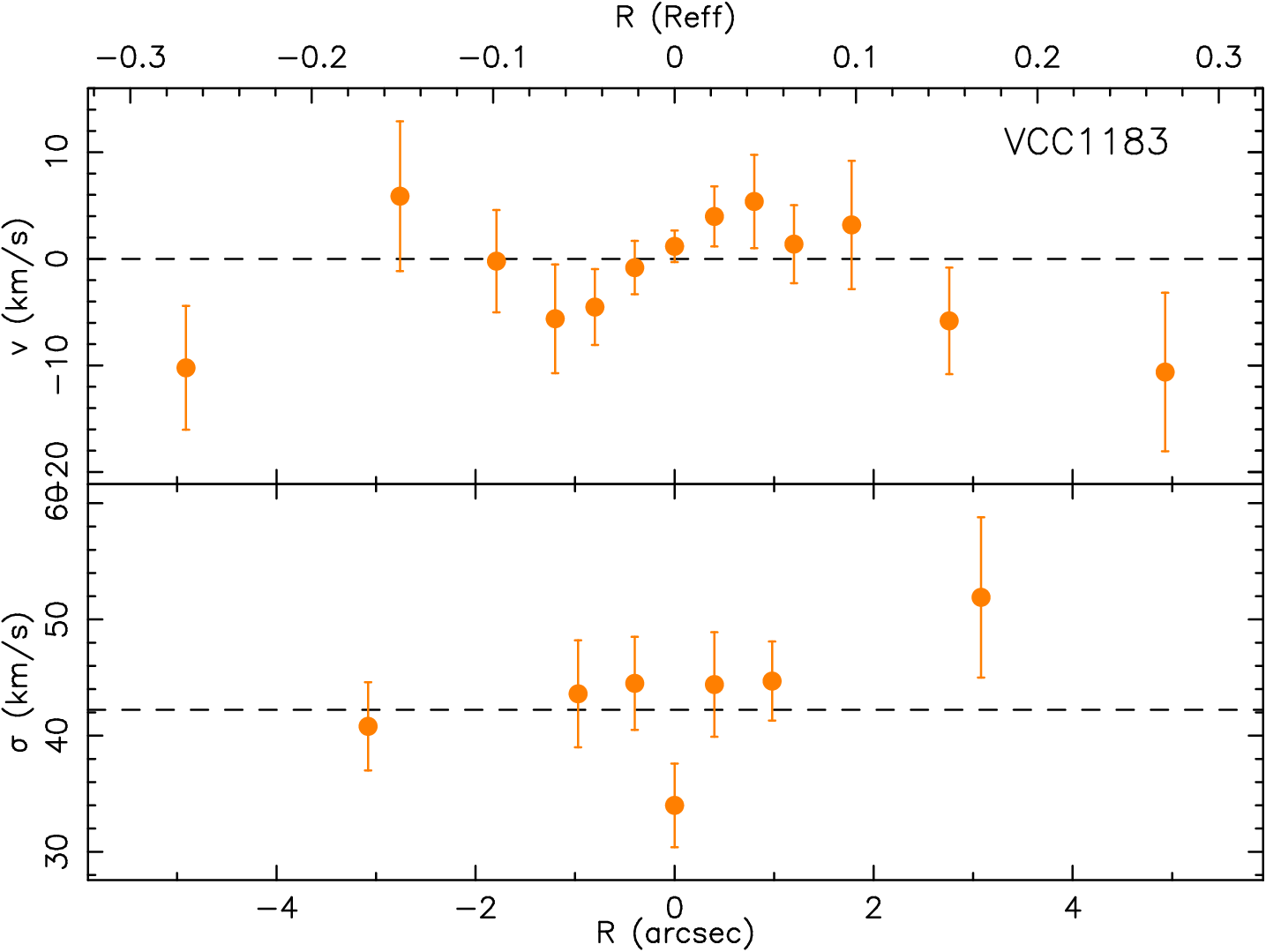}}
\resizebox{0.4\textwidth}{!}{\includegraphics[angle=-90]{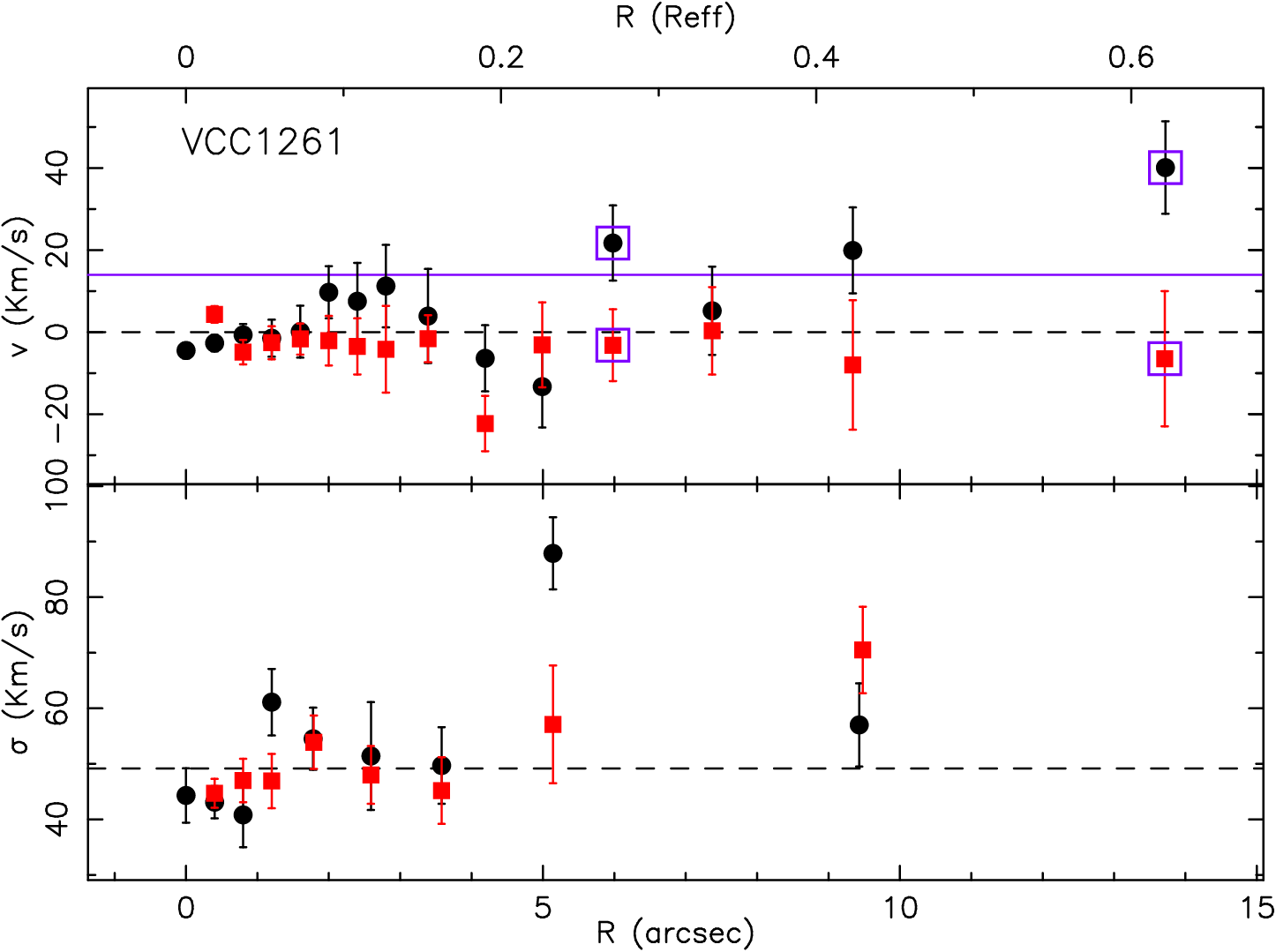}}
\resizebox{0.4\textwidth}{!}{\includegraphics[angle=-90]{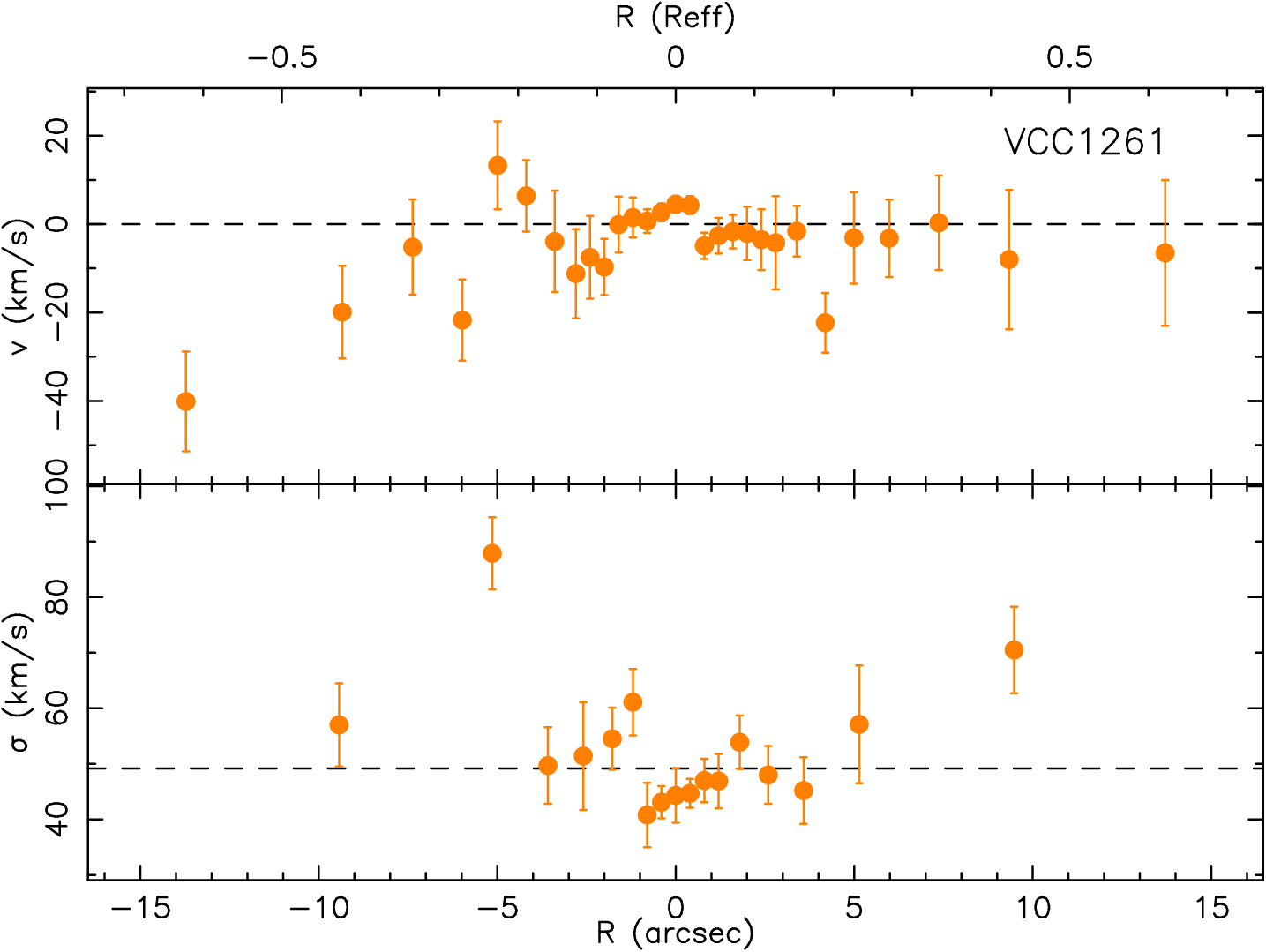}}
\resizebox{0.4\textwidth}{!}{\includegraphics[angle=-90]{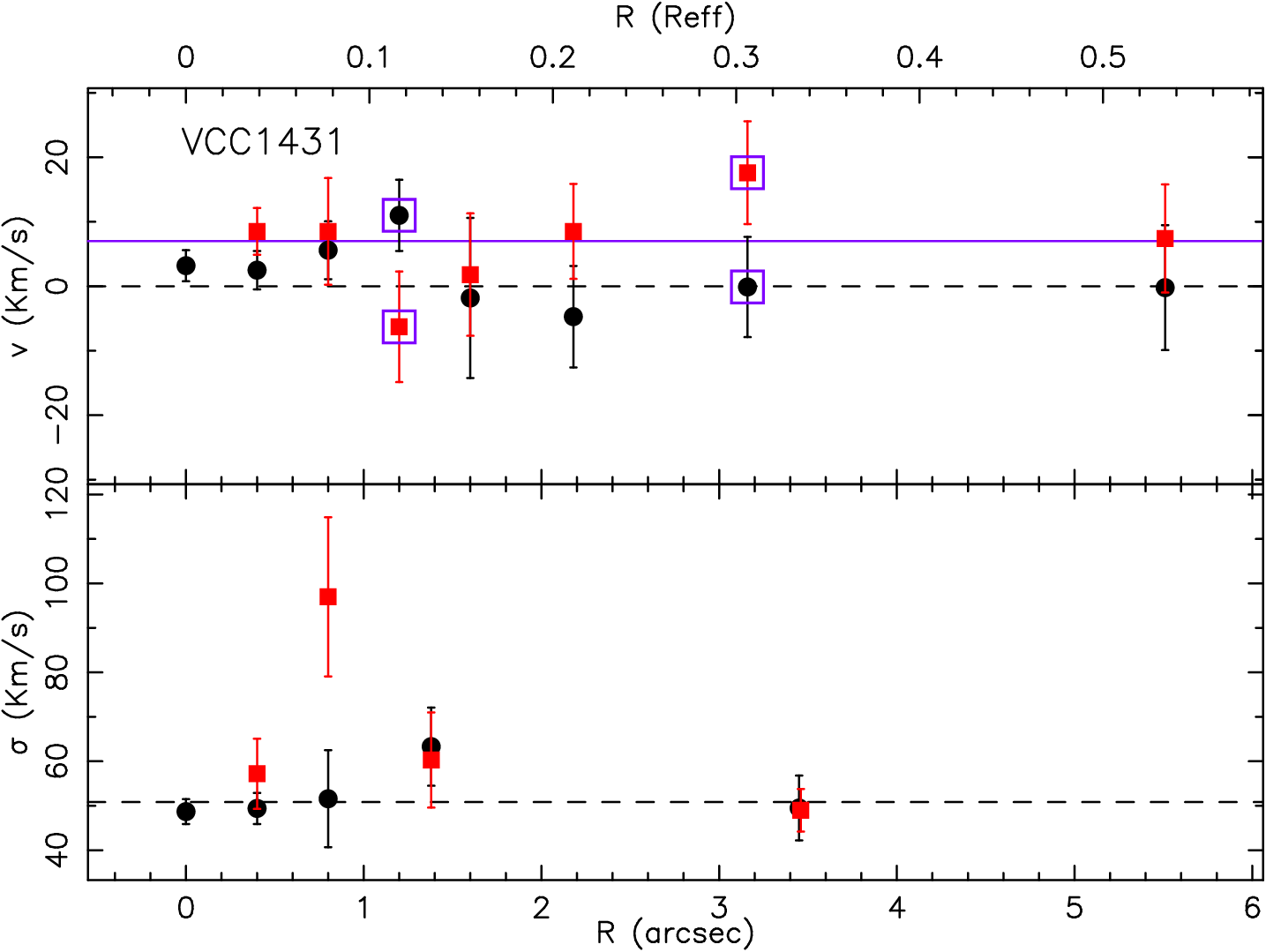}}
\resizebox{0.4\textwidth}{!}{\includegraphics[angle=-90]{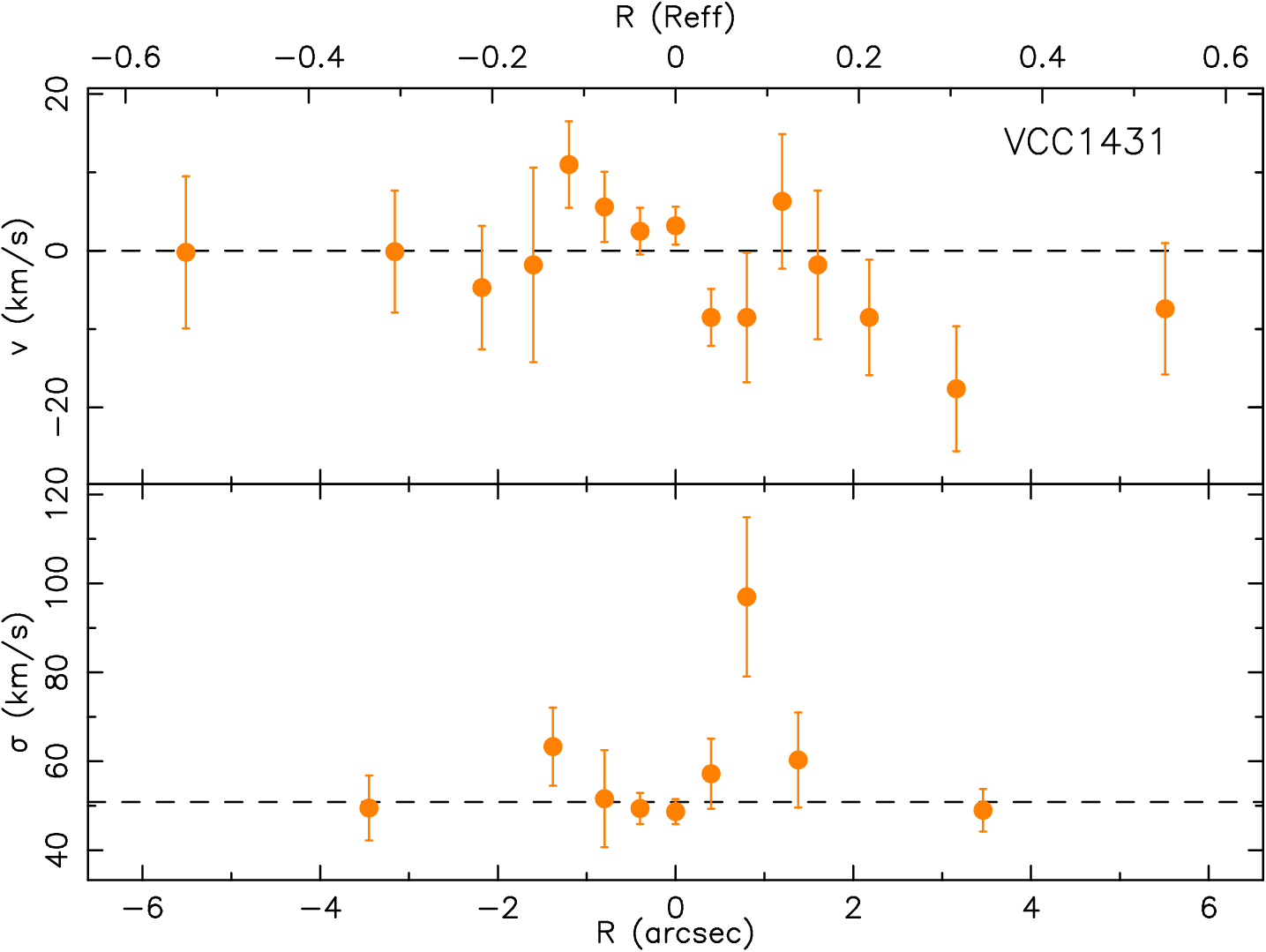}}
\caption{Continued}
\end{figure*}
\addtocounter{figure}{-1}
\begin{figure*}
\centering
\resizebox{0.4\textwidth}{!}{\includegraphics[angle=-90]{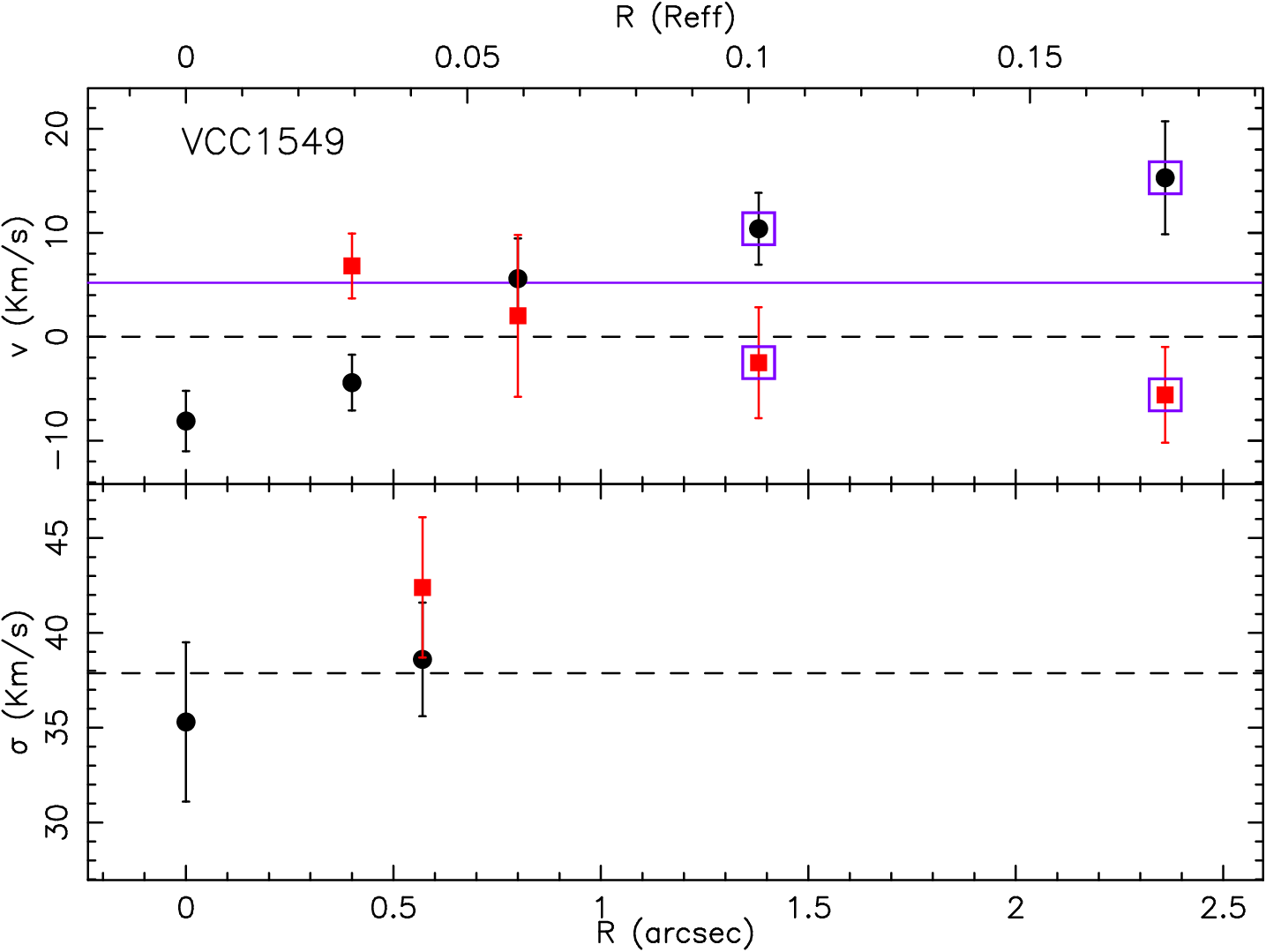}}
\resizebox{0.4\textwidth}{!}{\includegraphics[angle=-90]{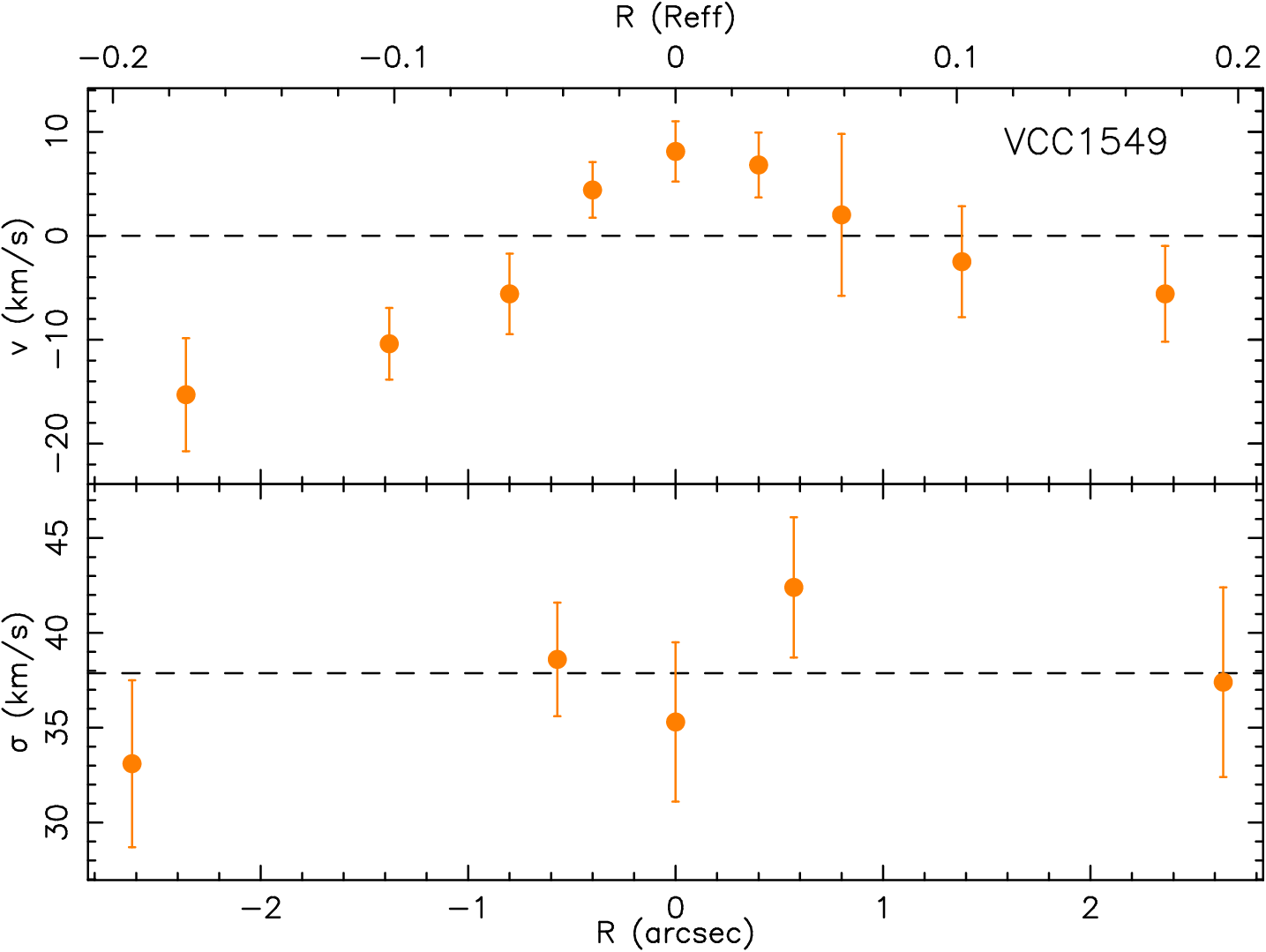}}
\resizebox{0.4\textwidth}{!}{\includegraphics[angle=-90]{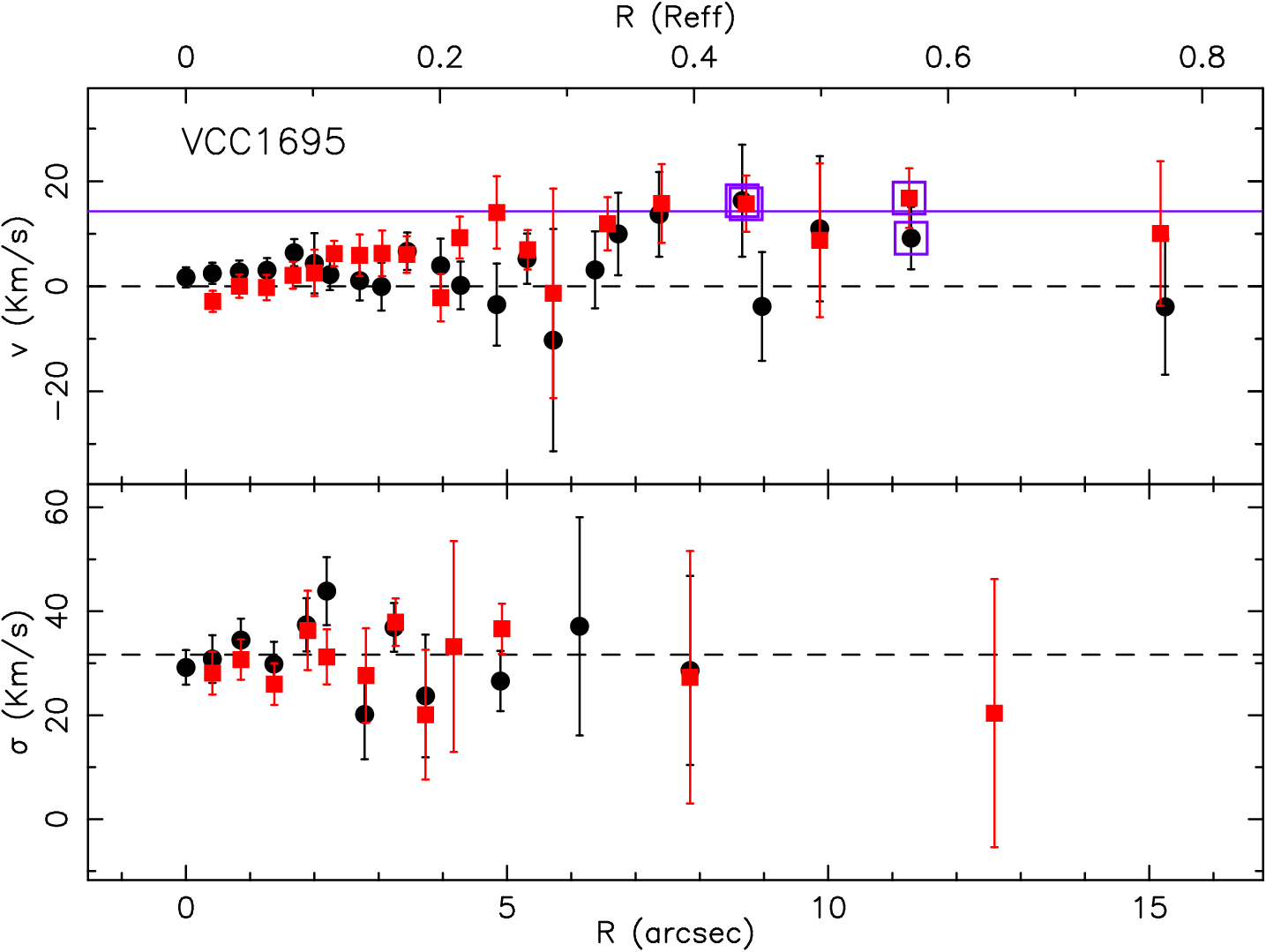}}
\resizebox{0.4\textwidth}{!}{\includegraphics[angle=-90]{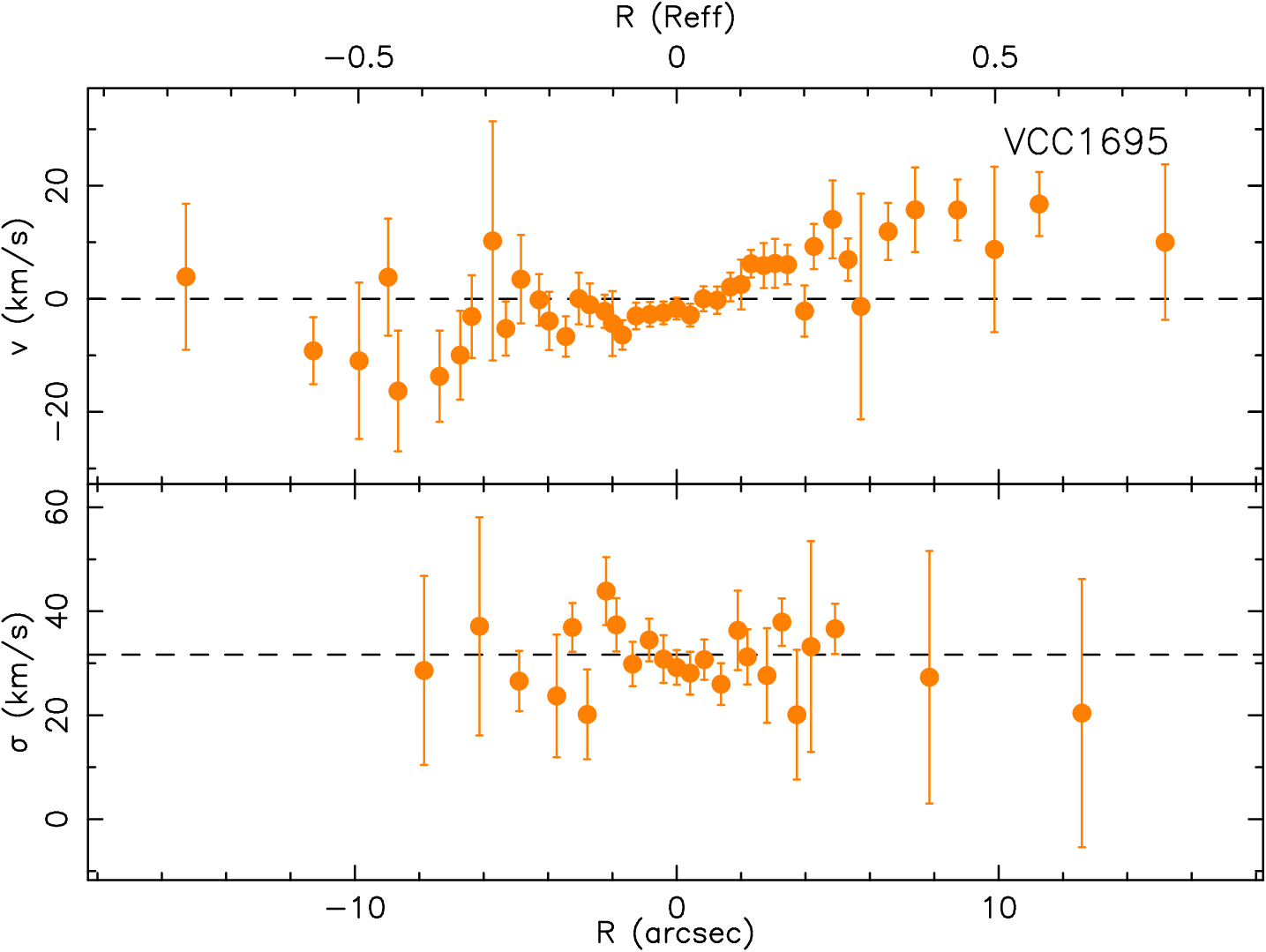}}
\resizebox{0.4\textwidth}{!}{\includegraphics[angle=-90]{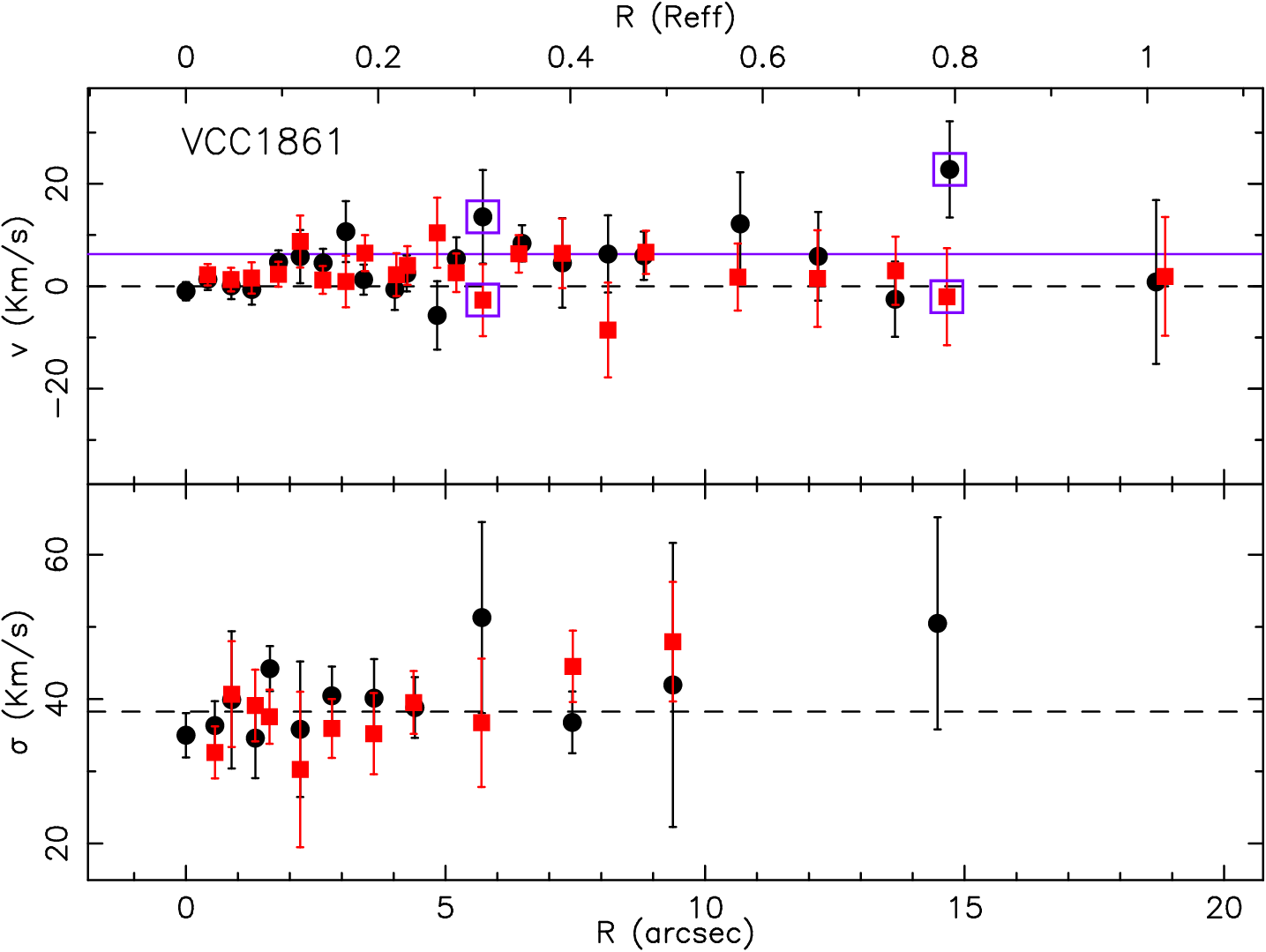}}
\resizebox{0.4\textwidth}{!}{\includegraphics[angle=-90]{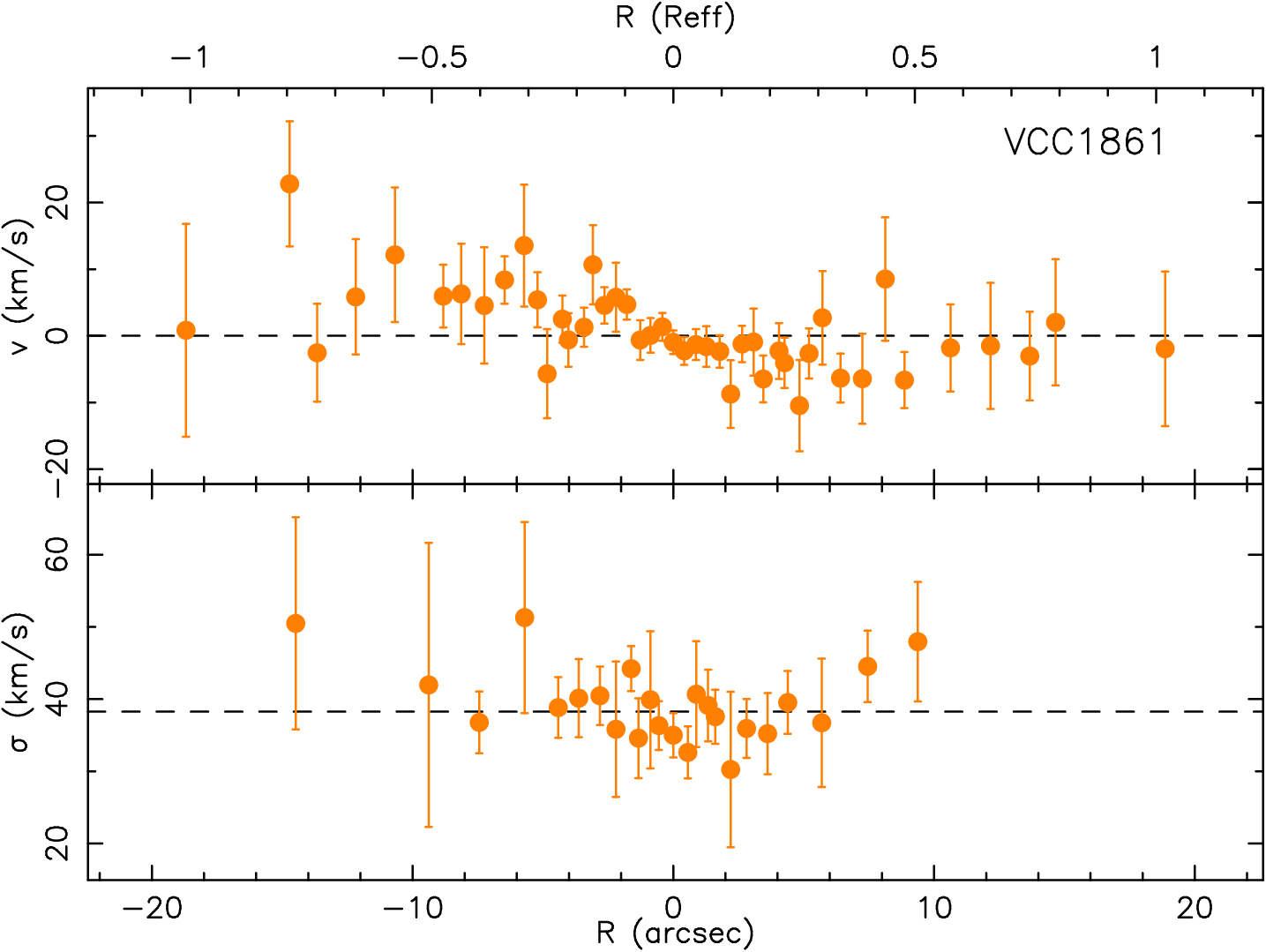}}
\resizebox{0.4\textwidth}{!}{\includegraphics[angle=-90]{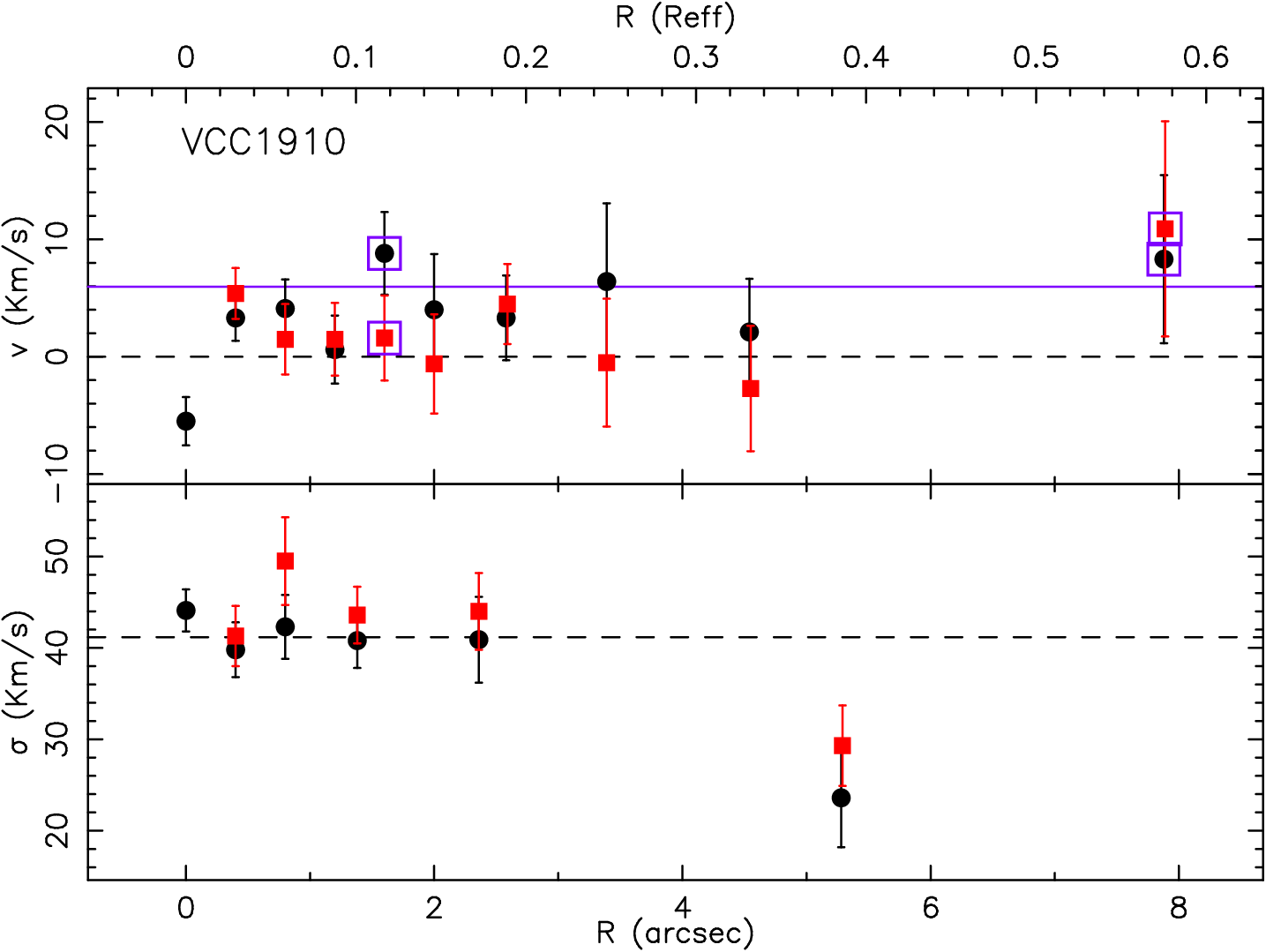}}
\resizebox{0.4\textwidth}{!}{\includegraphics[angle=-90]{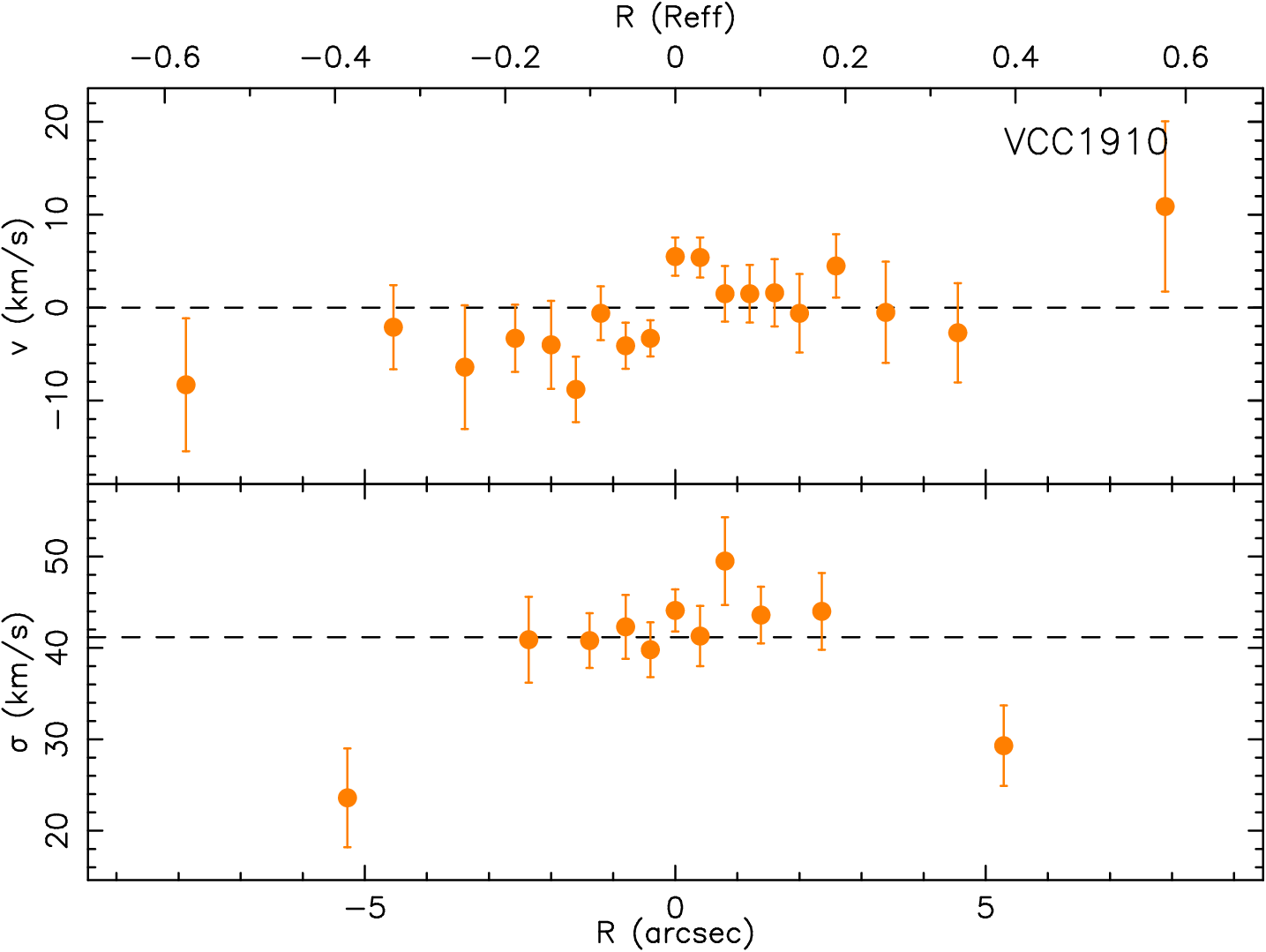}}
\caption{Continued}
\end{figure*}
\addtocounter{figure}{-1}
\begin{figure*}
\centering
\resizebox{0.4\textwidth}{!}{\includegraphics[angle=-90]{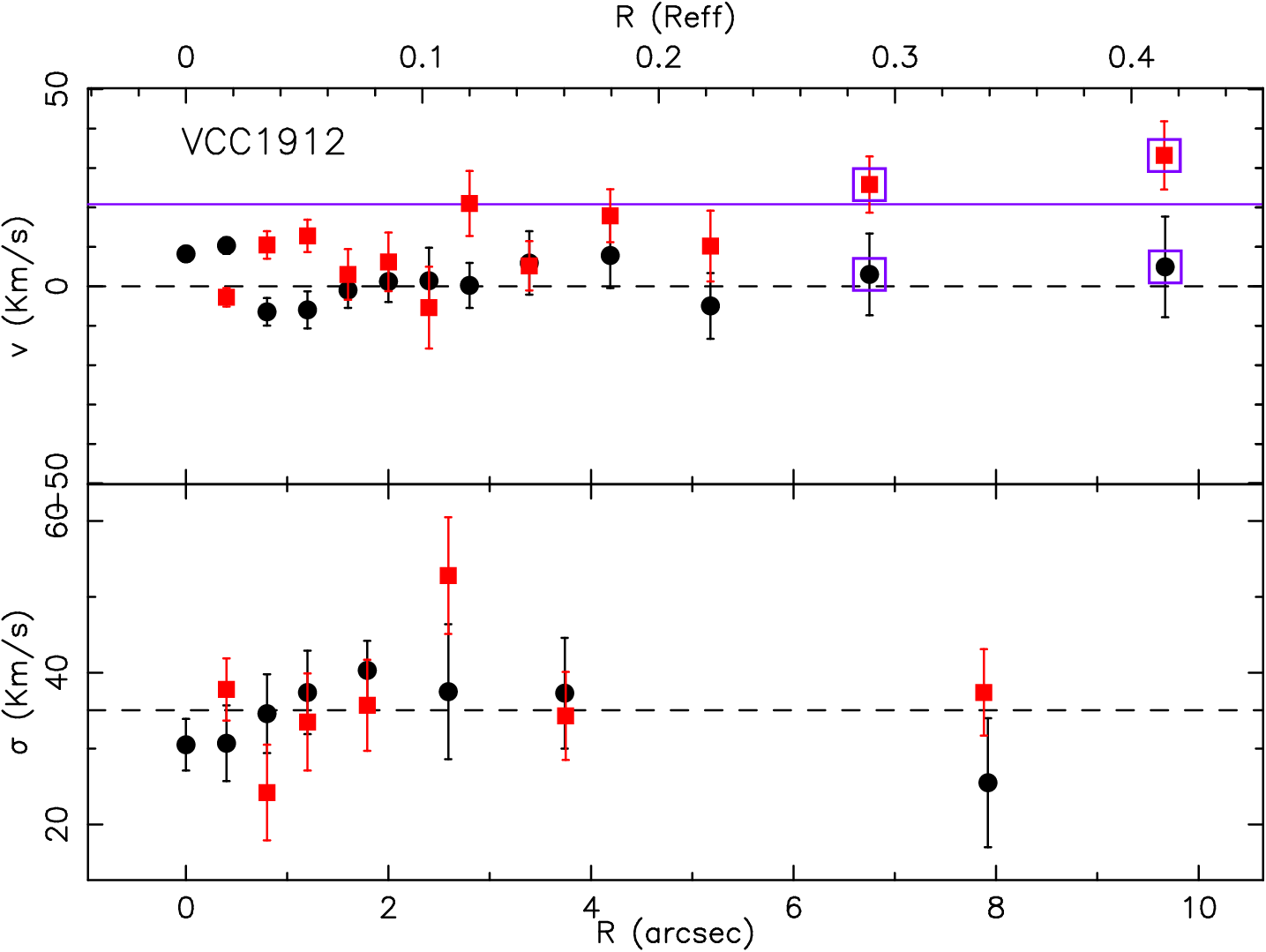}}
\resizebox{0.4\textwidth}{!}{\includegraphics[angle=-90]{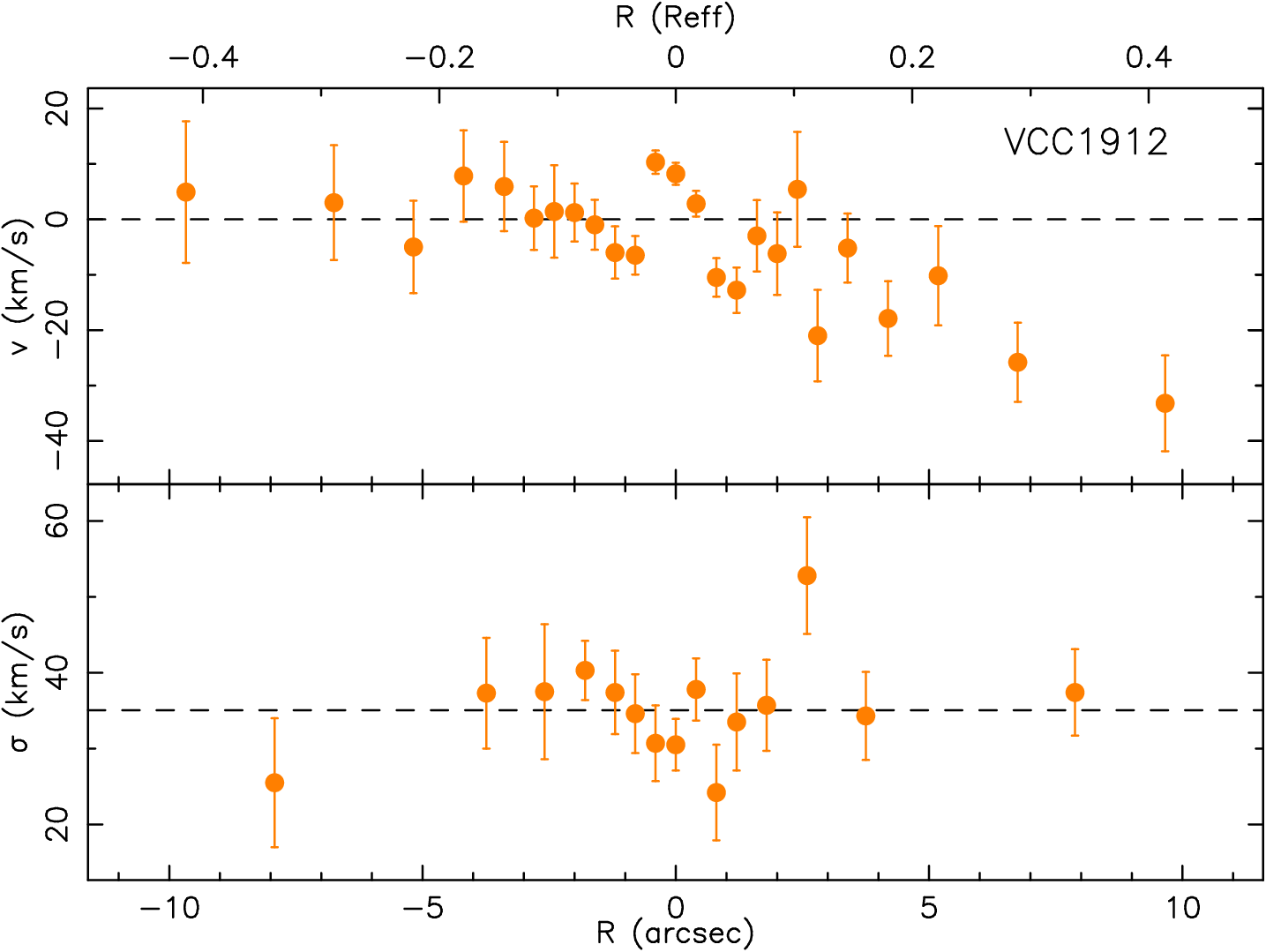}}
\resizebox{0.4\textwidth}{!}{\includegraphics[angle=-90]{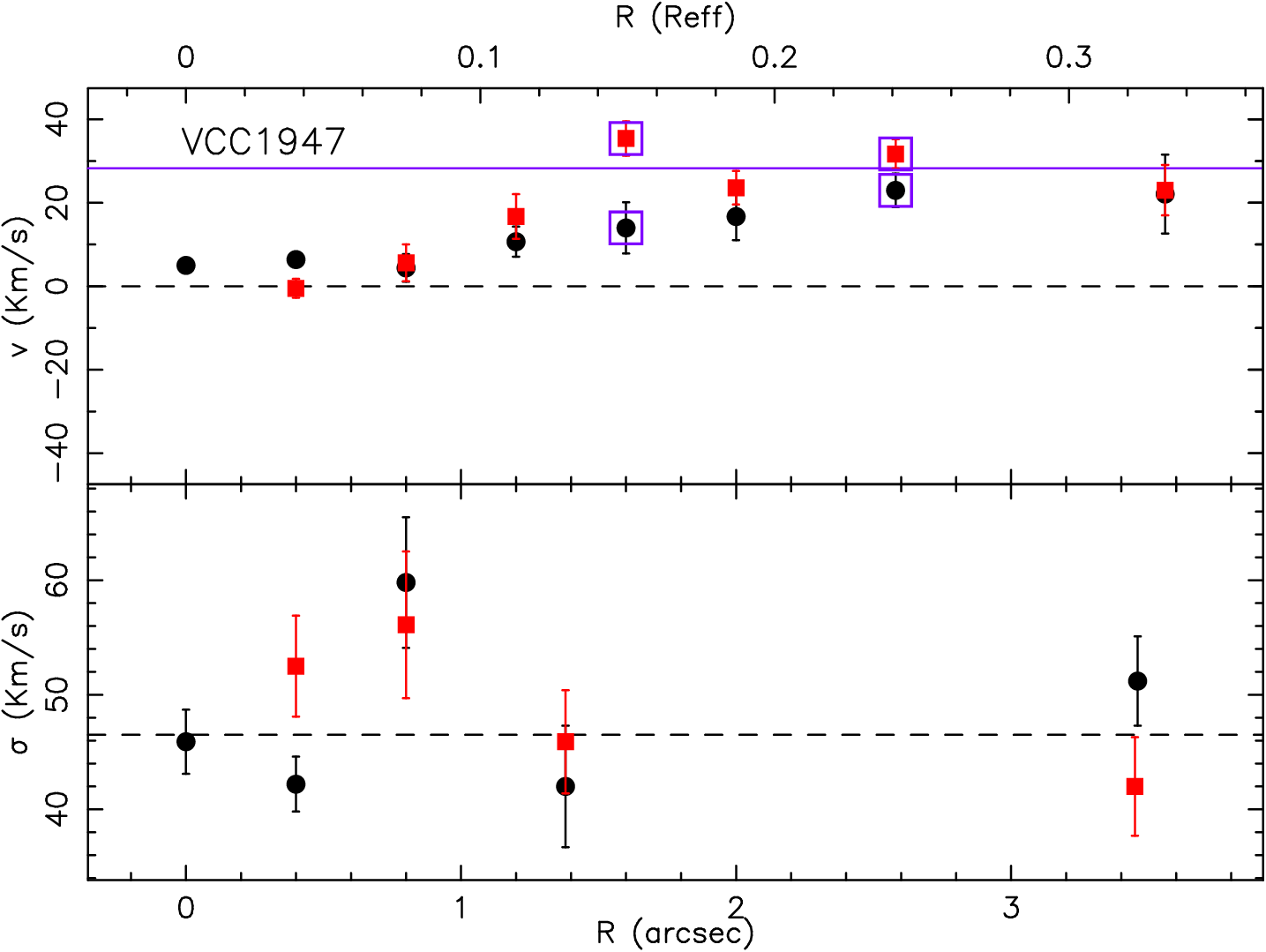}}
\resizebox{0.4\textwidth}{!}{\includegraphics[angle=-90]{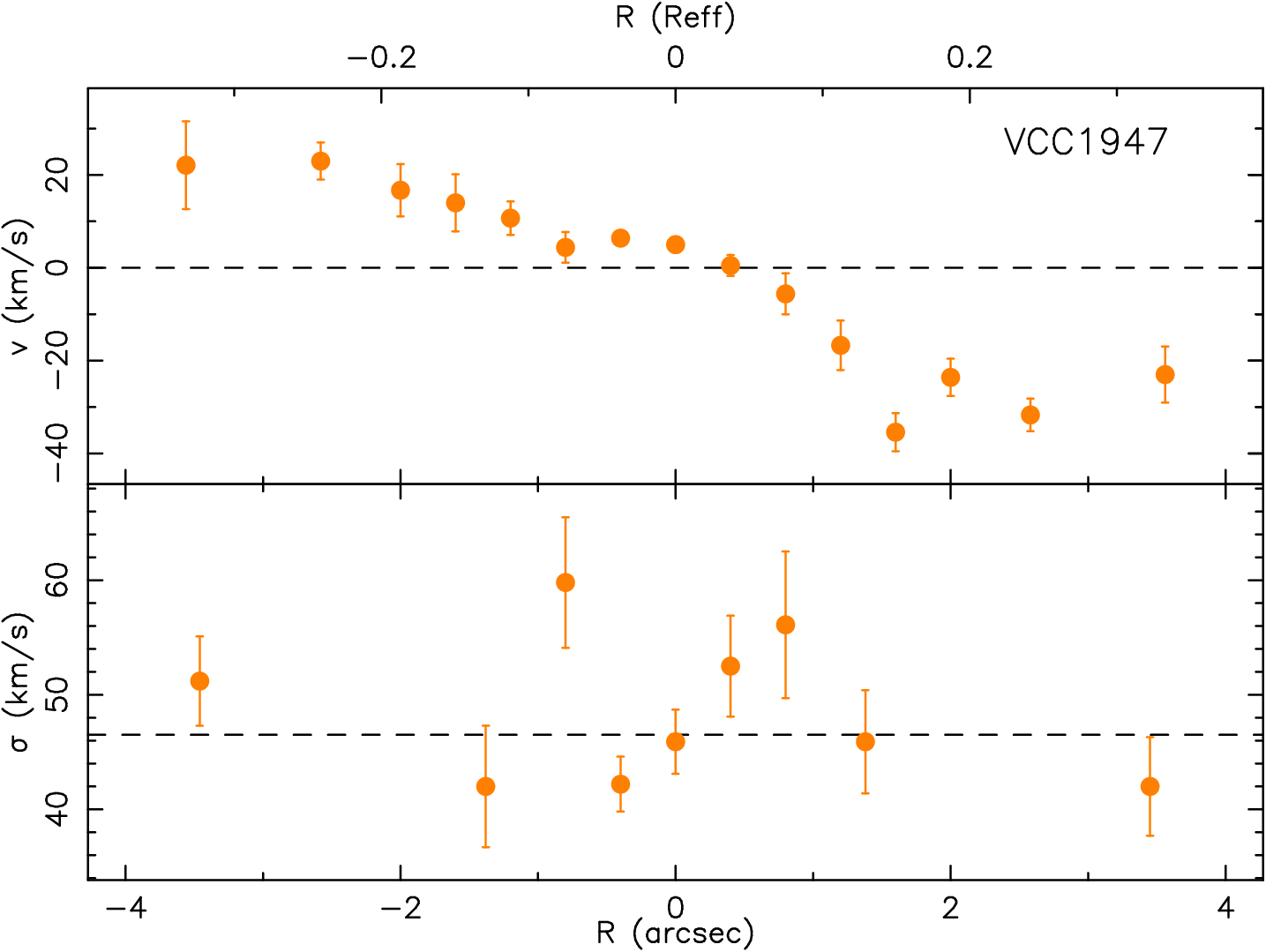}}
\caption{Continued}
\end{figure*}

\subsection{Central velocity dispersion and maximum rotational velocity}\label{central_values}

To compute the central velocity dispersion ($\sigma$) we shifted all the spectra to the same wavelength scale using the rotation curves displayed in Figure \ref{curves1}, and we coadded all the individual spectra up to one effective radius. The typical S$/$N ratio for the spectrum where the central $\sigma$ is computed is $\sim$60 $\AA^{-1}$. These results are shown in Table \ref{t4}.

The maximum rotational velocity ($v_{max}$) was calculated as the weighted average of the two highest velocities along the major axis on both sides of the galaxy at the same radius (for non-symmetric profiles at least three values were required). As a consequence values with larger errors weight less than those with smaller errors. For each galaxy these values are presented in Figure \ref{curves1} as purple squares.
We show in Appendix \ref{vmax} that, given the uncertainty, all galaxies with $v_{max}<$ 9 km s$^{-1}$ can be considered non-rotators.

The ratio between these two kinematic measurements, the maximum rotation velocity $v_{max}$ and the velocity dispersion $\sigma$, is called anisotropy parameter, $v_{max}/\sigma$, and it is used to study the rotational/pressure support of the galaxies. In Table \ref{t4} we show $(v_{max}/\sigma)^*$, the anisotropy parameter corrected from the inclination. This correction is done following the expression $(v_{max}/\sigma)^*=\frac{v_{max}/\sigma}{\sqrt{\epsilon/(1-\epsilon)}}$, where $\epsilon$ is the ellipticity. Note that for those galaxies with ellipticity close to zero, no correction can be done because they are nearly face on.  We choose a conservative value of $(v_{max}/\sigma)^*=0.8$ as the limit between pressure and rotationally supported systems in order to include those objects that, within the errors, are consistent with being flattened by rotation. This assumption is justified by the fact that the measured $v_{max}$ is a lower limit since the rotation curves are still rising.

\begin{table}
\begin{center}
\caption{Kinematic parameters.\label{t4}}
\begin{tabular}{|c|c|c|c|c|}
\hline \hline
  Galaxy      & $\sigma$ (km s$^{-1}$) & $v_{max}$ (km s$^{-1}$) & $(v_{max}/\sigma)^*$ & $v_{rad}$ (km s$^{-1}$)\\
\hline
PGC1007217  & 35.2$\pm$    0.9 &  30.6$\pm$  5.0  & 2.3 $\pm$ 0.4 & 1592.3 $\pm$ 3.3\\
PGC1154903  & 23.1$\pm$    4.1 &    8.6$\pm$  2.6  & 0.5 $\pm$ 0.2 & 1156.3 $\pm$ 6.4\\
NGC3073       & 39.8$\pm$    0.3&   17.1$\pm$  6.7  & 1.1 $\pm$ 0.4 & 1168.9 $\pm$ 2.5\\
VCC21            &26.1$\pm$    4.0 &  18.4$\pm$  5.9  & 0.9 $\pm$ 0.3  & 463.6 $\pm$ 9.2\\
VCC308          &31.7$\pm$    1.2&   30.4$\pm$  8.6  & 1.0 $\pm$ 0.3  & 1515.3 $\pm$ 3.2\\
VCC397          &29.9$\pm$    1.1 &  52.0$\pm$ 11.6 & 2.5 $\pm$ 0.6  & 2434.9 $\pm$ 2.4\\
VCC523          &45.8$\pm$    0.7 &  39.6$\pm$  5.7  & 1.5 $\pm$ 0.2  & 1515.7 $\pm$ 2.6\\
VCC856          &29.6$\pm$    2.5 &   9.7$\pm$  1.9  &  1.0 $\pm$ 0.2  &  980.2 $\pm$ 1.9\\
VCC917          &31.4$\pm$    1.4 &  21.6$\pm$  7.5 &  0.8 $\pm$  0.3 & 1236.2 $\pm$ 2.4\\
VCC990          &40.6$\pm$    1.0 &  26.3$\pm$  1.6  & 0.9 $\pm$ 0.1  & 1691.5 $\pm$ 2.1\\
VCC1087        &48.3$\pm$    0.7 &   7.1$\pm$  6.4  &  0.2 $\pm$ 0.2  &  644.6 $\pm$ 3.2\\
VCC1122        &37.2$\pm$    0.8 &  17.3$\pm$  7.7  &  0.5 $\pm$ 0.2 &  447.8 $\pm$ 2.7\\
VCC1183        &41.3$\pm$    1.2 &  10.1$\pm$  2.8  &  0.5 $\pm$ 0.1 &  1290.7 $\pm$ 2.3\\
VCC1261        &51.8$\pm$    0.9 &  13.9$\pm$  5.2  &  0.4 $\pm$ 0.1 & 1806.4 $\pm$ 2.1\\
VCC1431        &54.1$\pm$    1.2 &   7.0$\pm$  3.6  &   0.1 $\pm$ 0.1 & 1472.8 $\pm$ 3.1\\
VCC1549        &38.9$\pm$    1.9 &   5.2$\pm$  2.2  &   0.3 $\pm$ 0.1 &  1377.3 $\pm$ 2.9\\
VCC1695        &28.7$\pm$    1.1 &  14.3$\pm$  3.1 &   0.9 $\pm$ 0.2 &  1706.0 $\pm$ 3.0\\
VCC1861        &40.4$\pm$    0.9 &   6.3$\pm$  4.3  &   0.2 $\pm$ 0.1 &  617.2 $\pm$ 2.6\\
VCC1910        &39.0$\pm$    1.1 &   6.0$\pm$  2.3  &   0.4 $\pm$ 0.2 &  178.6 $\pm$ 2.2\\
VCC1912        &37.1$\pm$    1.0 &  20.8$\pm$  4.5 &   0.5 $\pm$ 0.1 & -110.5 $\pm$ 2.0\\
VCC1947        &45.3$\pm$    1.1 &  28.3$\pm$  2.1 &   1.1 $\pm$ 0.1 &  956.3 $\pm$ 1.8\\
\hline
\end{tabular}
\end{center}
NOTES:  Column 1: galaxy name.
             Column 2: central velocity dispersions computed within the $R_{eff}$. 
             Column 3: maximum rotation velocities. The $v_{max}$ adopted for VCC0856, VCC0990 and VCC1183 have been measured in the rotation curves of \citet{Chil09} due to their larger extent (see Section 5.3).
             Column 4: anisotropy parameter corrected from inclination. 
            Column 5: mean radial velocity observed. Values in agreement with those of NED database.

\end{table}

\subsection{Comparison to the literature}

Displaying simultaneously the kinematic profiles measured in this work with those of other authors (Figure \ref{curves_comparison1}), one sees that the radial extent of the kinematic curves varies from one work to another. In addition, the offsets found in the velocity dispersions are not always consistent within the errors (Figure \ref{f1}). These two differences are important, because different radial extents lead to different maximum rotation velocities and offsets in the velocity dispersion profiles lead to different central $\sigma$ values.

\subsubsection{Velocity dispersions} \label{sigmas_comparison}

\begin{figure*}
\centering
\resizebox{0.4\textwidth}{!}{\includegraphics[angle=-90]{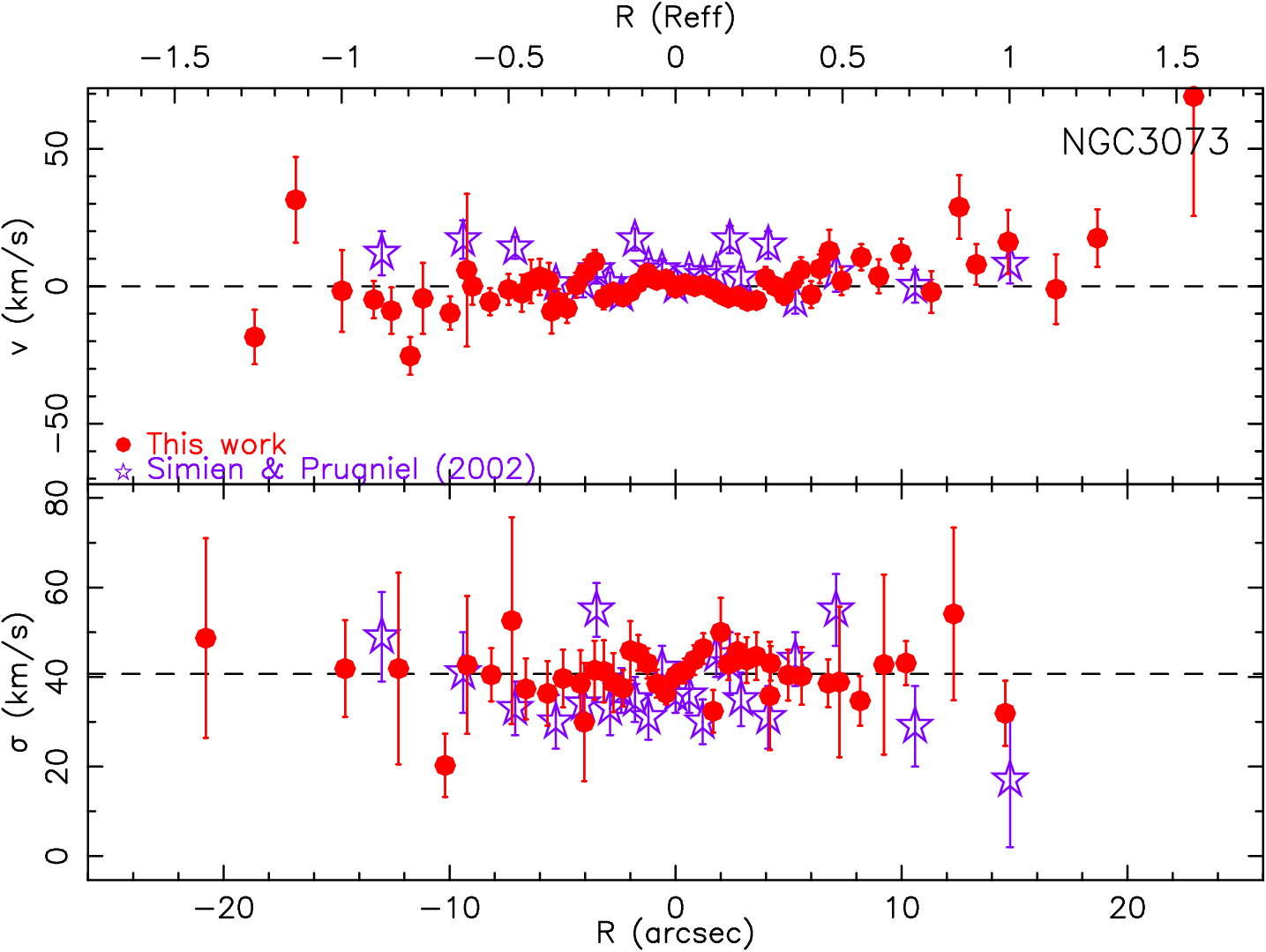}}
\resizebox{0.4\textwidth}{!}{\includegraphics[angle=-90]{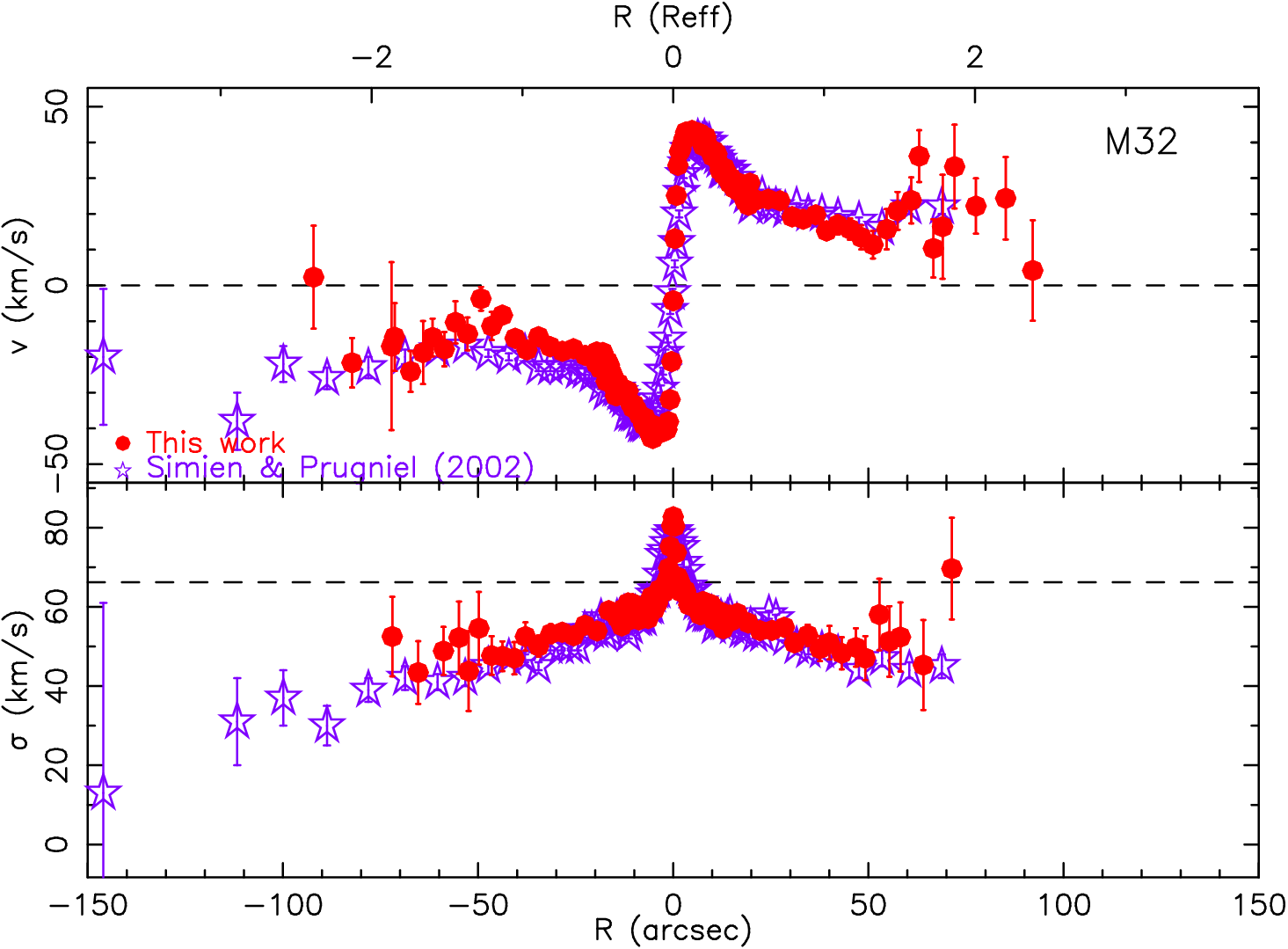}}
\resizebox{0.4\textwidth}{!}{\includegraphics[angle=-90]{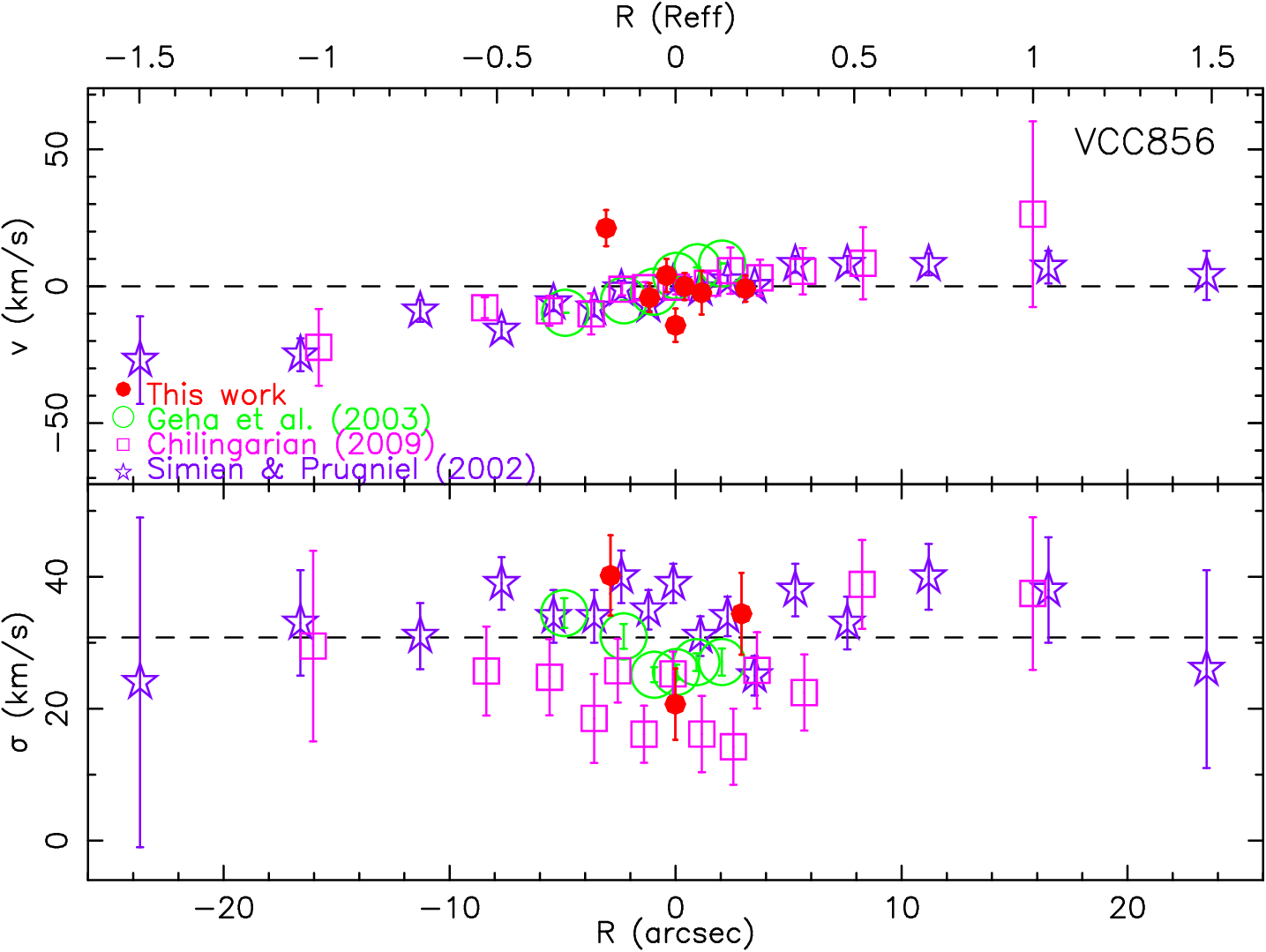}}
\resizebox{0.4\textwidth}{!}{\includegraphics[angle=-90]{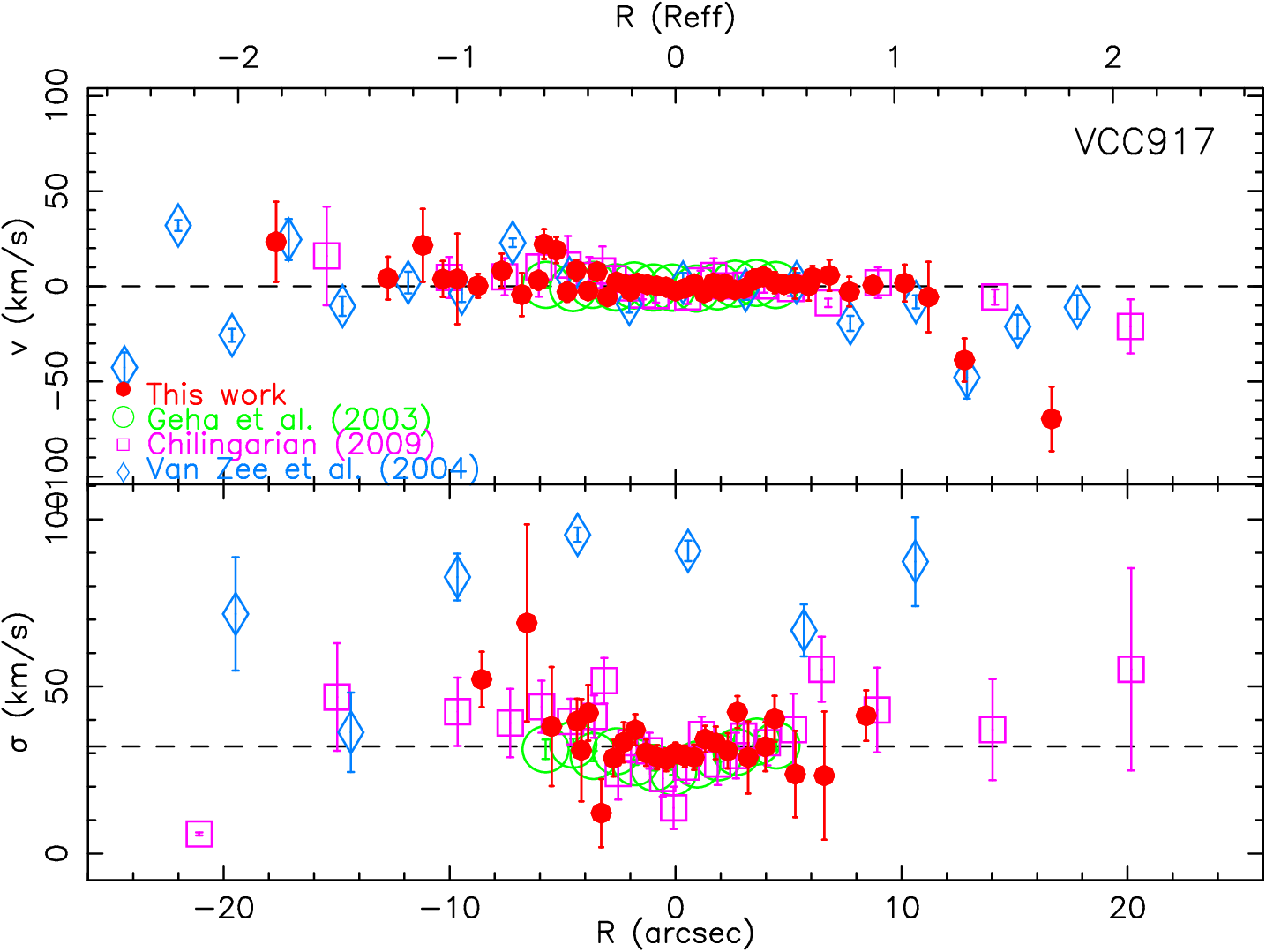}}
\caption{Comparison of our kinematic profiles with other works. In each diagram it is shown the rotation curve (upper panel) and the velocity dispersion profile (lower panel). The bottom X-axis is measured in arcseconds and the upper X-axis is measured as a fraction of the effective radius ($R_{eff}$) of each galaxy in i band (see Section 6). For Van Zee et al. (2004) we only present their velocity  profiles for MgbI, more similar in wavelength to our data.}
              \label{curves_comparison1}
\end{figure*}
\addtocounter{figure}{-1}
\begin{figure*} 
\centering
\resizebox{0.4\textwidth}{!}{\includegraphics[angle=-90]{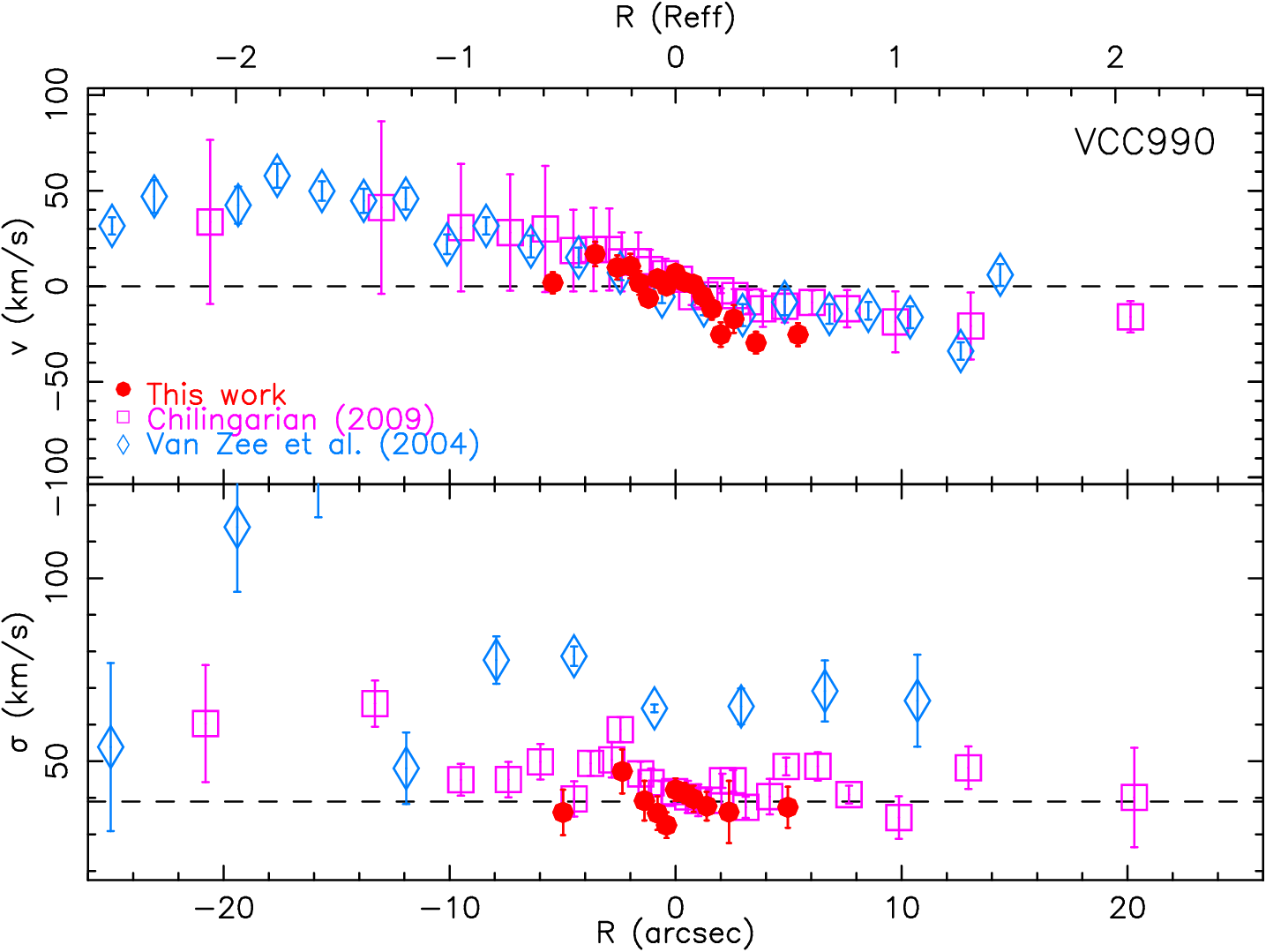}}
\resizebox{0.4\textwidth}{!}{\includegraphics[angle=-90]{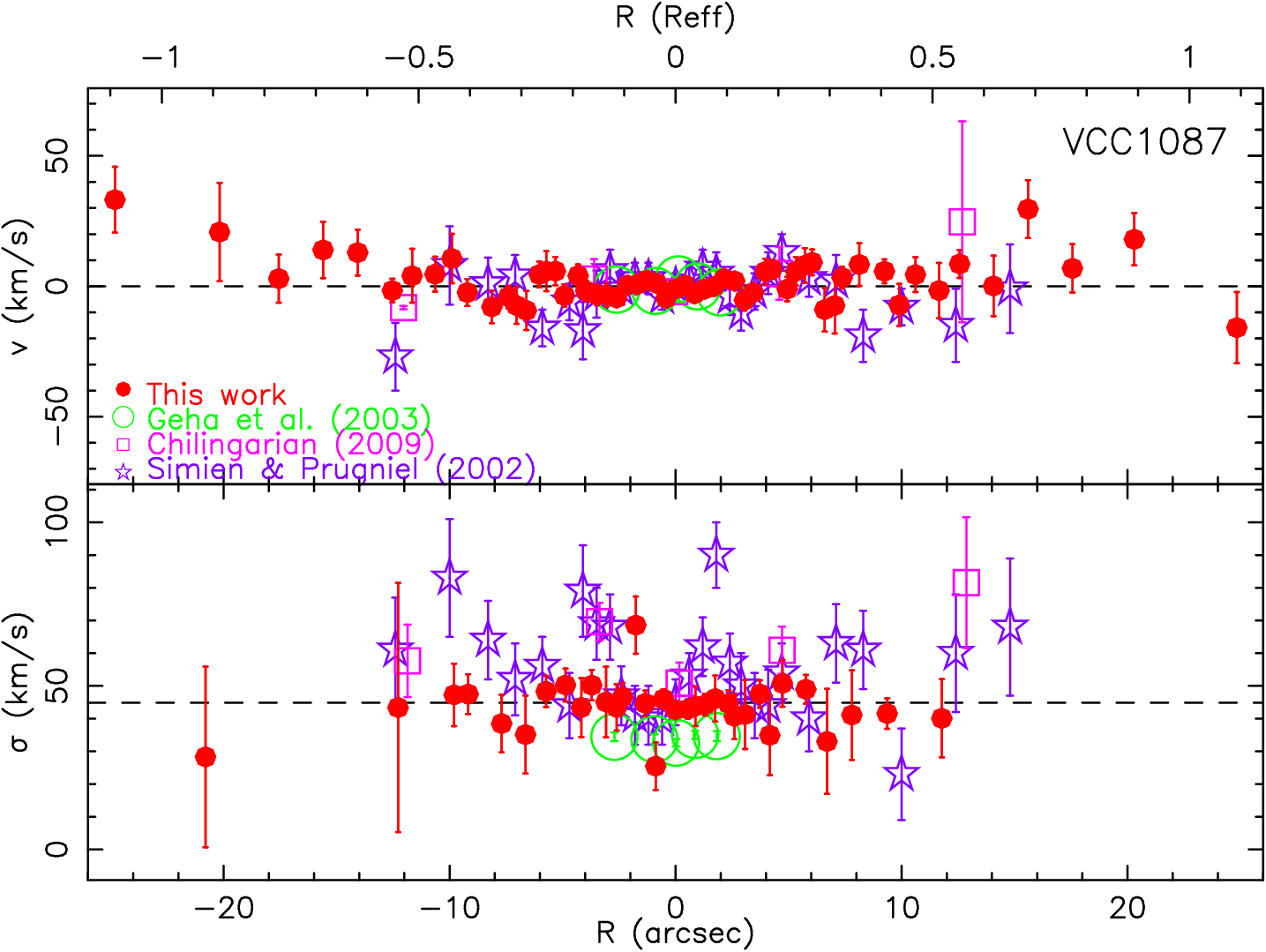}}
\resizebox{0.4\textwidth}{!}{\includegraphics[angle=-90]{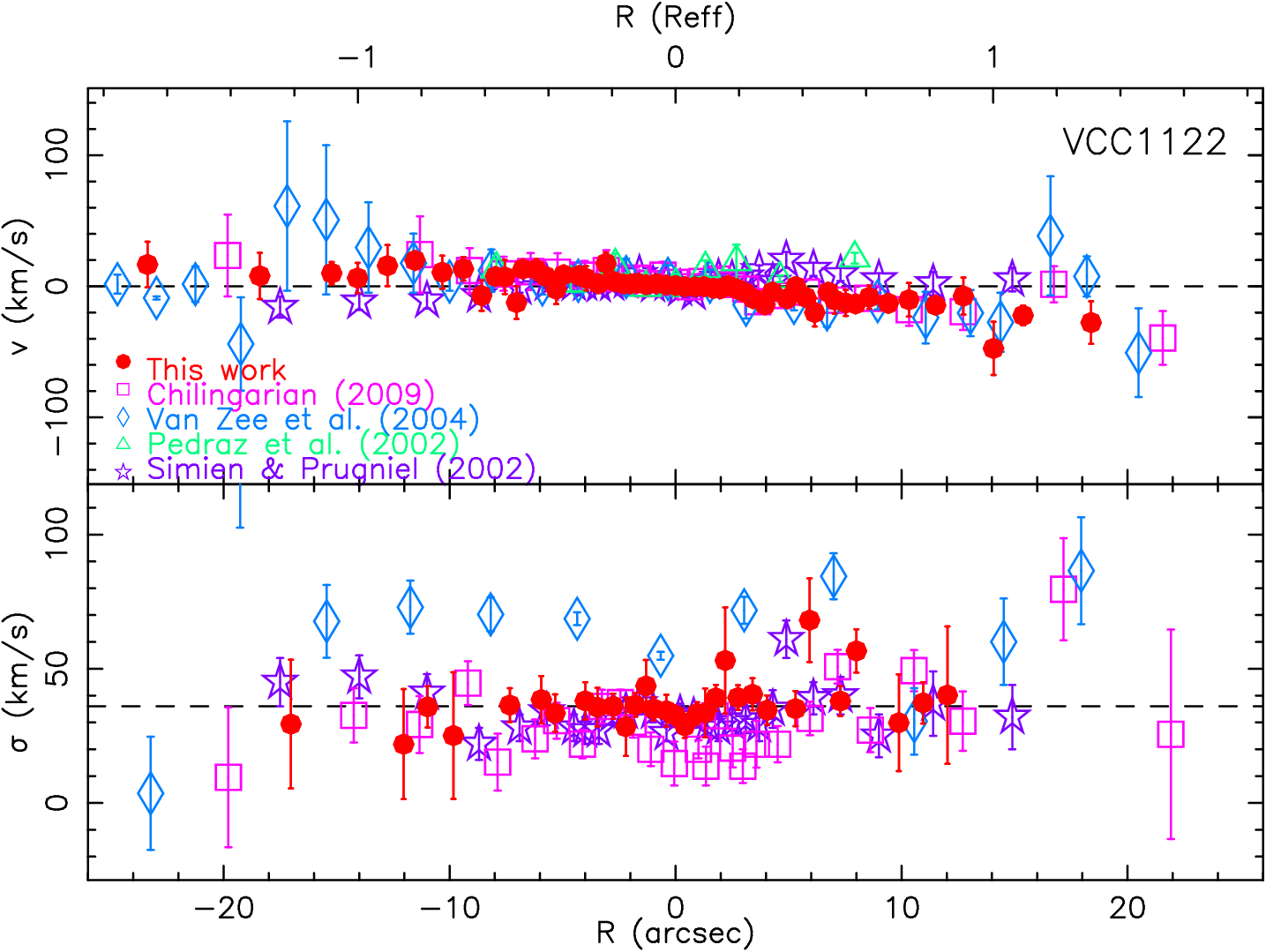}}
\resizebox{0.4\textwidth}{!}{\includegraphics[angle=-90]{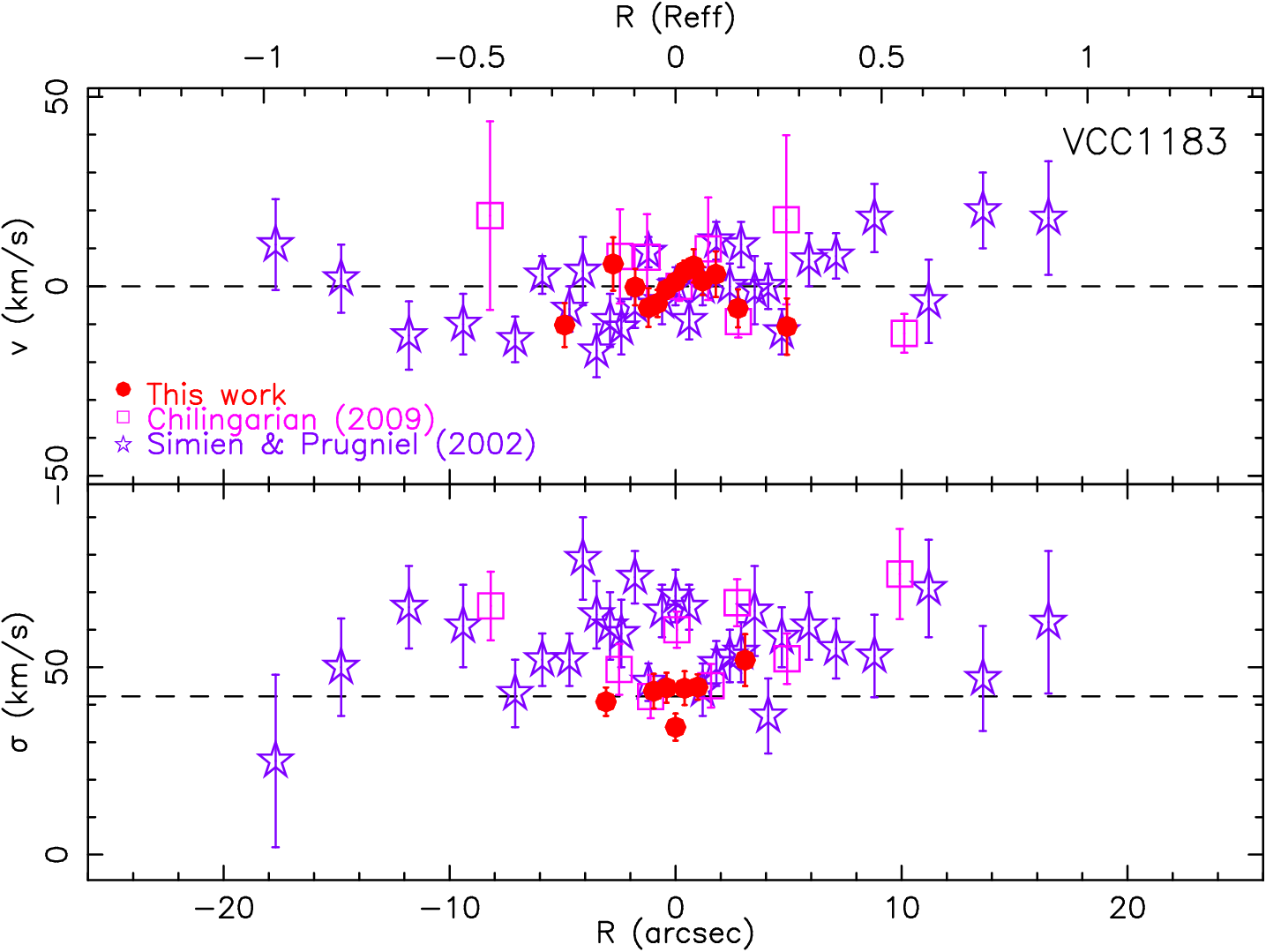}}
\resizebox{0.4\textwidth}{!}{\includegraphics[angle=-90]{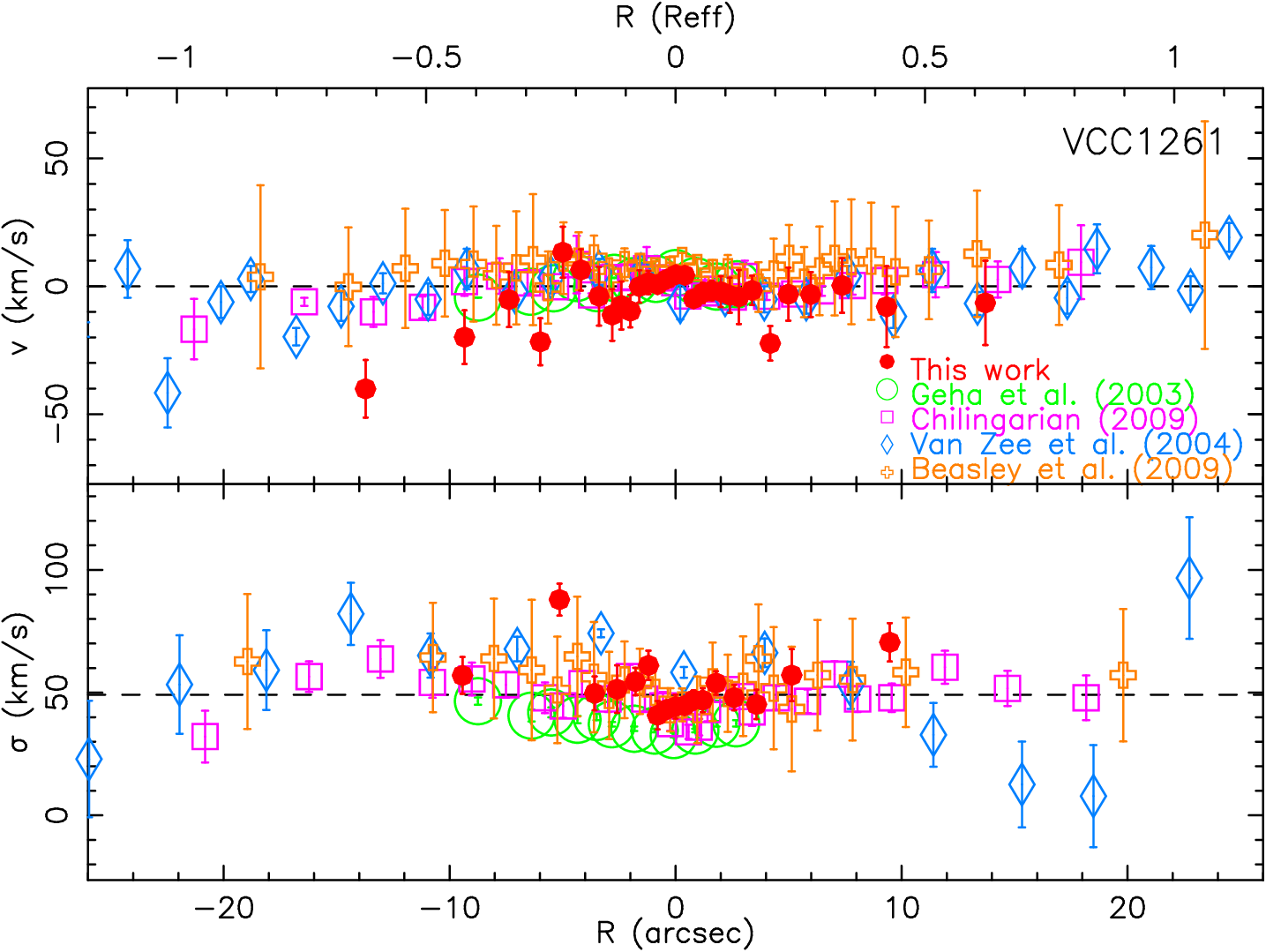}}
\resizebox{0.4\textwidth}{!}{\includegraphics[angle=-90]{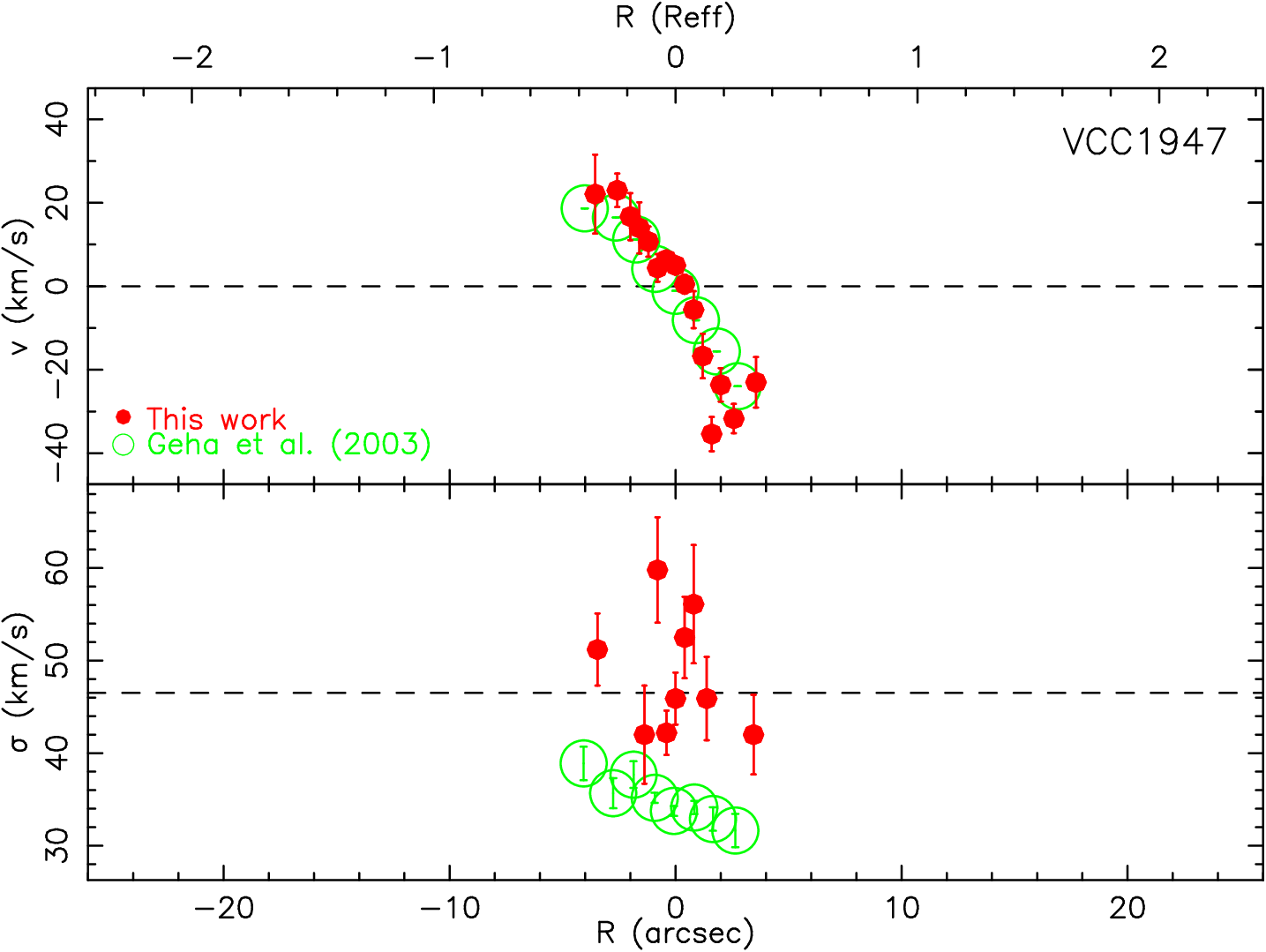}}
 \caption{Continued}
\end{figure*}

There are two factors that critically affect the measurements of the velocity dispersion of the galaxies:  the not identical instrumental $\sigma$ for the stars and the galaxies and the spectral types of the stars used to fit the width of the galaxy lines.

\citet{Caldwell} made their observations with a $\sigma_{inst}=$ 100 km s$^{-1}$, too high to accurately measure velocity dispersions of typical dwarf galaxies, below 50 km s$^{-1}$.

Concerning the equality of the instrumental resolution in the templates and in the galaxies \citet{Chil09} used simple stellar population (SSP) models of PEGASE.HR \citep{Lb03},  based on ELODIE \citep{Prug07}  ($R=10000$). In this case, the broadening of the SSP models to $\sigma_{inst}$ of the galaxies must be cautiously done because $\sigma_{inst}$ of the models is based on the mean value of the spectral resolution of the library they are based on, therefore small differences after the broadening between the SSP models and the galaxies can arise. 

Continuing with the same effect, in the works of \citet{Ped02} and \citet{VZ04} the stars observed were not filling the slit because they were not defocused and the seeing was smaller than their slit width. But in the case of \citet{Geha03} and \citet{Beasley09}, where their slit widths were 0.75$"$ and 1.0$"$ respectively, it is possible that some stars were filling the slit thanks to the seeing. However, in none of these works a correction was made to assure that the instrumental profile in the stars was the same as in the galaxies. 

In reference to the templates used to perform $\sigma$, some of the authors mentioned above used only one star to fit the galaxy spectrum, and in such a case the fact that the template is not representative of the stellar population of the galaxy might lead to large errors.
We have computed Monte-Carlo simulations to see the differences between fitting the galaxy spectrum with only one star and a linear combination of stars of spectral types from B to M, with different luminosity classes. The simulations consisted of a selection of PEGASE.HR models of 3 different ages (1 Gyr, 4 Gyr and 10 Gyr) and 3 different metallicities ($Z=$ 0.0, $Z=$ $-$0.4, $Z=$ $-$0.7)  with Salpeter IMF \citep{Lb03}; a total of 9 models. The stars used as templates were from MILES library \citep{SB06lib}. First of all we checked that after broadening the models (originally at FWHM$=$0.55 \AA) to the MILES resolution (FWHM$=$2.3 \AA) we obtained $\sigma=$ 0 km s$^{-1}$ when running MOVEL, thereby showing that there was no zero point offset. Secondly, we broadened the models to 40 km s$^{-1}$ to simulate dwarf galaxies of different stellar populations.  And finally, we ran MOVEL using 3 different kinds of templates: only one K1III star (as in \citet{Geha03} ), only one G8III star (as in \citet{VZ04}), and a linear combination of B to M stars with different luminosity classes (as in this work). To measure $\sigma$ we have masked the Balmer lines, especially those bluer than $H_{\delta}$ to avoid possible problems due to emission lines. Our simulations show that a linear combination of different stars is the most accurate method to obtain the velocity dispersion of the galaxies, never finding an error above 25$\%$, independent of the stellar population considered. Note that if a young population dominates the light of the galaxy (for ages of 1 Gyr and below), and a single G or K star is used as a template, an error up to 70$\%$ can be done.  When a single star is used as template, a dependence on metallicity for young populations (1 or 4 Gyr) is also found, in the sense that decreasing the metallicity increases the uncertainty. This dependence is likely due to offsets introduced by the method employed to compute $\sigma$. The results obtained have been summarised in Table \ref{t3}.

\begin{table}
\begin{center}
\caption{Uncertainties introduced when different templates are used to calculate the velocity dispersion of a dwarf galaxy with different stellar populations. $\Delta\sigma/\sigma$, defined in Figure \ref{sig}, uses as $\sigma_i$ 40 km s$^{-1}$. \label{t3}}
\begin{tabular}{|c|c|c|c|}
\hline \hline
                       & \multicolumn{3}{|c|}{$\frac{\Delta \sigma}{\sigma}$x100}  \\
\hline \hline
   Age (Gyr)     &     Z=+0.0   &  Z=-0.4  &  Z=-0.7  \\
 \hline
\multicolumn{4}{|c|}{Linear combination of B to M stars}\\
 \hline
1                     &    15$\%$    &  19$\%$  &  1$\%$   \\
4                     &      8$\%$    &   18$\%$ &  19$\%$ \\
10                   &      7$\%$    &   14$\%$ &  23$\%$ \\
\hline
\multicolumn{4}{|c|}{G8III}\\
\hline
1                     &   32$\%$     &   54$\%$ & 71$\%$\\
4                     &   15$\%$     &   18$\%$ & 20$\%$\\
10                   &   17$\%$     &   13$\%$ & 11$\%$\\
\hline
\multicolumn{4}{|c|}{K1III}\\
\hline
1                     &    29$\%$    &   47$\%$ & 59$\%$\\
4                     &    19$\%$    &   18$\%$ & 20$\%$\\
10                   &    17$\%$    &   13$\%$ & 11$\%$\\
\hline
\end{tabular}
\end{center}
\end{table}

   \begin{figure}
   \centering
  \resizebox{0.5\textwidth}{!}{\includegraphics[angle=-90]{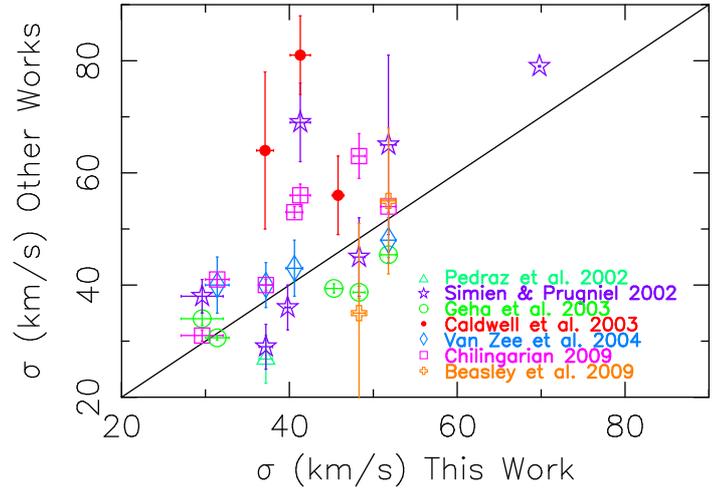}}
  \caption{Comparison between the velocity dispersions measured in this work versus those of other authors. The colours and symbols are the same as in Figure \ref{curves_comparison1}. We add a comparison with \citet{Caldwell}  (solid red points), who only measured central values and no kinematical profiles.}
              \label{f1}
    \end{figure}

\subsubsection{Rotation curves}

\begin{figure}
\centering
\resizebox{0.5\textwidth}{!}{\includegraphics[angle=-90]{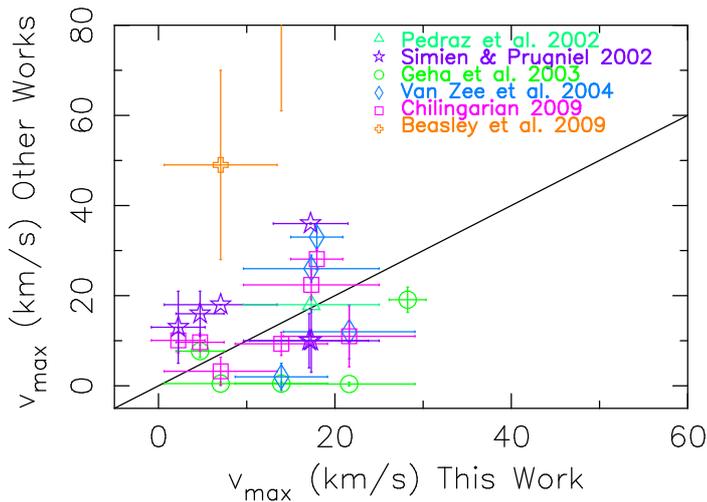}}
\caption{Comparison between the maximum rotational velocity measured in this work and those measured by other authors. Symbols and colours are the same as in Figure \ref{curves_comparison1}. VCC1261 from \citet{Beasley09} is nearly outside the plot due to its enormous rotation: 105 $\pm$ 44 km s$^{-1}$.}
              \label{f2}
\end{figure}

Different criteria have been used in the literature to measure the maximum rotation velocity. The main difficulty here is to have an extended rotation curve that reaches a clear plateau where the maximum rotation can be measured. As this is not so easy for dEs, an objective criteria, independent of the shape of the rotation curve in each case, must be adopted. The different criteria used by the various authors have led to maximum rotational velocities that are nevertheless in the majority of the cases nearly consistent within the errors (Figure \ref{f2}).
The criterion used by \citet{Ped02} is the same as the one adopted here. Although we achieved a radial extent of 23$''$ for VCC1122 and \citet{Ped02} only 8$''$, the latter value was enough to reach the flat part of the rotation curve, and, as a consequence, both measurements of the maximum rotation are identical. The explanation for the differences found with \citet{SimPrugVI} are mainly based on the fact that only two points were considered to obtain the maximum rotation in that paper. The differences with \citet{Geha03}  are due to the limitation of their data to the core of the galaxies, never reaching radii larger than 6$''$ (see Figure \ref{curves_comparison1}). The differences with \citet{VZ04} and \citet{Chil09} are related to a different radial extent of the rotation curves. Note that \citet{Chil09} does not calculate the maximum rotation, but we have applied our criterion to his rotation curves. Finally, the differences found with \citet{Beasley09} are due to the fact that $v_{max}$ is obtained from the analysis of Globular Clusters located up to $\sim$ 7$R_{eff}$. When their kinematic determined from the stellar component using long slit spectroscopy along the major axis of the galaxy is compared to our data, the agreement is evident (Figure \ref{curves_comparison1}). The maximum rotation values adopted for the analysis (see Table \ref{t4}) are our own values, except for VCC856, VCC990 and VCC1183, where the data come from \citet{Chil09} since he obtained larger radii than us. Note that the values from \citet{Beasley09} can not be adopted here because our work is dedicated to the analysis of the stellar component of dEs.

The comparison with \citet{Beasley09}, who finds rotation speeds much larger at 7$R_{eff}$ (100 and 50 km s$^{-1}$ higher for VCC1087 and VCC1261, respectively) than our data at the $R_{eff}$, indicate that the rotation curves of these galaxies are still rising.

\section{Photometric parameters}\label{phot}

In order to make a complete analysis of the kinematics, comparison with some photometric parameters is needed. For our study, we require I-band (Johnson-Cousins) total magnitudes and optical radii ($R_{opt}$, radius containing 83$\%$ of the total I-band luminosity \citep{Cat06}) to study the shape of the rotation curves. Effective radii ($R_{eff}$, radius containing 50$\%$ of the total light) is needed to measure the extent of the radial profiles in physical units of the galaxy. Ellipticities ($\epsilon$) are needed to make the appropriate corrections due to inclination. A parameter to measure the boxyness/diskyness of the isophotes ($C_4$) is also required to study the possible late-type origin of these dwarf early-type galaxies.

All these parameters have been obtained from $i$-band Sloan Digital Sky Survey (SDSS, \citet{SDSS}) data release 6 (DR6, \citet{SDSS_DR6}) photometry. They have been calculated using the IRAF\footnote{IRAF is distributed by the National Optical Astronomy Observatory, which is operated by the Association of Universities for Research in Astronomy, Inc., under the cooperative agreement with the National Science Foundation.} task {\sc ellipse} as described in Appendix \ref{photometry}. The transformation from $i$-band (SDSS) to $I$-band (Johnson-Cousins) has been done assuming $m_I=m_i-0.52\pm0.01$ mag \citep{Fukugita}.

\subsection{$C_4$: boxyness/diskyness parameter} \label{C4_subsection}

The boxyness/diskyness ($C_4$) parameter measures the deviations of  the isophotes from a perfect ellipse. If $C_4 > 0$ the isophotes are disky, indicating that some disk
substructure is present in the galaxy, and when $C_4 \le 0$ the isophotes are
boxy \citep{Carter78,Kormendy96}.  This parameter is independent from the surface brightness profile of the galaxy.
The C$_4$ parameter, determined for our galaxies as described in Appendix \ref{appC4}, is provided in  Table \ref{t5}.

   \begin{figure}
   \centering
  \resizebox{0.5\textwidth}{!}{\includegraphics[angle=-90]{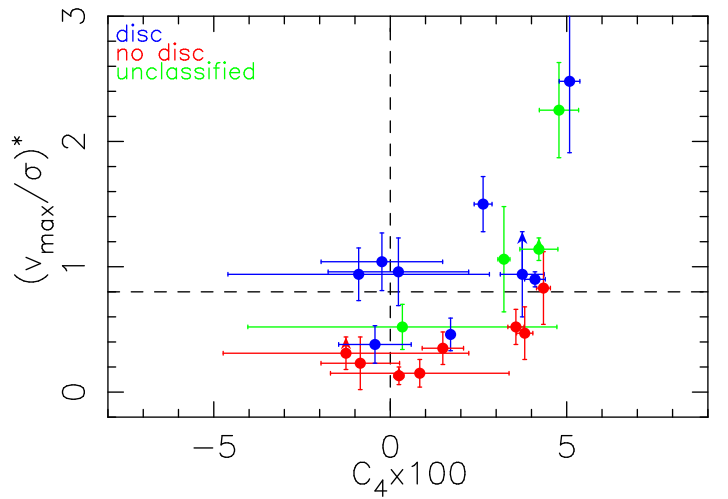}}
\caption{ Correlation between the anisotropic parameter $(v_{max}/\sigma)^*$ and $C_4$x100. Colours indicate \citet{Lisk06a} classification. The galaxies with $v_{max}$ measured inside the central 6$"$ are considered lower limits and are indicated with arrows. The horizontal dashed line at $(v_{max}/\sigma)^*=$ 0.8 indicates a rough boundary between pressure and rotationally supported galaxies. The blue points are classified as having a disk in \citet{Lisk06a}, whereas for the red points no disk is found in that paper. See Appendix \ref{appC4} for a discussion.}
              \label{vs_C4}
    \end{figure}

Our $C_4$ classification for disky isophotes agrees in
general with the morphological classification of \citet{Lisk06b} as can be seen in Figure \ref{vs_C4}.  Note that \citet{Lisk06b} classified a galaxy as disky when disk features (spiral arms, edge-on disks, or bars) were detected after
subtracting an axisymmetric light distribution from the original image or after unsharp masking.
In this Figure we can see that the red dots, those galaxies classified as being without underlying structures in \citet{Lisk06b}, are grouped around $C_4$ $\le$ 0 (boxy or elliptical isophotes), while the blue dots (galaxies with some disky structure in \citet{Lisk06b}) are all consistent with $C_4$ $>$ 0 (disky isophotes). This justifies the use of the $C_4$ parameter to detect the presence of an underlying disk. Three exceptions are found and discussed in Appendix \ref{appC4}. 

In this respect, it is important to underline the correlation between $C_4$ and the anisotropic parameter evident in Figure \ref{vs_C4}, first found by \citet{Bender88} for more massive elliptical galaxies.


\begin{table*}
\begin{center}
\caption{Derived parameters. \label{t5}}
\begin{tabular}{|c|c|c|c|c|c|c|}
\hline \hline
  Galaxy            & d (Mpc)                   &      $M_I$ (mag)         &   $\epsilon$         & $R_{eff}$ ($"$)         &  $C_4$x100          &Disk classification (L06b)\\
\hline
PGC1007217    & 20.23   $\pm$ 1.40 &  -17.39 $\pm$ 0.15 & 0.13 $\pm$ 0.05 & 11.11 $\pm$  0.02 &  4.8$^{(*)}$  $\pm$    0.6 &  ---      \\
PGC1154903    & 15.00   $\pm$ 1.10 &  -15.91 $\pm$ 0.16 & 0.34 $\pm$ 0.02 &  8.36 $\pm$  0.04 &  0.3  $\pm$    4.4 &  ---   \\
NGC3073       & 33.73   $\pm$ 14.44&  -20.43 $\pm$ 0.93 & 0.14 $\pm$ 0.02 & 14.78 $\pm$  0.03 &  3.2$^{(*)}$  $\pm$    0.2 &  ---      \\
VCC21         & 16.74   $\pm$ 0.15 &  -17.78 $\pm$ 0.03 & 0.36 $\pm$ 0.03 & 15.48 $\pm$  0.11 &   3.7$^{(*)}$  $\pm$    0.6 &  Disk      \\
VCC308        & 16.41   $\pm$ 0.32 &  -18.78 $\pm$ 0.05 & 0.04 $\pm$ 0.03 & 19.22 $\pm$  0.04 &  0.2  $\pm$    2.0 &  Disk      \\
VCC397        & 16.41   $\pm$ 0.32 &  -17.62 $\pm$ 0.05 & 0.33 $\pm$ 0.03 & 13.75 $\pm$  0.02 &  5.1$^{(*)}$  $\pm$    0.3 &  Disk      \\
VCC523        & 16.74   $\pm$ 0.15 &  -19.16 $\pm$ 0.03 & 0.25 $\pm$ 0.01 & 20.90 $\pm$  0.03 &  2.6$^{(*)}$  $\pm$    0.3 &  Disk      \\
VCC856        & 16.83   $\pm$ 0.46 &  -18.49 $\pm$ 0.06 & 0.09 $\pm$ 0.03 & 15.82 $\pm$  0.06 & -0.2  $\pm$    1.7 &  Disk      \\
VCC917        & 16.74   $\pm$ 0.15 &  -17.39 $\pm$ 0.03 & 0.41 $\pm$ 0.02 &  9.68 $\pm$  0.04 &   4.3$^{(*)}$  $\pm$    0.2 &  No Disk      \\
VCC990        & 16.74   $\pm$ 0.15 &  -18.27 $\pm$ 0.03 & 0.34 $\pm$ 0.02 &  9.73 $\pm$  0.02 &  4.1$^{(*)}$  $\pm$    0.3 &  Disk      \\
VCC1087       & 16.67   $\pm$ 0.46 &  -18.94 $\pm$ 0.06 & 0.28 $\pm$ 0.03 & 22.74 $\pm$  0.05 & -0.9  $\pm$    1.1 &  No Disk   \\
VCC1122       & 16.74   $\pm$ 0.15 &  -17.94 $\pm$ 0.03 & 0.50 $\pm$ 0.04 & 14.06 $\pm$  0.05 &  3.8$^{(*)}$  $\pm$    0.2 &  No Disk   \\
VCC1183       & 16.74   $\pm$ 0.15 &  -18.59 $\pm$ 0.03 & 0.22 $\pm$ 0.12 & 18.23 $\pm$  0.01 &  1.7$^{(*)}$  $\pm$    0.0 &  Disk      \\
VCC1261       & 18.11   $\pm$ 0.50 &  -19.39 $\pm$ 0.06 & 0.37 $\pm$ 0.05 & 22.07 $\pm$  0.03 &  1.5$^{(*)}$  $\pm$    0.6 &  No Disk   \\
VCC1431       & 16.14   $\pm$ 0.45 &  -18.47 $\pm$ 0.06 & 0.03 $\pm$ 0.01 & 10.32 $\pm$  0.01 &  0.2  $\pm$    0.1 &  No Disk   \\
VCC1549       & 16.74   $\pm$ 0.15 &  -18.18 $\pm$ 0.03 & 0.16 $\pm$ 0.01 & 13.56 $\pm$  0.05 & -1.3  $\pm$    3.5 &  No Disk   \\
VCC1695       & 16.52   $\pm$ 0.61 &  -18.13 $\pm$ 0.08 & 0.22 $\pm$ 0.05 & 19.78 $\pm$  0.10 & -0.9  $\pm$    3.7 &  Disk      \\
VCC1861       & 16.14   $\pm$ 0.45 &  -18.57 $\pm$ 0.06 & 0.04 $\pm$ 0.02 & 18.52 $\pm$  0.04 &  0.8  $\pm$    2.5 &  No Disk   \\
VCC1910       & 16.07   $\pm$ 0.44 &  -18.63 $\pm$ 0.06 & 0.14 $\pm$ 0.04 & 13.70 $\pm$  0.02 & -0.4  $\pm$   1.0 &  Disk      \\
VCC1912       & 16.74   $\pm$ 0.15 &  -18.62 $\pm$ 0.03 & 0.54 $\pm$ 0.06 & 23.34 $\pm$  0.02 &  3.6$^{(*)}$  $\pm$    0.2 &  No Disk   \\
VCC1947       & 16.74   $\pm$ 0.15 &  -18.46 $\pm$ 0.03 & 0.23 $\pm$ 0.01 & 10.70 $\pm$  0.02 &  4.2$^{(*)}$  $\pm$    0.5 &  ---      \\
\hline
\end{tabular}
\end{center}
NOTES: Column 1: galaxy name.
            Column 2: distances in Mpc from Surface Brightness Fluctuations (SBF)
for individual Virgo galaxies from \citet{Mei07} when available, or the mean A/B
Cluster distance from \citet{Mei07} for the rest of them (note that E, N and S
Clouds are East, North and South areas of Cluster A). The distance for NGC3073
from SBF by \citet{SBF01} and PGC1007217 and PGC1154903 distances are from
NED/IPAC Database derived from redshift with $H_0=$73 $\pm$ 5 km s$^{-1}$
Mpc$^{-1}$ (these two distances must be used cautiously). 
            Column 3: Absolute magnitudes in $I$-band (Johnson-Cousins in AB
system) converted from $i$-band measured in SDSS images using $m_I=m_i-0.52$ mag.
            Column 4: ellipticities from $i$-band (SDSS) images. The quoted errors indicate
the RMS scatter in the ellipticity between 3$''$ and the $R_{eff}$.  
            Column 5: Effective radius from $i$-band (SDSS) images. 
            Column 6: Diskyness/Boxyness parameter from $i$-band (SDSS) images. The asterisks indicate $C_4$ measured as the maximum in the region 3$''$- 3$R_{eff}$ if prominent disky features are found; in the rest of the cases the quoted values are the average in this same radial range and the errors the RMS scatter (see Appendix \ref{appC4}). 
            Column 7: Disk/No Disk classification by \citet{Lisk06a} (L06b) (no values indicate that these galaxies are not included in their analysis).
\end{table*}


\section{Analysis} 

The analysis presented in this work is primarily focused on the rotationally supported systems.
Although the majority of the dwarf galaxies (15 out of 21) show some rotation ($v_{max}>$ 9 km s$^{-1}$, Table \ref{t4}), only 11 are rotationally supported ($(v_{max}/\sigma)^*>$ 0.8) \citep{etj09b}. 
Here we try to understand whether the observed kinematic properties of the rotationally supported systems are consistent with those of star forming galaxies of similar luminosity.

\subsection{Shape of the rotation curves}

\citet{Cat06} made a systematic study of the shape of the rotation curves of late-type spiral galaxies as a function of luminosity  based on the method described in \citet{Persic96}. They fitted the rotation curves following the Polyex model \citep{GH02} which has the form:
\begin{equation}
V_{PE}(r)=V_0\Big(1-e^{-r/r_{PE}}\Big)\bigg(1+\frac{\alpha r}{r_{PE}}\bigg)
\end{equation}
This analytical function depends on 3 parameters: $V_0$, $r_{PE}$ and $\alpha$, which represent
the amplitude, the exponential scale of the inner region and the slope of the outer part of the
rotation curve, respectively. The mean fitted rotation curves from \citet{Cat06} are normalised
to the optical radius ($R_{opt}$, radius containing 83$\%$ of the total $I$-band luminosity), and
the velocities are corrected from inclination. To compare them with our rotationally supported
galaxies, we have calculated the inclinations as in \citet{Giov97a}:
\begin{equation}
cos^2i=\frac{(1-\epsilon)^2-q_0^2}{1-q_0^2}
\end{equation}
where $i$ is the inclination, $\epsilon$ is the ellipticity and $q_0$ is a constant value that depends on the thickness of the disk. Here we assume $q_0=0.3$, a conservative value for dwarf galaxies shaped as thick disks \citep{Lisk07}.\footnote{For early-type spirals $q_0=0.2$ \citep{Giov97a}. The most recent measurements of the thickness of dwarf galaxies gives $q_0=0.3-0.35$ \citep{SJ10}. The difference in $v_{max}$ after the correction for inclination between using $q_0=0.2$ and $q_0=0.35$ is of 4.4$\%$, insignificant.}  

The comparison between the mean rotation curves of \citet{Cat06} and those of our rotationally supported objects must be done in the same luminosity regime since the derived parameters of the Polyex model are luminosity dependent. As the dwarf galaxies analysed in this work have magnitudes below the minimum magnitude in \citet{Cat06} ($M_I=$ $-$19.4), the reference Polyex model of low luminosity star forming systems has been determined extrapolating linearly the three faintest values of the Polyex parameters to $M_I=$ $-$18.49, the mean $I$-band magnitude of the galaxies here analysed. The parameters used to construct this curve are $V_0=$ 74.58 km s$^{-1}$, $r_{PE}=$ 0.35$"$ and $\alpha=$ 0.03.

\begin{figure}
\centering
\resizebox{0.5\textwidth}{!}{\includegraphics[angle=-90]{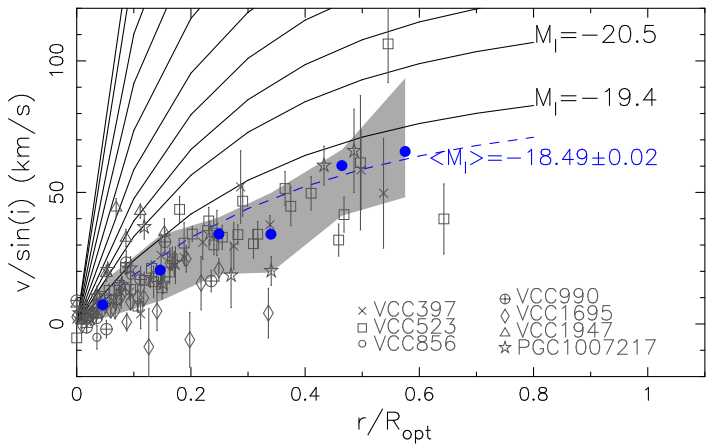}}
\caption{The observed  rotation curves of rotationally supported dEs (grey symbols) are compared to the mean rotation curves of late-type spiral galaxies (black solid and blue dashed lines) from \citet{Cat06}. Blue filled dots represent the median observed rotation curve of rotationally supported dE in bins of $r/R_{opt}=0.1$. The last bin contains all data for $r/R_{opt}\ge0.5$. The grey area indicates rotation velocities within 1$\sigma$ from the median. }
       \label{catinella}
\end{figure}

In Figure \ref{catinella} we compare the mean fitted rotation curves of
\citet{Cat06} for late-type spirals (obtained from emission lines, black solid curves)
 with the rotation curves of our rotationally
supported dEs determined from absorption lines  (grey symbols).  Of the 11 rotationally supported dEs, only 7 have been
considered for this analysis because 3 of them, VCC21, VCC917 and NGC3073, have poor quality
rotation curves ($\frac{\Delta v_{max}}{v_{max}}$ larger than 25$\%$), and VCC308 has 
$\epsilon$ lower than 0.1, implying that the galaxy is
nearly face on.  The blue dots in Figure \ref{catinella} show the median rotation curve of our dEs
in bins of $\frac{r}{R_{opt}}=0.1$. The grey  area contains the 1$\sigma$ deviation from
this median value (68$\%$ of the values are inside this area). The blue dashed line is the
extrapolated Polyex model for $M_I=$ $-$18.49.

It is evident from Figure \ref{catinella} that our rotationally supported galaxies are
characterised by rotation curves that are similar to those of late-type spiral galaxies of equal
luminosity. Or, in other words, galaxies with similar rotation curves 
have similar absolute magnitudes despite their morphological type. We see that rotationally 
supported dEs dynamically behave
like small late-type spiral galaxies.

It is interesting to see that, despite their similar exponential radial light distribution, dEs have two different kinematic behaviour, they can be either pressure or rotationally supported. Furthermore, rotationally supported dEs have rotation curves similar to those late-type spirals despite the fact that these latter objects are gas dominated systems.

\subsection{Tully-Fisher relation}

Given the similarity in the kinematic properties of rotation supported dEs with those of late-type spirals we expect that these systems follow the Tully-Fisher relation, as firstly proposed by \citet{VZ04}.
The Tully-Fisher relation is a typical scaling relation valid for star forming, rotating systems, linking the total luminosity to the maximal rotation velocity of the galaxy.

\begin{figure}
\centering
\resizebox{0.5\textwidth}{!}{\includegraphics[angle=-90]{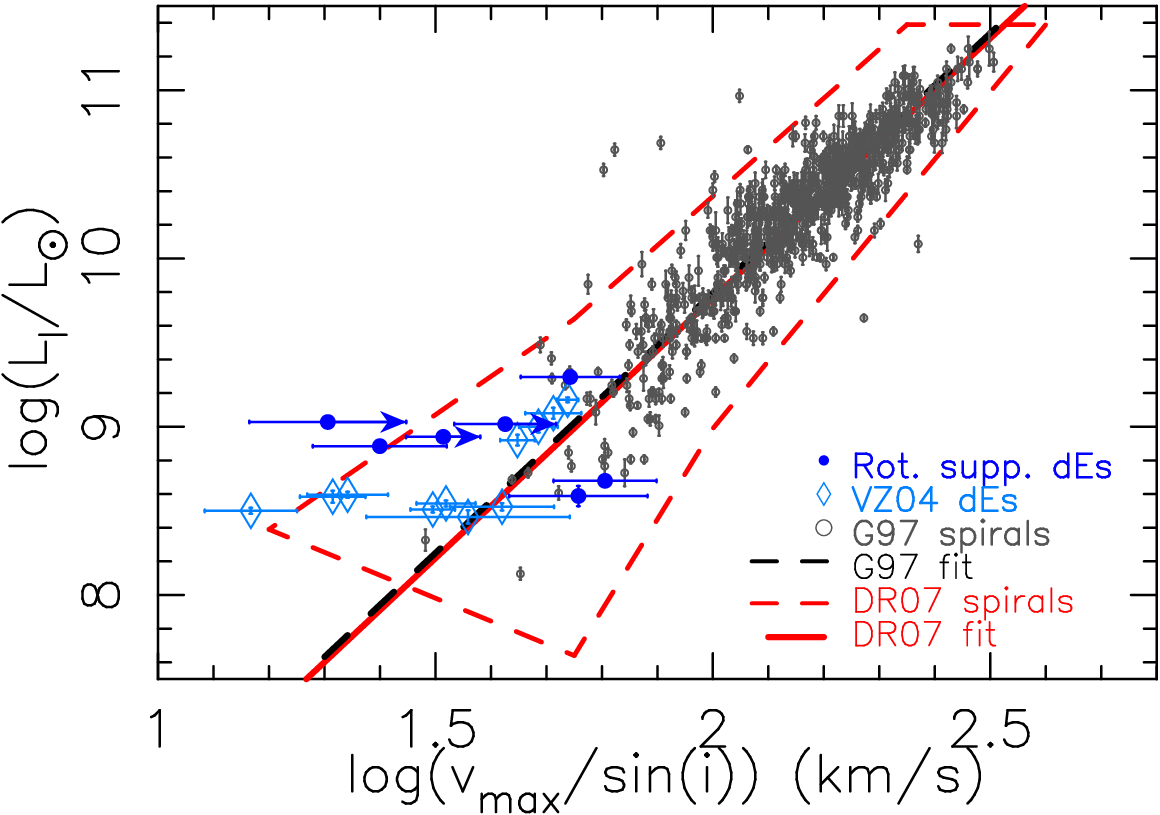}}
\caption{ Tully-Fisher relation for 7 of our rotationally supported dEs (in dark blue),  the dEs from \citet[][VZ04, in light blue]{VZ04b,VZ04}, the normal spirals from \citet[][G97, in grey]{Giov97b} and \citet[][DR07, red limited area]{DR07}. Absolute magnitudes of dEs have been obtained using distances from \citet{Mei07} (criterion described in Table \ref{t5}). For Giovanelli et al. data we use $H_0=73$ km s$^{-1}$ Mpc$^{-1}$ (\citet{Mei07}). The arrows indicate lower limits of $v_{max}$ (those obtained in the inner 6$"$). Fits of the Tully-Fisher relation are indicated in black for the normal spirals of \citet{Giov97b} and in red for \citet[][DR07]{DR07}. DR07 fit has been transformed to I band using the colour-morphology relation from \citet{Fukugita} and using $H_0=73$ km s$^{-1}$ Mpc$^{-1}$ and $M_{I{\odot}}=4.08$ and $M_{B{\odot}}=5.48$ from \citet{BinnMerr98}.}
              \label{TFR}
\end{figure}

In Figure \ref{TFR} we compare the Tully-Fisher relation for our (dark blue)\footnote{No asymmetric drift is applied neither to our dEs nor to those of  \citet{VZ04}. This correction would increase $v_{max}$ by 2.5 $\pm$ 0.9 in those rotationally supported systems plotted in Figure \ref{TFR}.} and \citet{VZ04} (light blue) dEs to that of normal late-type galaxies of \citet{Giov97b} (grey symbols) and \citet{DR07} (red dashed area), respectively.
Figure \ref{TFR} clearly shows that these rotationally supported dEs follow the Tully-Fisher relation with a similar scatter as the normal spirals of \citet{DR07} and thus kinematically behave as late-type spirals. The $v_{max}$ of dEs plotted in Figure \ref{TFR} is probably a lower limit since it is generally measured where the rotation curve is still rising, as suggested by the kinematics of the globular clusters. It is thus conceivable that the agreement between the Tully-Fisher relation of rotationally supported dEs and late-type spirals of similar luminosity is even better than that depicted in Figure \ref{TFR}.

\subsection{Dark matter content}

The shape of the rotation curves gives information about the dark matter content and distribution of late-type galaxies \citep[e.g.,][]{Cat06}. Similarly, $\sigma$ can be used to measure the dark matter content of pressure supported systems. Following \citet{Beasley09} we estimate the total dynamical mass of our sample galaxies using the relation: 

\begin{equation}\label{Mtot}
M_{tot}=M_{press}+M_{rot}
\end{equation}

where $M_{press}$ is the mass inferred from the velocity dispersion after the contribution from rotation has been removed and $M_{rot}$ is the mass deduced by the intrinsic rotation velocity of the galaxies. $M_{press}$ inside the half-light radius is defined  as in \citet{Cappellari06}:

\begin{eqnarray}\label{wolf}
M_{press} &\simeq& 2.5G^{-1}\sigma^2R_{eff} \nonumber \\
               &\simeq& 580 \bigg ( \frac{\sigma^2}{{\rm km^2 s^{-2}}} \bigg) \bigg ( \frac{R_{eff}}{{\rm pc}} \bigg)  {\rm M_{\odot}}
\end{eqnarray}

The rotation curves of rotationally supported systems are characterised by an approximately constant gradient suggesting solid body rotation up to the $R_{eff}$. In this case $M_{rot}$ is given by the relation:

\begin{eqnarray}\label{beasley}
M_{rot}&=&\frac{R_{eff}v_{max}^2}{G} \nonumber \\
          &=&\bigg( \frac{R_{eff}}{{\rm pc}} \bigg) \bigg( \frac{v_{max}^2}{{\rm km^2 s^{-2}}}\bigg) \bigg( \frac{1}{4.3\times 10^{-3}} \bigg)  {\rm M_{\odot}}
\end{eqnarray}

Dynamical mass-to-light ratios ($\Upsilon_I$, Table \ref{t6}) are then measured using the $I$-band luminosities and equation \ref{wolf} for pressure supported systems and the sum of equations \ref{wolf} and \ref{beasley} for rotationally supported objects \footnote{Note that this method to obtain $M_{tot}$ is equivalent to introducing the asymmetric drift.}.
Stellar mass-to-light ratios ($\Upsilon_I^*$, Table \ref{t6}) are computed using  the models of single stellar populations (SSP) of \citet{Vaz10} and the ages and metallicities from \citet{Mich08}.

\begin{table}
\begin{center}
\caption{Dynamical and stellar mass-to-light ratios in I-band in solar units.\label{t6}}
\begin{tabular}{|c|c|c|c|}
\hline \hline
Galaxy & $(\Upsilon_I)_{\odot}$    & $(\Upsilon_I^*)_{\odot}$\\
\hline 
PGC1007217   & 4.0 $\pm$ 0.8 & 1.9  \\
PGC1154903   & 1.8 $\pm$ 3.1 & 1.2 \\
NGC3073        &  0.6 $\pm$ 0.4 & 0.2\\
VCC21            &  1.5 $\pm$ 0.4 & 0.5\\
VCC308          &  1.5 $\pm$ 0.3 & 1.1\\
VCC397          &  4.8 $\pm$ 1.2 & 0.9\\
VCC523          &  2.2 $\pm$ 0.2 & 1.3\\
VCC856          &  1.1 $\pm$ 0.2 & 2.0 \\
VCC917          &  2.5 $\pm$ 0.4 & 2.4 \\
VCC990          &  1.9 $\pm$ 0.1 & 2.8 \\
VCC1087      &    2.3 $\pm$ 0.1 & 2.4 \\
VCC1122      &    2.4 $\pm$ 0.1 & 2.2 \\
VCC1183      &    1.9 $\pm$ 0.1 & 1.4 \\
VCC1261      &    2.3 $\pm$ 0.1 & 1.4 \\
VCC1431      &    2.3 $\pm$ 0.1 & 3.4 \\
VCC1549      &    2.1 $\pm$ 0.2 & 3.1 \\
VCC1695      &    1.7 $\pm$ 0.3 & 1.2  \\
VCC1861      &    1.6 $\pm$ 0.1 & 2.8  \\
VCC1910      &    1.6 $\pm$ 0.1 & 3.2 \\
VCC1912      &    2.1 $\pm$ 0.1 & 0.6 \\
VCC1947      &    1.8 $\pm$ 0.1 & 1.7 \\
\hline
\end{tabular}
\end{center}
NOTES: $(\Upsilon_I^*)_{\odot}$ is only indicative, the large errors in the stellar populations \citep{Mich08} make the uncertainties of $(\Upsilon_I^*)_{\odot}$ of the same order as the values.
\end{table}

\begin{figure}
\centering
\resizebox{0.5\textwidth}{!}{\includegraphics[angle=-90]{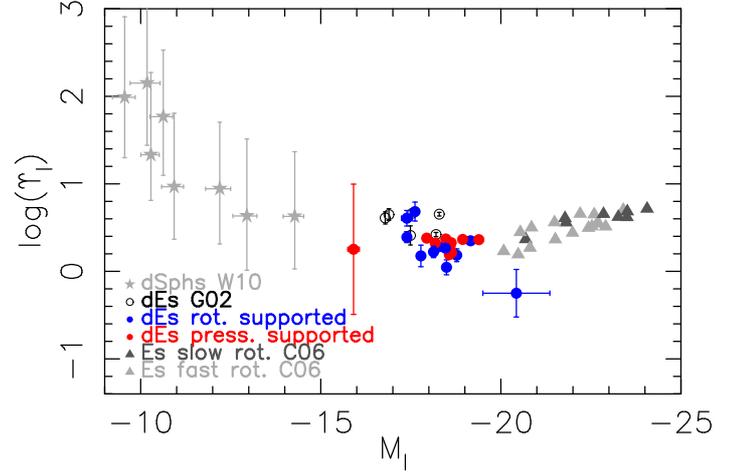}}
\caption{Dynamical mass-to-light ratio as a function of the absolute magnitude in I-band.  Red and blue dots are our pressure and rotationally supported dEs respectively. For comparison open circles are dEs by \citet[][G02]{Geha02}, grey asterisks are dSphs from \citet[][W10]{Wolf10} and dark and light grey triangles are slow and fast rotators respectively from \citet[][C06]{Cappellari06}. The transformation to I band has been performed using the colour-morphology relation by \citet{Fukugita} for Es. For the dwarf galaxies we have used $V-I=1.03\pm0.04$ as calculated by \citet{VZ04b}. 
Note that  NGC3073 in not a dE given its high luminosity ($M_I=-20.4$ mag).}
              \label{MLI}
\end{figure}

Figure \ref{MLI} presents the relation between the dynamical mass-to-light ratio and the absolute I-band magnitude for our galaxies (red and blue dots), the sample of classical elliptical galaxies from \citet[][$\Upsilon_{Jeans}$ in I-band]{Cappellari06}, the dEs from \citet{Geha02} and the Milky Way dwarf spheroidals (dSphs) from \citet{Wolf10}\footnote{The $\Upsilon_I$ values of \citet{Wolf10} have been converted to equation \ref{wolf} for consistency.}. Consistently with  \citet{Zaritsky06} and \citet{Wolf10}, we observe that the total mass-to-light ratio within the $R_{eff}$ of dEs is the lower limit of the decreasing and increasing $\Upsilon_I$ vs. luminosity relations observed in giant ellipticals \citep{Cappellari06} and dSphs \citep{Wolf10}, respectively. Dwarf early-type galaxies have on average $\Upsilon_I=2.00 \pm 0.04$ ${\rm \Upsilon_{I \odot}}$, with a slightly higher dispersion in rotating systems ($RMS=0.06$ ${\rm \Upsilon_{I \odot}}$) than in pressure supported systems ($RMS=0.04$ ${\rm l\Upsilon_{I \odot}}$) .

Our sample of dEs have on average $M_{dyn}/M^*=1.6 \pm 1.2$ (obtained as the ratio between the dynamical and stellar mass-to-light ratios, $\Upsilon_I$ and $\Upsilon_I^*$, respectively, Table \ref{t6}), thus they are not dominated by dark matter within the $R_{eff}$  (as previously suggested by \citet{Geha02,Forbes08}), consistently with what is found is massive ellipticals \citep{Cappellari06} but contrary to dSphs \citep{Wolf10}.

\section{Discussion}

How does this observational evidence compare with the different scenarios of galaxy formation? In a more general context we recall that in the most recent hierarchical models of galaxy formation only the most massive ellipticals have been formed through major merging events \citep{deLucia06}.
The strong morphological segregation observed in high density environments \citep{Sand85,FB94,Blant05} indicates that the cluster environment plays a major role in the formation of dEs.

In \citet{etj09b} we have shown that rotationally supported dEs, characterised by a disky structure, are preferentially located in the field and in the periphery of the cluster, while pressure supported systems are closer to the center. We also found that rotationally supported dEs have, on average, younger stellar populations than pressure supported systems.
This evidence suggests that rotationally supported dEs are low luminosity late-type galaxies which recently entered the cluster and lost their gas because of the interaction with the hostile environment, being transformed, on short time scales, into dEs. It seems thus clear that not all dwarf early-type galaxies are the low luminosity extension of massive ellipticals.

The new kinematic data in our hand support this scenario: rotationally supported systems have rotation curves similar to those of late-type galaxies of similar luminosity and follow the Tully-Fisher relation, the most representative scaling relation for late-type systems. Since the angular momentum of these objects is conserved, the most plausible scenario for gas stripping is the ram pressure exerted by the dense and hot IGM on the fragile ISM of the low luminosity star forming galaxies freshly entering the cluster environment (see \citet{Boselli08a,Boselli08b} for an extensive discussion). 

For these rotationally supported objects gravitational interaction with the cluster potential or with other cluster members (galaxy harassment) can be excluded since they would, on relatively short time scales, reduce the angular momentum of the perturbed galaxies, leading to the formation of pressure supported systems. This process, however, could still be invoked to explain the kinematic and structural properties of the remaining pressure supported dEs populating the core of the cluster (half of our sample), whose statistical analysis of their scaling relations (Fundamental Plane) will be the subject of a future communication.

It is indeed possible that, as for the massive galaxies, gravitational interactions played a major role at early epochs, when the velocity dispersion of the cluster was lower since galaxies were accreted through small groups (preprocesing) than at the present epoch \citep{BG06}.

\section{Conclusions}

We present medium resolution (R$\sim$3800) spectroscopy for 21 dwarf early-type
galaxies, 18 located in the Virgo cluster and 3 in the field. These spectra has been obtained
with the IDS at the INT(2.5m) and with ISIS at the WHT(4.2m) at El Roque de los Muchachos Observatory (La Palma,
Spain). We have used these data to measure the kinematic profiles of these systems and
calculate their maximum rotation velocity as well as their central velocity dispersion. In order to
guarantee the reliability of the data, we have compared our observations with the data of 972 simulated galaxies in 5 different
wavelength ranges corresponding to those used during the observations, and we have run, for
each simulated galaxy, 100 Monte-Carlo simulations.
The comparison between observed data and simulations shows that the adopted data extraction
technique is appropriate for measuring the kinematic parameters of the target galaxies. We have
also shown that velocity dispersions can not be measured for S$/$N ratios below 15, while for
radial velocities with S$/$N$\geq$10 accurate results are obtained.

Our analysis has shown that dEs have on average dynamical M$/$L ratios within the effective radius smaller than those of massive ellipticals and dSphs (in average $log(\Upsilon_I)=0.3 \pm 0.0$ ${\rm log(\Upsilon_{I \odot})}$). We thus confirm that, within the effective radius, dEs are not dark matter dominated objects.

We have found that rotationally supported dEs have rotation curves similar to those of star forming systems of similar
luminosity and  follow the same Tully-Fisher relation. Combined with the evidence that these systems are young objects with disk-like structures generally located in the outskirts of the cluster \citep{etj09b}, these observations 
are consistent with a picture where these rotationally supported dEs result from the transformation of star
forming systems that recently entered the cluster and lost their gas through their interaction
with the environment. The observed conservation of the angular momentum in the rotationally
supported dEs suggests that a milder
ram pressure stripping event as the responsible of the gas removal has to be preferred to more
violent gravitational interactions (harassment) which would rapidly heat up the perturbed
systems. Therefore, all these evidences suggest that dEs are not the low luminosity end of massive early-types because if that was the case all dEs would be  rotating with $v_{max}/\sigma$ higher than those of Es, but a population of non-rotators has also been found and, in addition, the evidences of being stripped late-type spirals are strong enough as to consider it as a possible origin of dEs in clusters.

\begin{acknowledgements}
We thank the MAGPOP EU Marie Curie Training Network for financial support for the collaborating research visits and observations that allowed to make this paper. ET thanks the financial support by the Spanish research project AYA2007-67752-C03-03. We thank Consolider-GTC project for partial financial support.
This paper made use of the following public databases: SDSS, NED, HyperLEDA, GOLDMine.
We are grateful to the anonymous referee for a critical report that has improved the quality of the paper.

\end{acknowledgements}

\bibliographystyle{aa}
\bibliography{references}{}

\begin{appendix}
\section{Quantification of the bias introduced in $v_{max}$}\label{vmax}

\begin{figure}
   \centering
  \resizebox{0.4\textwidth}{!}{\includegraphics[angle=-90]{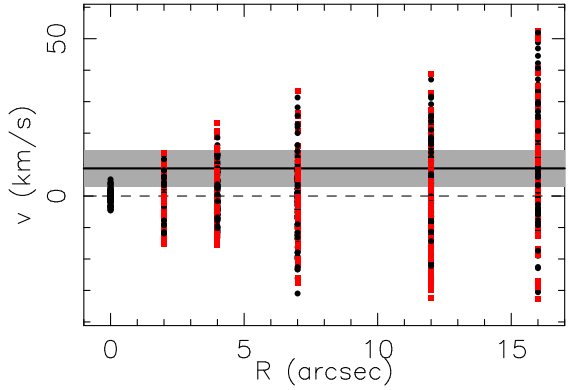}}
  \caption{Simulations performed to quantify the bias that can be introduced in the measurement of $v_{max}$. Plotted are the 100 folded rotation curves computed with random numbers distributed as a Gaussian with width the typical errors of the target galaxies for that radius. Red squares and black dots are the left and right arms of the unfolded rotation curves respectively. The thick black line is the mean $v_{max}$ for all the simulations. The grey shaded area is the scatter for this mean.}
              \label{simul_folded}
    \end{figure}

Using the technique described in Section \ref{central_values} to measure the maximal rotation speed, it seems that none of the galaxies have zero rotation (look at the $v_{max}$ presented in Table \ref{t4}). However, rotation curves like PGC1154903 or VCC1087 (Figure \ref{curves1}) appear to be statistically consistent with no rotation. The fact that we always find positive maximum velocities is a consequence of the method used to measure it. To quantify the bias  introduced by using this technique we have run some simulations. We have taken a zero-rotation object with errors typical of those galaxies that statistically are non-rotators. We have taken a typical rotation curve with 11 bins, at 0, 2$''$, 4$''$, 7$''$, 12$''$ and 16$''$, with symmetrical errors of 2, 5, 7, 10, 12 and 15 km s$^{-1}$ respectively.  
From these errors we have generated 100 simulated rotation curves assigning to each radius a random number with a Gaussian distribution, of which the width is the error associated to that radius. After folding these simulated rotation curves, shown in Figure \ref{simul_folded}, we calculated $v_{max}$ following exactly the same technique as the one we used for the target galaxies. We then obtained a mean value for the 100 values of $v_{max}$ and its scatter, 
9 $\pm$ 6 km s$^{-1}$, shown as a thick black line and a grey shaded area in Figure \ref{simul_folded}. 
As a result, we consider that those galaxies with $v_{max}<9$ km s$^{-1}$ are not rotating, based on our data. For the other galaxies the rotation is significant (three times above the standard deviation, except for VCC1122 and VCC1261, see Table \ref{t4}, 3rd column), so that the systematic bias
described here is much less relevant. For VCC1122 (17.3 $\pm$ 7.7 km s$^{-1}$) and VCC1261 (13.9
$\pm$ 5.2 km s$^{-1}$) the rotation is marginal, as also shown by their low $v_{max}/\sigma$.

\section{Absolute I band magnitudes, optical and half-light radii and ellipticity} \label{photometry}

The $I$-band images for our sample of 21 dwarf galaxies were drawn from Sloan Digital Sky Survey (SDSS, \citet{SDSS}) data release 6 (DR6, \citet{SDSS_DR6}) and converted to Johnson-Cousins systems following Appendix \ref{errors}. 
The photometric parameters were calculated using the IRAF task {\sc ellipse}.

To remove the stars from the images we used the IRAF task {\sc fixpix}. This task allows us to remove the stars interpolating the surrounding galaxy area. To improve the final outcome, we averaged the results of interpolating along the horizontal and vertical directions for each star.
For those galaxies that were on the edge of the FITS images or had a very bright nearby star, a more careful procedure was implemented. Taking advantage of the fact that the galaxies are ellipticals, so that they have smooth and axisymmetric surface brightness profiles, the {\sc bmodel} task of IRAF can be used to replace the affected areas of the galaxy by the azimuthal average of the unaffected ones.  The output from {\sc bmodel} was used only to replace a small fraction of pixels, so this procedure was not affected by the possible presence of more subtle features, such as bars or spiral arms, which could not be reproduced by the model provided by the {\sc bmodel} task. Once extremely bright nearby stars had been removed the same procedure as above was followed with {\sc ellipse} and {\sc fixpix}.

The procedure followed to run {\sc ellipse} is dependent on the parameters we want to measure. First of all we run {\sc ellipse} fixing only the center of the galaxy assuming a step between isophotes of 1 pixel, the rest of the parameters were left free. We also made the masks for the stars to be removed as described above. With the aim of measuring the absolute magnitude, $R_{opt}$ and $R_{eff}$ we run {\sc ellipse} again fixing, the center of the galaxy, the ellipticity and the position angle (PA) to avoid overlap between consecutive isophotes. The adopted $\epsilon$ and PA in this case are the typical values in the outer parts of the galaxy (beyond 1.5-2$R_{eff}$, region where these two parameters stabilise). To measure $\epsilon$ and $C_4$ we run again {\sc ellipse} after removing the stars leaving fixed only the center of the galaxy.

Asymptotic magnitudes and the radii were derived as in \citet{GdP07}. We first computed the accumulated flux and the gradient in the accumulated flux (i.e., the slope of the growth curve) at each radius, considering as radius the major-axis value provided by {\sc ellipse}. After choosing an appropriate radial range, we performed a linear fit to the accumulated flux as a function of the slope of the growth curve. The asymptotic magnitude of the galaxy was the Y-intercept, or, equivalenly, the extrapolation of the growth curve to infinity. Once  the asymptotic magnitude was known, the optical and effective radii of each galaxy were obtained as the major-axis of an elliptical isophote containing 83$\%$ and 50$\%$ of the total flux respectively. For the asymptotic magnitudes different sources of error have been considered (see Appendix \ref{errors}). The resulting uncertainty is $\sim$0.02 mag.

The ellipticities ($\epsilon$) were measured as the mean value between 3$"$ and
the $R_{eff}$, the galaxy region covered by our spectroscopic observations.

\section{Errors in magnitudes}\label{errors}

The zero points (ZP) and the errors in the $i$-band magnitudes have been computed as described in SDSS documentation. The ZP have been obtained from $F_0$, the flux a source produces in counts per second in the image, calculated as a function of three parameters ($aa$, $kk$ and $airmass$) defined as:
\begin{equation}
F_0=\frac{t_{exp}}{10^{0.4(aa+kk\times airmass)}}
\end{equation}
where the exposure time ($t_{exp}$) is the same for all the SDSS images (53.91 seconds).
The uncertainties in the $i$-band magnitudes are affected by different sources of error: firstly, the errors in the flux, that can be calculated following the equation:
\begin{equation}
\Delta F=\sqrt{\frac{F+sky}{gain}+N_{pix}(dark~variance+\Delta sky)}
\end{equation}
where $F$ is the total flux in counts, the $sky$ and $\Delta sky$ are the background sky and its error (in counts), the $gain$ and the $dark~variance$ are given in the header and $N_{pix}$ is the number of pixels in the largest aperture where the flux is measured. This error was typically $10^{-3}$mag. Other error sources are the error introduced in the fit to the growth curve (between $10^{-3}$mag and $6\times10^{-3}$mag), and the error due to photometric zero point differences between the different scans of SDSS, which might lead to an error of 0.01 mag. 
SDSS $i$-band magnitudes are not exactly in the AB system, so an error of 0.01 mag might also be introduced (see SDSS documentation about the photometric flux calibration). And finally, we have transformed our data from the SDSS $i$-band to the Johnson-Cousins $I$-band assuming $m_I=m_i-0.52\pm0.01$ mag \citet{Fukugita} given that their $r-i$ colour ranges from 0.23 mag to 0.57 mag.

Adding quadratically all these sources of error, the final estimated error is 0.02 mag for the apparent $I$-band magnitudes.

\section{The $C_4$ Boxyness/Diskyness parameter}\label{appC4}

The boxyness/diskyness parameter is defined as the fourth moment 
in the Fourier
series as follows 
\begin{equation} \label{C4}
I(\Phi)=I_0+\sum_{k}[S_ksin(k\Phi)+C_kcos(k\Phi)]
\end{equation}
$I(\Phi)$ is the intensity measured in each isophote. The first two moments in
this series describe completely an ellipse. Higher order moments ($k$ $\ge$ 3)
define deviations of the isophotes from ellipses. The third order moments
($S_3$ and $C_3$) represent isophotes with three fold deviations from ellipses
(e.g. egg-shaped or heart-shaped), while the fourth order moments ($S_4$ and
$C_4$) represent four fold deviations. Rhomboidal or diamond shaped isophotes
have nonzero $S_4$. For galaxies that are not distorted by interactions, $C_4$
is the most meaningful moment indicating the disky/boxy shapes of the
isophotes  (see Figure 1 from \citet{Pel90} for an example of these different shapes).

   \begin{figure}
   \centering
  \resizebox{0.4\textwidth}{!}{\includegraphics[angle=-90]{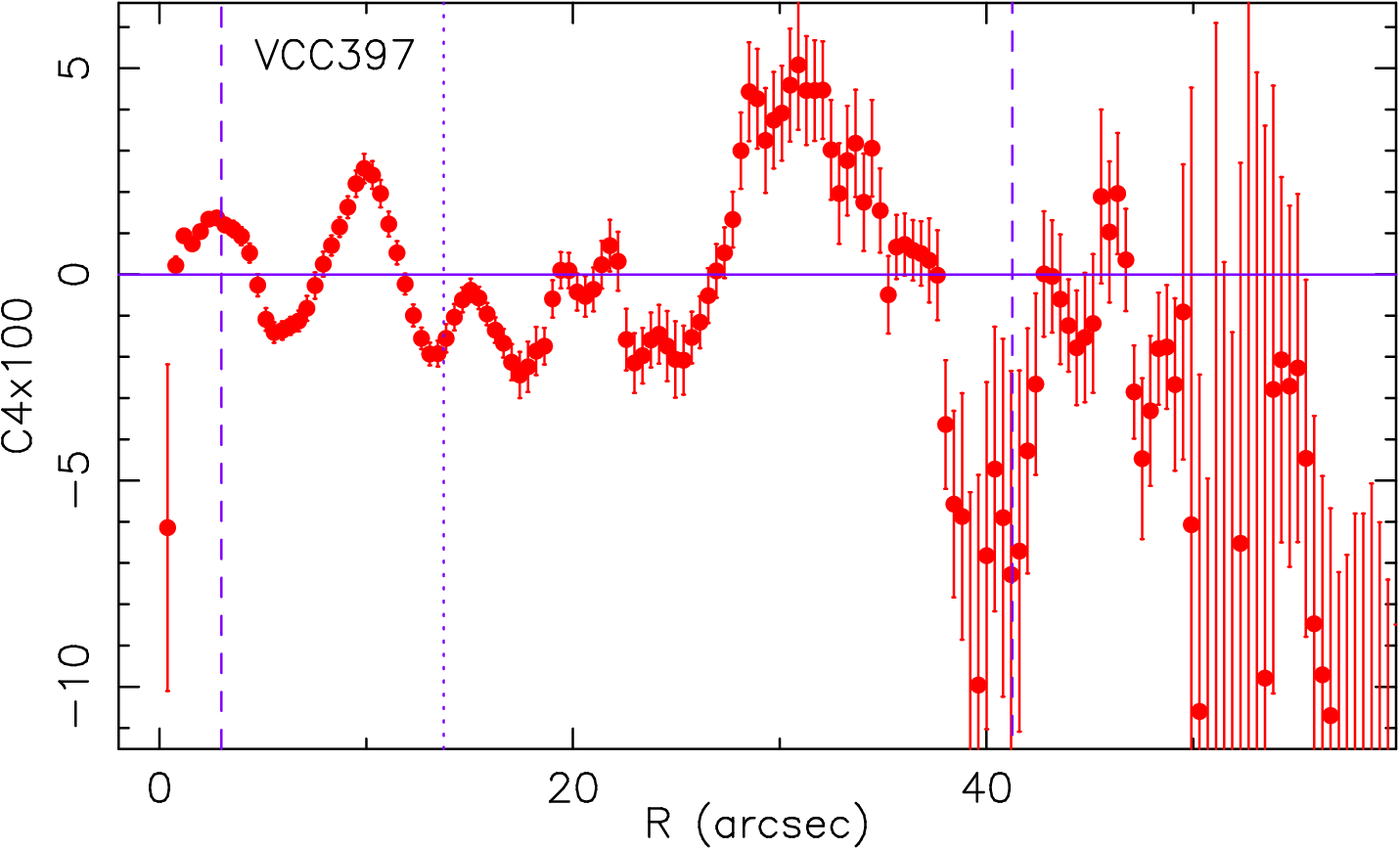}}
  \resizebox{0.4\textwidth}{!}{\includegraphics[angle=-90]{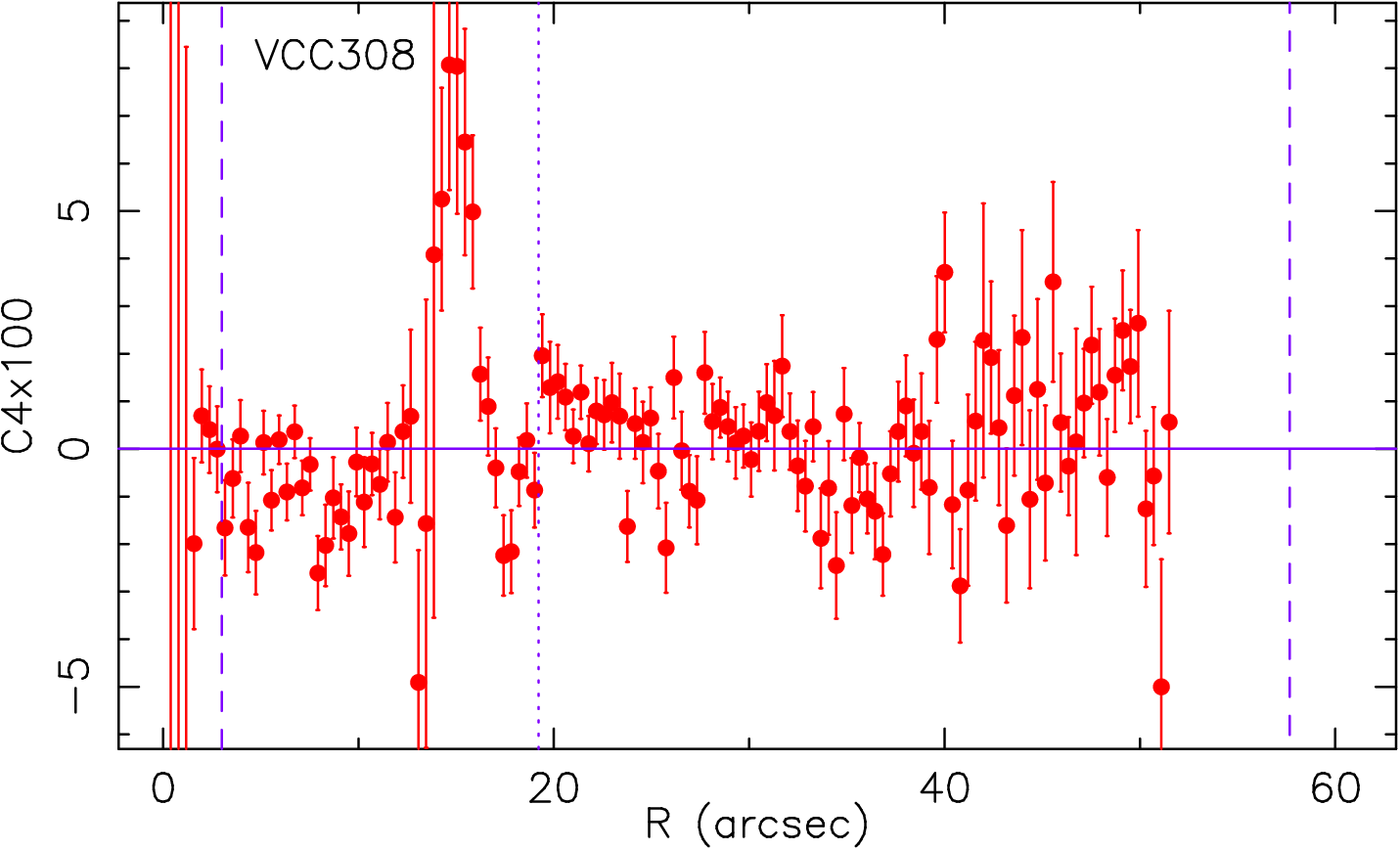}}
 \resizebox{0.4\textwidth}{!}{\includegraphics[angle=-90]{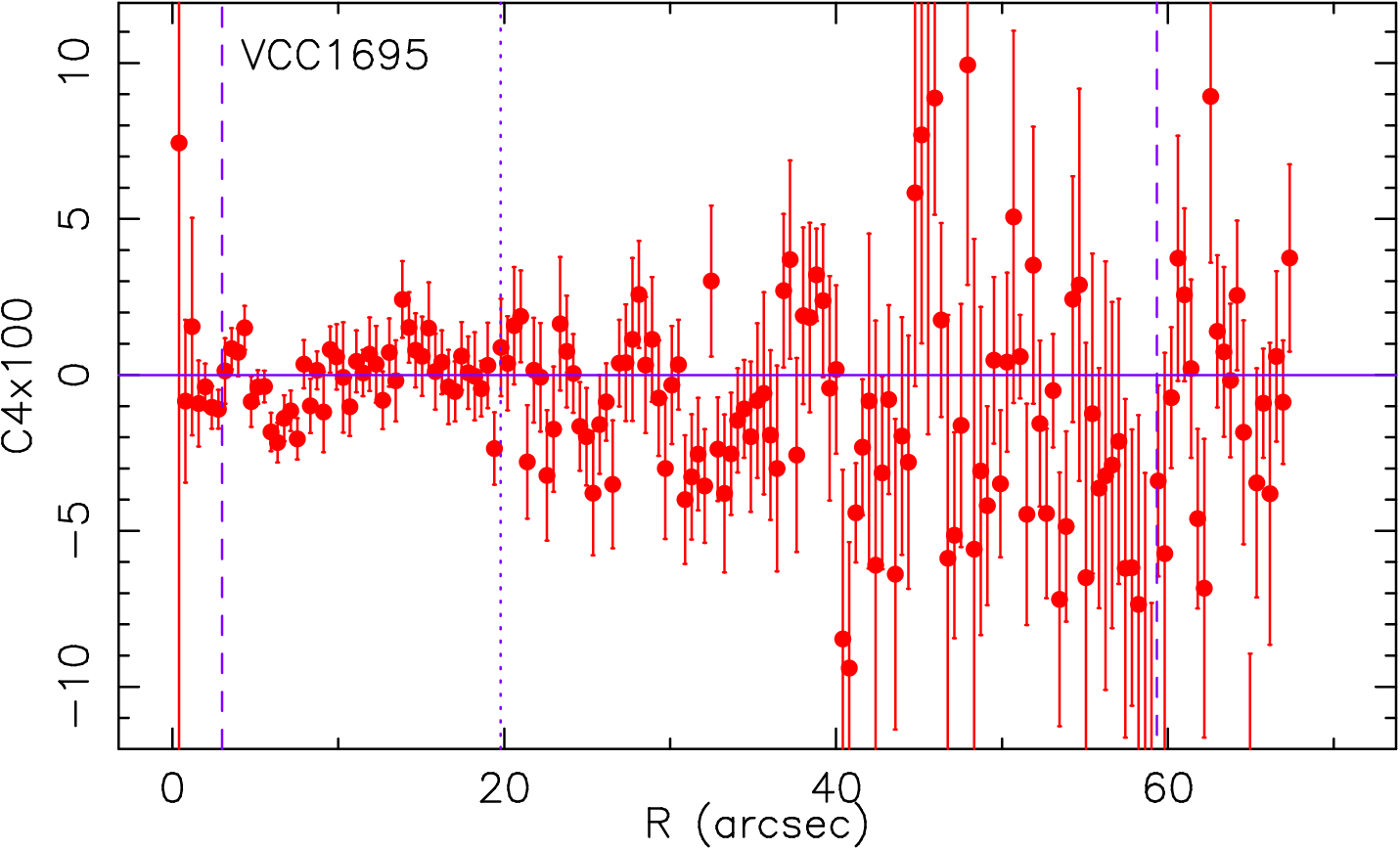}}
 \caption{ Examples of $C_4$ vs. radius for three galaxies: VCC397, VCC308 and VCC1695. The dashed purple lines indicate the region between
3$"$ and the 3$R_{eff}$. The dotted purple line shows the $R_{eff}$. In the upper panel one can see that if an average value is used between the dashed lines, $C_4$ will be compatible with zero and as a consequence, the prominent disky structure will be smeared out by the adjacent regions. In contrast, in the middle panel a galaxy with no clear disky structures is shown. In this latter case the errors are larger.  The bump in $C_4$ only covers $\sim$ 3$"$, while the rest of the galaxy is boxy. In this situation a mean value of the $C_4$ and its scatter is more representative. In the lower panel VCC1695 is an example of a typical boxy shaped galaxy.}
              \label{C4_examples}
    \end{figure}

$C_4$ is measured in $i$-band SDSS images using {\sc ellipse} that performs equation \ref{C4} along the radius of the galaxy fixing only the center of the galaxy and leaving the rest of {\sc ellipse} parameters free, as described in Section \ref{C4_subsection}. Figure \ref{C4_examples} shows $C_4$ as a function of radius for 3 dEs.  Due to the large changes of $C_4$ with radii taking an averaged value is therefore not the best way to detect disks in these galaxies, especially if they cover only a limited range in radius. We have thus adopted the following procedure:  

If at least one prominent bump is detected, which has a width larger than $\sim$6$"$ inside three effective radii (above this radius the scatter of $C_4$ and its error becomes too large as to be reliable), we consider the galaxy to be disky, and assign the maximum $C_4$ between 3$"$ and 3$R_{eff}$ to the global $C_4$. The error in this measurement has been estimated by dividing the photometric error of the maximum of $C_4$ by the square root of the number of points that describe the disky structure in order to quantify the reliability of the bump considered. If the bump is described by a large number of points it is highly likely that the bump is truly there and as a consequence the error will be small, but if the number of points is small but the photometry is of high quality then the error will be small again. In any other case the error will be large and the result must be used cautiously.

Otherwise, if the values oscillate around $C_4$ $\sim$ 0 (lower panel of Figure \ref{C4_examples}), are 
always negative or if there is a bump with a 
small radial coverage (see middle panel Figure \ref{C4_examples}), we assign a mean value and its scatter between 3$"$ and  3$R_{eff}$ to the global $C_4$. In this case, the RMS quantifies simultaneously the quality of the photometry and the possible presence of small bumps (as it is the case of VCC308, middle panel of Figure \ref{C4_examples}).

The results obtained for this parameter are listed in Table \ref{t5} and plotted vs. the anisotropic parameter, $(v_{max}/\sigma)^*$, in Figure \ref{vs_C4}. The agreement between the $C_4$ classification in disky/boxy galaxies and the morphological classification from \citet{Lisk06a} is evident but apart from three red dots. These filled circles 
correspond to VCC917, a rotationally supported galaxy (above the horizontal dashed line) with strong disky isophotes but no structure found by \citet{Lisk06b}); VCC1122, a not rotationally supported dE but with an appreciable rotation in Figure \ref{curves1} and very important disky structures in the inner $R_{eff}$ not detected by \citet{Lisk06b}); and VCC1912,  inside half the effective radius a moderate rotation is found in this system with a clear disky feature that peaks at 1.5$R_{eff}$. For this galaxy, however, no underlying structure was found by \citet{Lisk06b}. More importantly VCC308 and VCC856, two rotationally supported galaxies with boxy $C_4$, present prominent spiral arms in \citet{Lisk06b}. In \citet{Ferrarese06}, based on ACS-HST images, VCC856 also shows spiral arms but their analysis of the isophotes' shapes shows that they are boxy too. Looking at Table \ref{t5} we see that both galaxies are nearly face-on, which means that the isophotes are boxy since face-on disks are round and not disky. As a consequence we emphasise the fact that boxy isophotes could miss disk features (mainly if the galaxies are face-on), but not the other way round (see VCC917).

\end{appendix}

\end{document}